\documentclass[aps,nofootinbib,showpacs,preprintnumbers]{revtex4}
\usepackage{epsfig}

\begin{document}
\preprint{USM-TH-155, hep-ph/0406028, v3}
\title{Pad\'e-related resummations of the pressure of 
quark-gluon plasma by approximate inclusion of $g_s^6$-terms}

\author{G.~Cveti\v{c}}
  \email{gorazd.cvetic@usm.cl}
\affiliation{Department of Physics, Universidad T\'ecnica
Federico Santa Mar\'{\i}a, Valpara\'{\i}so, Chile}
\author{R.~K\"ogerler}
 \email{koeg@physik.uni-bielefeld.de}
\affiliation{Department of Physics, Universit\"at Bielefeld, 
33501 Bielefeld, Germany}

\date{\today}

\begin{abstract}
We perform various resummations of the hot QCD pressure
based on the actual knowledge of the perturbation
series which includes the $\sim g_s^6 \ln(1/g_s)$
and part of the $\sim g_s^6$ terms. Resummations
are performed separately for the short- and long-distance 
parts. The $\sim g_s^6$ term of the short-distance
pressure is estimated on the basis on the known
UV cutoff dependence of the long-distance part.
The resummations are of the Pad\'e and
Borel-Pad\'e type, using in addition the
(Pad\'e-)resummed expression for the squared
Debye screening mass $m_{\rm E}^2$ and,
in some cases, even for the EQCD coupling
parameter $g_{\rm E}^2$.
The resummed results depend weakly on the
yet unknown $\sim g_s^6$ terms and on the
the short-range renormalization scale,
at all temperatures. The dependence on the
long-range renormalization scale is appreciable
at low temperatures $T \alt 1$ GeV.
The resulting dependence of pressure on temperature $T$
is compatible with the results of the lattice calculations
at low $T$.

\vspace{0.5cm}

\end{abstract}

\maketitle

\section{Introduction and Description of the Approach}
\label{sec:introduction}

Today we have at our disposal a well-elaborated technique for perturbatively
treating field theory at finite temperature and/or chemical potential
\cite{Kraemmer}. This
formalism goes far beyond the ordinary $(T = 0)$ perturbation theory insofar
as a correct treatment of infrared divergences (specifically those which 
are connected with $T \not= 0$) requires partial resummation of infinitely
many specific diagrams. This implies -- among other things -- 
that the infrared
convergent perturbative expressions come out as series in $g_s$ (the
fundamental coupling constant) rather than in $g_s^2$ [or $a = 
g_s^2/(4 \pi^2)$].

During the last decade, several physical quantities, the most prominent example
being the free energy density $F$ of the quark-gluon plasma, 
have been calculated within this formalism
up to ${\cal O}(g_s^5)$ \cite{Arnold,Braaten:1996ju} and (partially) even to 
${\cal O}(g_s^6)$ \cite{Kajantie:2002wa,Kajantie:2003ax}
Disappointingly, in spite
of the relatively high orders available, the results are of very limited
applicability even at very high temperatures 
($T \agt 1$ TeV) where $g_s$
certainly becomes small. In fact, if successive terms in the perturbative
series are added, the corresponding truncated sum changes dramatically,
jumping up and down. Furthermore, the unphysical dependence on the
renormalization scale is strong and seems to become even stronger
with increasing order (the renormalization scale in
${\overline {\rm MS}}$ scheme will be denoted as 
${\overline \mu}$, and in a general scheme as $\mu$).  
Both effects considerably reduce the reliability
of the perturbative results for representing physical quantities and
have therefore been object of intensive theoretical studies. 

Ways out
of the convergence dilemma are in general looked for by either 
performing some clever resummations or by
reorganizing perturbation theory in some way (actually,
in some cases, this amounts to the same thing) \cite{Kraemmer}. 
Several specific 
approaches have been applied up to now. Among them are quasi-particle
models \cite{Biro}, $\Phi$-derivable 
approximations \cite{Blaizot}, screened \cite{Karsch} 
or optimized \cite{Chiku}
perturbation theory, hard thermal loop perturbation theory \cite{Andersen}.

Within the present paper we concentrate mainly on the question of 
renormalization scale (RS ${\overline \mu}$) dependence of the
resummed results, in ${\overline {\rm MS}}$ scheme, as well
as on the dependence of the results on the yet unknown
part of the  $g_s^6$-term in truncated perturbation series.
The resummation approaches mentioned above do have
some residual renormalization scale dependence.
Furthermore, the conventional choice for the 
RS ${\overline \mu}$ at finite $T$ (${\overline \mu} \simeq 2 \pi T$) 
is not natural since different energy scales get involved in all 
calculations (see later).

In order to improve this situation, we replace the (partially resummed) 
perturbation series by approximants which are more stable under the 
variation of ${\overline \mu}$. The Pad\'e approximants (PA's) 
\cite{Baker,Gardi:1996iq,Brodsky:1997vq,PA} and other
Pad\'e-related resummations, such as Borel-Pad\'e (BPA's) \cite{BPA} 
and modified Baker-Gammel approximants (mBGA's) 
\cite{Cvetic:1997ca,Cvetic:2000mh},
are known in general to reduce the unphysical ${\overline \mu}$-dependence 
significantly (mBGA's even entirely).
Usually these approximants are applied to a given 
truncated perturbation series 
(TPS) of a given order, and they fulfill in addition the
``minimal'' requirement: upon re-expanding them in powers
of the coupling parameter, they reproduce the TPS to the given order.
In this sense they are sometimes considered to be
some sort of mathematical artifice without any deeper physical motivation.
This opinion is delusive, however, because of several reasons: 

\begin{enumerate}
\item
The basis of all field theoretic approaches is the path
integral expression for the generating functional which has to be attributed a
specific mathematical meaning. In ordinary perturbation theory this is
achieved by Taylor expanding the corresponding exponential of the
interaction Lagrangian leading to the well-known power series. But this is
by no means better or more natural than by approximating it in terms of other
approximants (e.g., Pad\'{e}) -- we are only less used to it. In some
sense Pad\'{e} or Pad\'{e}-related sequences seem even more adapted than
power series since they allow for (pole) singularities, 
and we know that most physical amplitudes do include singularities.

\item
The asymptotically divergent nature of the expansion in powers of 
the coupling parameters represents a practical problem
concerning the actual evaluation, and calls for, among other
things, nonperturbative information in order to fix the 
renormalon ambiguities and to produce unique predictions. On the other
hand, due to the generally better convergence of Pad\'{e}
and Pad\'e-related approximants,
one hopes that the convergence of the corresponding sequence 
of the (Pad\'e-)resummed TPS's will
improve and that it would converge to a
specific prediction. The Pad\'e sequence, in fact, with increasing
order shows convergence under rather general 
conditions even when the corresponding TPS is (asymptotically) divergent  
\cite{Baker}.

\item
It can be shown that PA's (and mBGA's) can be represented by
weighted averages of running coupling parameter (one-loop running in the
case of PA, exact running in the case of mBGA) at specific values of
reference momenta \cite{Doering,Brodsky:1997vq}, 
and are thus a realization of the 
fact that each physical quantity in field theory is characterized 
by specific values of the momentum flow. 
In this sense they represent different approximations to the Neubert's
formalism \cite{Neubert:1994vb} of renormalization improved 
perturbation theory.
\end{enumerate}

{}From the diagrammatic point of view, PA and Pad\'e-related
methods represent a resummation of those (infinitely many) 
diagrams which -- when added -- approximately (or exactly)
cancel the ${\overline \mu}$-dependence of the given quantity.

PA's, BPA's and mBGA's have been applied to TPS for several QCD or QED
quantities at zero temperature 
(see, for example, Refs.~\cite{PA,BPA,Cvetic:2000mh}) 
and have often led to significant improvement of the RS-dependence problem.
Recently, same types of approximation have also been used for similar
purposes in finite temperature gauge theory: PA's in
Refs.~\cite{Kastening:1997rg,Cvetic:2002ju};
somewhat related Borel methods in Refs.~\cite{Parwani:2000rr},
by discerning some information on renormalons (for renormalon properties
in $\phi^4$ theory at finite $T$, see Ref.~\cite{Loewe:1999kw}).
The first PA resummations at finite-$T$ \cite{Kastening:1997rg} 
consisted in simply replacing the available power series
(in powers of $g_s$) for the entire QCD free energy by various
PA's. Although the results demonstrate a weakened
${\overline \mu}$-dependence, 
we regard some aspects of this approach as problematic.
Our reservation is due to the simple observation that two ingredients --
(a) the infrared stable TPS of thermal field theory, and (b)
the Pad\'{e}(-related) approximations applied to this TPS
-- imply resummation of infinitely many terms. The
corresponding classes of diagrams are, however, neither equivalent nor
disjunct. Therefore, care has to be taken to avoid double counting and
to disentangle the various resummations. 

Recently, we have developed a 
formalism to consistently treat this problem \cite{Cvetic:2002ju}.
The procedure is the following: Consider some physical quantity and its
TPS (to a given order) in thermal field theory. We specifically have in
mind the static pressure $p_{\rm QCD}$ of the QCD plasma,
i.e., the negative of the free energy density $F_{\rm QCD}$. It is
connected to the partition function ${\cal Z}$ by the relation
\begin{eqnarray}
p_{\rm QCD} &=& -F_{\rm QCD}  =  
\lim_{V \to \infty} \frac{T}{V} \ln {\cal Z} \ ,
\label{pQCD1}
\\
{\cal Z} &= & \int {\cal D} A_\mu^a {\cal D} \psi {\cal D} \bar \psi 
\exp \left( - \int^{1/T}_0 d \tau
\int d^3 x {\cal L}_{\rm QCD} \right) \ ,
\label{Z}
\end{eqnarray}
where ${\cal L}_{\rm QCD}$ is the (Euclidean) QCD-Lagrangian density,
$T$ is the temperature, $V$ is the three-dimensional volume,
and the renormalization convention $p_{\rm QCD}(T \! = \! 0) = 0$ is taken;
$g_s$ is the QCD gauge coupling parameter. 
The corresponding TPS has the generic structure
\begin{equation}
p = p_{\rm ideal} \left[ C_0 + C_2 g_s^2 + C_3 g_s^3 + C_4 g_s^4 + 
C_5 g_s^5 + C_6 g_s^6 + ...\right] \ ,
\label{pexp}
\end{equation}
where the coefficients $C_i$ may include contributions of order $\ln g_s$.
The terms proportional to odd powers of $g_s$ [or, equivalently, fractional
powers of $a \equiv g_s^2/(4 \pi^2)$] are exclusively due to resummation
of specific (ring, or daisy) diagrams, but the latter also contribute to terms
proportional to even powers of $g_s$. Clearly $g_s$ and the coefficients $C_i$
depend on the chosen renormalization scheme and, in particular, on the
renormalization scale ${\overline \mu}$ ($\propto T$). 
Due to the asymptotic freedom we expect that $g_s$
becomes sufficiently small at large temperature $T$. In the following, we
concentrate on such a large-$T$ situation. 

The first step needed for the application of the formalism of
Ref.~\cite{Cvetic:2002ju} is the separation in
Eq.~(\ref{pexp}) of the purely perturbative part from the
contributions stemming from the (ring-)resummations.
This can be done in an unambiguous 
and consistent way, since the resummation terms
represent exclusively the contributions of the (bosonic) zero modes to
the considered physical quantity. Note in this respect that resummation is
needed purely for taming the finite-$T$ infrared divergences, and only
the infrared parts of the higher order diagrams contribute to the
(finite) order terms. Therefore, the identification of the
contributions from the zero modes (long-range contributions) is needed. 
This can be achieved either by integrating out directly the 
high momentum regime or, more elegantly, by utilizing the effective field 
formalism \cite{Braaten:1995cm,Braaten:1996ju}
based on dimensional reduction method
Refs.~\cite{Ginsparg:1980ef,Kajantie:1995dw,Braaten:1995cm,Braaten:1996ju}.
The idea behind is that (at high enough temperature, with $g_s$ 
sufficiently small) the four-dimensional thermal QCD 
at length scales $\agt 1/(g_s T)$
is equivalent to an effective
three-dimensional field theory of (static) boson fields.
This effective theory represents the physics of the 
zero bosonic modes and thus reproduces all static correlations of the original
QCD at the aforementioned distances. Consequently, the contribution of the 
zero modes to the partition functions or to the pressure can
be calculated  by means of the effective theory. 
Within QCD the long-distance
part can be further subdivided corresponding to the two regimes
$R \sim 1/(g_s T)$ (determined by color-electric
screening) and $R \simeq 1/(g_s^2 T)$ (color-magnetic screening). 
Denoting the corresponding contributions to $p$ (or $F$) by the
subscripts M and G, respectively,  the final decomposition 
corresponding to the three energy regions is
\begin{equation}
p = p_{\rm E} + p_{\rm M} + p_{\rm G} \ .
\label{pdecomp}
\end{equation}
Here, $p_{\rm E}$ denotes the short-distance,
and $p_{\rm M} + p_{\rm G} \equiv p_{\rm M+G}$ 
the zero-mode long-distance contribution.

Now comes one of the main points of the formalism of 
Ref.~\cite{Cvetic:2002ju}: Having the separation (\ref{pdecomp})
of the pressure into $p_{\rm E}$ and $p_{\rm M+G}$,
we can argue that both are separately physical quantities,
and thus both, when calculated to all orders, are 
separately ${\overline \mu}$-independent. The reason for it is shown by the following
indirect argument: Consider any (static) two-point-correlator within the
given field theory. It is a measurable quantity and so is its long-range
behaviour. But the latter is determined solely by the exchange of zero modes
(note that the exchange of a mode with frequency $\omega_n$ contributes a
term to the correlator which vanishes like $\exp(-|\omega_n|R)$ at 
$R \to \infty$, thus
the only surviving contribution at large distances is the one with
$\omega_0 = 0$). Thus the contribution of $\omega_0$ 
to each correlator is measurable
(and therefore ${\overline \mu}$-independent) once we have fixed,
by convention, the minimal value $R_{\rm min}$ of what
we consider large distances (or: factorization scale
$\Lambda_{\rm E} \sim 1/R_{\rm min}$).
Since these correlators are derived
from the partition function ${\cal Z}$ by means of functional derivatives,
even the zero-mode contribution to ${\cal Z}$, and 
thus also to $p$, is ${\overline \mu}$-independent (physical). Consequently,
the remaining part of the measurable pressure --
the short-range contribution $p_{\rm E}$
represented by the ordinary perturbative terms
-- must also be ${\overline \mu}$-independent. 
The RS-independence of both parts
separately has been demonstrated analytically in Ref.~\cite{Cvetic:2002ju}
by using the known perturbation series (up to order $g^5$).

The separation and the ${\overline \mu}$-independence of the separate terms
allows a more consistent application of Pad\'e and Pad\'e-related
approximations: they are applied to TPS's of each quantity
$p_{\rm E}$ and $p_{\rm M+G}$ separately.\footnote{
Since only little information is available about $p_{\rm G}$,
we will find it more convenient to apply Pad\'e(-related)
resummations to $p_{\rm M+G}$ and not separately to
$p_{\rm M}$ and $p_{\rm G}$.} We can illustrate the importance of
using separate approximations  with 
a simple example -- the Pad\'e-resummation
of a quantity $S \equiv (S_1\!+\!S_2)$,
where $S_1$ and $S_2$ are separately physical quantities,
each of them available as power series of $a$ up to 
next-to-leading order (NLO):
\begin{equation}
S_j  = a \left( 1 + r_{(j)} a \right) + {\cal {O}}(a^3) 
\qquad (j\!=\!1,2)
\label{Sjs} \ .
\end{equation}
When we apply to the TPS of the sum $S$
a PA, say $[1/1](a)$, and then expand this back in
powers of the coupling $a$, we obtain
\begin{eqnarray}
S^{[1/1]} &=&  2~a \left[ 1 - \frac{1}{2}(r_{(1)}\!+\!r_{(2)}) a
\right]^{-1}
 = 2~a \left[ 1 + \frac{1}{2}(r_{(1)}\!+\!r_{(2)}) a
+ \frac{1}{4} ( r_{(1)}^2 + r_{(2)}^2 + 2 r_{(1)} r_{(2)} ) a^2 
\right] + {\cal {O}}(a^4) \ .
\label{SP11}
\end{eqnarray}
The coefficient at $\sim a^3$ here
has a term $2 r_{(1)} r_{(2)}$. Therefore, the Pad\'e-resummed
$S$ contains mixing effects at $\sim a^3$, i.e.,
an interference effect between the two amplitudes for the 
physical observables $S_1$ and $S_2$. 
This is not acceptable, because $S$ is
the (incoherent) sum of $S_1$ and $S_2$.
Therefore, the PA should not be applied to the entire
sum $S$, but separately to $S_1$ and $S_2$.
This argument holds also when different PA's
or Pad\'e-related resummations are applied,
and when the order of the TPS is higher.

One of the consequences of the separate treatment of
$p_{\rm E}$ and $p_{\rm M+G}$ is that the 
natural renormalization scale ${\overline \mu}$ used 
in these two quantities should be of the order of the 
energy of the modes contributing to them: 
${\overline \mu} \sim 2 \pi T$ in $p_{\rm E}$, and
${\overline \mu} \alt g_s T$ in $p_{\rm M+G}$. 

Within the present paper we will apply Pad\'e (PA) and 
Borel-Pad\'e approximants (BPA's) to the evaluation of
$p_{\rm QCD}$. The modified Baker-Gammel
approximants (mBGA's), which are ${\overline \mu}$-independent
\cite{Cvetic:1997ca} or even renormalization scheme
independent \cite{Cvetic:2000mh}, are technically more involved.
An analysis with mBGA's, which at least could serve as some kind of
quality control of the general procedure, will be presented elsewhere. 
On the other hand, BPA's are also applied here and they represent
an extension of the Pad\'e analysis: 
PA's are applied to the corresponding TPS's of the
Borel transforms, and then the resummed quantity is obtained
by Borel integration. This procedure gets its
motivation from the hope that the Borel summation might defuse 
the notorious divergence problem of perturbation theory,
as well as from the fact that the
power expansion of the Borel transform
has significantly better convergence properties
than the original series and is thus more amenable
to the Pad\'e-type resummations.

One technical remark should be added here -- the
separation of the energy range into a high and a low
energy region requires introduction of a
factorization scale $\Lambda_E$ which
defines a boundary between the two:
$2 \pi T >  \Lambda_E   >    g_s T$.
Consequently, the two contributions separated in the described way
acquire an artificial dependence on $\Lambda_{\rm E}$,
although the sum of the two terms $p_{\rm E} + p_{\rm M+G}$ has to be 
$\Lambda_{\rm E}$-independent.
Similarly, for the individual terms $p_{\rm E}$ and
$p_{\rm M+G}$ to have physical meaning themselves, 
the factorization scale should be suitably chosen.
Technically this means that the approximants for $p_{\rm E}$ and
$p_{\rm M+G}$, although independent of each other,
should be chosen such that the $\Lambda_E$-dependence is minimized.

In our previous paper \cite{Cvetic:2002ju} we applied 
this procedure to calculation of the 
free energy (pressure) both in QCD with $n_f$ (massless) quarks and in a 
$\phi^4$-theory. At that time the relevant TPS's had been calculated only
up to terms of order $g_s^5$ with the following implications for the 
perturbative structure of the relevant contributions: 
The short-distance term
$F_{\rm E}$ ($= p_{\rm E}$) is determined perturbatively up to NLO 
[in $a \equiv a({\overline \mu}) \equiv g_s^2({\overline \mu})/(2 \pi)^2$] 
\begin{eqnarray}
F_{\rm E}/F_{\rm ideal}& =& p_{\rm E}/p_{\rm ideal} =
1 - B(n_f) \widetilde{F}_{\rm E} \ ,
\label{FElow}
\\
\widetilde{F}_{\rm E} &=& a \left\{ 1 + 
C_{\rm E} (n_f, \Lambda_{\rm E}, {\overline \mu}) a \right\} \ ,
\label{tFElow}
\end{eqnarray}
and so is the electric (Debye) screening mass $m_{\rm E}$
\begin{equation}
\widetilde{m}_{\rm E}^2 \equiv
\frac{1}{ 4 \pi^2 T^2 } \frac{1}{(1 + n_f/6)} m_{\rm E}^2
= a \left\{ 1 + C_{\rm M} (n_f, {\overline \mu}) a \right\}
\label{tmE2}
\end{equation}
The long-distance part was a power series in 
$g_s$ [$\equiv g_s({\overline \mu})$] up to order $g_s^2$
\begin{eqnarray}
F_{\rm M} &=& - p_{\rm M} =
- \frac{2}{3 \pi} T m^3_{\rm E} \, \widetilde{p}_{\rm M} \ ,
\label{FMlow}
\\
\widetilde{p}_{\rm M} &=& 1 + 
C_{\rm M1} (n_f, \Lambda_{\rm E}, {\overline \mu}) g_s + 
C_{\rm M2} (n_f) g_s^2 \ .
\label{tpMlow}
\end{eqnarray}
We refer to Ref.~\cite{Cvetic:2002ju} for compilation of
explicit expressions for $B$, $C_{\rm E}$, $C_{\rm M}$,
$C_{\rm M1}$, $C_{\rm M2}$.     
These are rather short power series and they allow construction of only
low order Pad\'e approximants, e.g., PA $[1/1](a)$ for $F_{\rm E}$ and 
$m_{\rm E}^2$, and PA $[0/2](g_s)$ for ${\widetilde p}_{\rm M}$. 
In addition, the TPS for ${\widetilde p}_{\rm M}$ 
($= -{\widetilde F}_{\rm M}$)
is very strongly divergent, and the TPS's for ${\widetilde F}_{\rm E}$
and for $\widetilde{m}^2_E$ are divergent to a somewhat lesser extent.
Therefore, resummation results based on these TPS's should
be taken with care. Nevertheless, the application of these low order
approximants yielded results which were fairly stable under the variation of
${\overline \mu}$, although -- at low temperatures $T \alt 10$ GeV
-- they deviated substantially from the lattice results \cite{Boyd,Karsch:2000ps,Papa}.

In such a situation, an additional order in the perturbation series
(i.e., terms of order $g_s^6$) is much more than an additional tiny
correction, but constitutes a significant enlargement of the basis for 
Pad\'e approximations, 
since it adds additionally both to $F_{\rm E}$ and $F_{\rm M}$
and (for the first time non-vanishingly) to $F_{\rm G}$. 
Therefore, the calculation
of the $\sim g_s^6~\ln (1/g_s)$ contribution to the
long-range part of the pressure published in Ref.~\cite{Kajantie:2002wa} is
extremely gratifying. The full $O(g_s^6)$ contribution cannot be
achieved perturbatively because of the well-known breakdown of perturbation
theory due to incurable infrared divergences occurring at this order
\cite{Gross:1980br,Linde:px}. What
can be evaluated is the coefficient of the logarithmic ultraviolet divergence
contained in $F_{\rm G}$, because of the superrenormalizability of the 
corresponding effective three-dimensional field theory; and this is exactly the
coefficient of the ${\cal O}\left( g_s^6 \ln (1/g_s) \right)$-term 
in $F_{\rm G}$. The term purely
proportional to $g_s^6$ remains unspecified and has to be treated as a free
parameter unless information from nonperturbative methods (e.g., lattice
calculations) is inferred.

Within the present paper we will utilize the new results 
\cite{Kajantie:2002wa} on the perturbation expansion of 
the QCD-pressure as much as possible in order to find approximants 
which are reasonably stable under RS-variation. This gives us
the freedom of using higher order Pad\'e or Borel-Pad\'e
approximants and thus treating the
long-range part $p_{\rm M+G}$ ($=-F_{\rm M +G}$) 
in a more reliable way. 
Unfortunately, the full four-loop contribution to the short-range part 
$p_{\rm E}$ ($= - F_{\rm E}$),
which is in principle perfectly calculable within ordinary perturbation 
theory and would yield ${\cal O}(a^3)$ correction to 
Eq.~(\ref{tFElow}), is not yet 
available at the moment. Therefore, we have no direct basis for improving
the approximants to $p_{\rm E}$. 
Nevertheless, we can obtain some restricted
information about the ${\cal O}(g_s^6)$-terms in $p_{\rm E}$
from the requirement that $p_{\rm E}$ be ${\overline \mu}$-independent
and $(p_{\rm E}+p_{\rm M+G})$ be $\Lambda_{\rm E}$-independent.
A constant (${\overline \mu}$- and $\Lambda_{\rm E}$-independent) term in
the coefficient at ${\cal O}(g_s^6)$ in $p_{\rm E}$ still remains
unspecified, but its value can roughly be estimated.
The new results of Ref.~\cite{Kajantie:2002wa}
also allow a better approximation to the parameter
$g_{\rm E}^2$ of the effective theory, 
which further improves the resulting predictions.

In Sec.~\ref{sec:PEs} we describe the separation of the QCD-pressure 
into contributions stemming from different energy regions and 
specify the known corresponding effective Lagrangians. 
We then present the available 
perturbation expansion of the long- and the short-range contributions and
work out the effects of the factorization scales $\Lambda_E$ and $\Lambda_M$
(which have not been explicitly disentangled 
in Ref.~\cite{Kajantie:2002wa} because of their simplified
treatment of the RS-dependence). In Sec.~\ref{sec:numscales},
the short-distance and the long-distance parts of the
pressure are resummed separately by Pad\'e and/or Borel-Pad\'e
approximants, and the selection of approximants is narrowed down
by requiring weak residual ${\overline \mu}$-dependence of
$p_{\rm E}$ and of $p_{\rm M+G}$, 
and weak $\Lambda_E$-dependence of $(p_{\rm E}+p_{\rm M+G})$.
In Sec.~\ref{sec:numres} the resummed results as a function
of temperature $T$ are presented, as well as arguments for
narrowing down further the selection of acceptable approximants.
In Sec.~\ref{sec:TPS}, results of TPS evaluations as a function of $T$
are presented, for comparison. In Sec.~\ref{sec:comp},
our results are compared with the predictions obtained
within other approaches, in particular with lattice results (which are
available only for rather low temperatures), and finish with some 
concluding remarks. A short compilation of the Pad\'e and Borel-Pad\'e
approximants is given in the Appendix.

\section{Perturbation expansion of long- and
short-distance pressure}
\label{sec:PEs}

The basic information about the physics of a quark-gluon system in thermal
equilibrium at temperature $T$ is contained in the expression for the
pressure $p_{\rm QCD} (T)$ or, equivalently, the free energy density
$F_{\rm QCD} (T)$. In $d+1$ dimensions $(d = 3 - 2 \epsilon)$ 
this is given by Eqs.~(\ref{pQCD1})-(\ref{Z}),
with $d^3 x$ replaced by $d^dx$. 
Boundary conditions over the finite time $\tau$
direction are periodic for bosons and anti-periodic for fermions.

When the temperature $T$ is above the masses $m_q$ of
active quarks, and the QCD gauge coupling $g_s$ is small
enough, there are three physically different scales involved: 
$\sim 2 \pi T$, $\sim g_s T$, $\sim g_s^2 T$.
The last two correspond to the color-electric and 
color-magnetic screening, respectively.
The decomposition (\ref{pdecomp}) 
$p_{\rm QCD}=p_{\rm E}+p_{\rm M}+p_{\rm G}$
reflects the contributions from the aforementioned
three energy regimes: $p_{\rm E}$ are contributions from
the modes with energies in the interval $[\Lambda_{\rm E}, \infty]$,
$p_{\rm M}$ from those in $[\Lambda_{\rm M}, \Lambda_{\rm E}]$, 
$p_{\rm G}$ from those in $[0, \Lambda_{\rm M}]$, where the factorization
scales $\Lambda_{\rm E}$ and $\Lambda_{\rm M}$
define the borders between the three energy regimes
\begin{equation}
g_s^2 T \ < \ \Lambda_{\rm M} \ < g_s T \ < \ \Lambda_{\rm E} 
\ < \ 2 \pi T \ .
\label{fscales}
\end{equation}
The long-distance part $p_{\rm M} + p_{\rm G} \equiv p_{\rm M+G}$ 
is due to the bosonic zero
(Matsubara) frequency mode, whereas $p_{\rm E}$ contains all higher modes.
For calculating the different parts analytically one most conveniently
uses the method of effective Lagrangians
\cite{Braaten:1994na,Braaten:1995cm,Braaten:1996ju}
based on the dimensional reduction method 
\cite{Ginsparg:1980ef,Kajantie:1995dw,Braaten:1995cm,Braaten:1996ju}:
Whereas the high-energy regime behavior, which is responsible for
$p_{\rm E}$, is determined by the original $(d+1)$-dimensional
QCD Lagrangian, the low-energy regime behavior, responsible
for $p_{\rm M+G}$, can be represented by a $d$-dimensional 
effective bosonic theory called electrostatic QCD (EQCD), such that
\begin{eqnarray}
p_{\rm QCD}&=& p_{\rm E} + \frac{T}{V} \ln \int
{\cal D} A^a_i {\cal D} A^a_0 
\exp \left( - \int d^d x {\cal L}_{\rm EQCD} \right) \ ,
\label{pE1}
\\
{\cal L}_{\rm EQCD}&=& \frac{1}{2} {\rm Tr} F^2_{i j} +
{\rm Tr} [D_i,A_0]^2 + m_{\rm E}^2 {\rm Tr} A_0^2 +
\lambda_{\rm E}^{(1)} \left( {\rm Tr} A_0^2 \right)^2 +
\lambda_{\rm E}^{(2)} {\rm Tr} A_0^4 + \ldots \ ,
\label{LagrE}
\end{eqnarray}
where $i = 1,\ldots, d$; $F_{i j} = (i/g_{\rm E}) [D_i,D_j]$, $D_i =
\partial_i - i g_{\rm E} A_i$. The notation $A_{\mu} = A_{\mu}^a {\bar
T}^a$ is used, with ${\bar T}^a$ being the Hermitean generators of
SU(3) normalized to 
${\rm Tr} \, {\bar T}^a {\bar T}^b = (1/2) \delta^{a b}$. 

The four parameters of the effective theory
are the Debye screening mass $m_{\rm E}$ ($\sim g_s T$),
the coupling parameters $g^2_{\rm E}$ ($\sim g_s^2 T$)
and $\lambda^{(1)}, \lambda^{(2)}$ ($\sim g_s^4 T$)
(the latter two are not independent if $d = 3$). 
These and the hard scale pressure $p_{\rm E}$ ($\sim T^4$)
are obtained by the well-known matching procedure: the parameters as 
functions of $g_s,T$ (and the UV-cutoff $\Lambda_{\rm E}$ of 
${\cal L}_{\rm EQCD}$) must be 
tuned in such a way that the effective theory reproduces the
physical effects of the full theory at large distances. The dots in
Eq.~(\ref{LagrE}) stand for higher dimensional effective interaction terms 
which yield corrections $\delta p \sim g_s^7 T^4$ 
and are thus neglected here. 

Due to the separate screening of the 
color-electric $(m_{\rm E} \sim g_sT)$ and the
color-magnetic $(m_{\rm M} \sim g^2_sT)$ degrees of freedom within QCD,  
the above EQCD-action can be further decomposed into two parts 
corresponding
to two physically different energy scales. The color-electric scales
$(\sim g_s T)$ are separated by integrating out $A_0$
\begin{eqnarray}
\frac{T}{V} \ln \int {\cal D}A^a_i {\cal D} A^a_0~~\exp \left(- \int d^d
x {\cal L}_{\rm EQCD}\right) &=& 
p_{\rm M} (T) + \frac{T}{V} 
\ln \int {\cal D}A^a_i \exp \left(- \int d^d x
{\cal L}_{\rm MQCD}\right) 
\label{pM1}
\\
& \equiv & p_{\rm M} (T) + p_{\rm G} (T) \ ,
\label{pG1}
\end{eqnarray}
where
\begin{equation}
{\cal L}_{\rm MQCD} = \frac{1}{2} {\rm Tr}~F_{ij}^2 + ...
\label{LagrM}
\end{equation}
represents the Lagrangian density
for magnetostatic QCD -- the effective theory for
the energy region below $\Lambda_{\rm M}$, typically
energies $\sim g_s^2 T$.
Here $F_{ij} = (i/g_{\rm M}) [D_i, D_j]$ with 
$D_i = \partial_i - i g_{\rm M} A_i$.

There are two matching coefficients now: the long-distance 
[$\sim 1/(g_s T)$] pressure 
$p_{\rm M}$ ($\sim m_{\rm E}^3 T \sim g_s^3 T^4$)
and the
coupling parameter $g^2_{\rm M}$ which is close to $g^2_{\rm E}$
\cite{Farakos:1994kx}
\begin{equation}
g^2_{\rm M} = g^2_{\rm E} \left( 1 + {\cal O}(g^2_{\rm E}/m_{\rm E})
\right) = g^2_{\rm E} \left( 1 + {\cal O}(g_s) \right) \ .
\label{gM}
\end{equation}

Now we will present the perturbation expansions
for $p_{\rm X}$ (X = E, M, G)
on the basis of the results of Ref.~\cite{Kajantie:2002wa}.
The authors of Ref.~\cite{Kajantie:2002wa}
present each $p_{\rm X}/(T {\widetilde \mu}^{-2 \epsilon})$ as\footnote{
$\widetilde \mu = {\overline \mu} (e^{\gamma_{\rm E}}/4 \pi)^{1/2}$,
where ${\overline \mu}$ is the renormalization scale in the
${\overline {\rm MS}}$ scheme.} 
an expansion in powers of the coupling parameters of the
respective effective theory.
We note that they used dimensional regularization in the
${\overline {\rm MS}}$ scheme, thereby invoking
a common renormalization scale ${\overline \mu}$
for all $p_{\rm X}$ and for other matching coefficients.
Therefore, their results do not contain explicitly the (physical) 
factorization scales $\Lambda_{\rm E}$ and $\Lambda_{\rm M}$, 
and contain the unphysical infinite terms $\propto 1/\epsilon$ 
which cancel in the sum $p_{\rm QCD}=p_{\rm E+M+G}$. 
We will obtain our basic formulae on the basis of their aforementioned 
expansions by replacing in the logarithms of expansion coefficients
their renormalization scale ${\overline \mu}$
by the IR and/or UV cutoffs of the respective effective theory,\footnote{
For $p_{\rm G}$ (MQCD), the UV cutoff is $\Lambda_{\rm M}$;
for $p_{\rm M}$ (EQCD), the IR cutoff is $\Lambda_{\rm M}$, and the
UV cutoff is $\Lambda_{\rm E}$. For $p_{\rm E}$ (QCD), the
IR cutoff is $\Lambda_{\rm E}$, and the UV cutoff will
be taken to be ${\overline \mu}_{\rm E}$ ($\sim 2 \pi T$), 
i.e., the (${\overline {\rm MS}}$) renormalization scale 
for $g_s({\overline \mu}_{\rm E})$ appearing in $p_{\rm E}$.} 
and by removing all the terms proportional to $1/\epsilon$
in each $p_{\rm X}/(T {\widetilde \mu}^{-2 \epsilon})$
and then taking $\epsilon=0$ in each $p_{\rm X}$.
While the aforementioned $1/\epsilon$ terms cancel in the sum 
$p_{\rm E+M+G}/(T {\widetilde \mu}^{-2 \epsilon})$,
the effect of the common factor $1/{\widetilde \mu}^{-2 \epsilon}$
also disappears (i.e., reduces to one) in this sum when $\epsilon \to 0$.
Which of the $\ln {\overline \mu}$-terms in the coefficients
get replaced by logarithms of the UV scale and which
by those of the IR scale of the effective theory
-- this is unambiguously determined by the requirement
that the entire $p_{\rm E+M+G}$ be independent of the
factorization scales. 
On the other hand, the aforementioned procedure to 
eliminate first the $1/\epsilon$ infinities from each 
$p_{\rm X}/(T {\widetilde \mu}^{-2 \epsilon})$ and then replace 
$\epsilon \mapsto 0$ to obtain $p_{\rm X}/T$ is certainly not unique.
Other procedures would result in expansions for $p_{\rm X}$ which
differ from ours by certain constant numbers 
(independent of scales) in some of the coefficients 
of the expansion in powers of the coupling parameters
of the respective effective theory, this ambiguity being a
manifestation of renormalization freedom.
The overall sum $p_{\rm E+M+G}$ would remain the same.
We note that our procedure leads to the decomposition
of Ref.~\cite{Braaten:1996ju}, at least to the order
available in that Reference. Further, we checked that
the re-expansion in powers of QCD coupling parameter
$g_s({\overline \mu})$ of each obtained $p_{\rm X}$
reproduces for the sum $p_{\rm E+M+G}$ the same expansion 
as the one obtained in Ref.~\cite{Kajantie:2002wa}. 

The expansion of  $p_{\rm G}$ starts at $\sim g_{\rm M}^6$
($\sim g_s^6$)
\begin{eqnarray}
p_{\rm G} &=& T g_{\rm M}^6 \frac{8 \cdot 3^3}{ (4 \pi)^4}
\left[ 8~\alpha_{\rm G} 
\ln \left( \frac{ \Lambda_{\rm M} }{ 2 m_{\rm M} } \right)
+ \delta_{\rm G} \right]
\label{pG2}
\\
\alpha_{\rm G} &=& \frac{43}{96} - \frac{157}{6144} \pi^2
\approx 0.195715  \ ,
\label{alG}
\end{eqnarray}
where the value (\ref{alG}) was obtained in
\cite{Kajantie:2002wa} and $m_{\rm M} \equiv 3 g_{\rm M}^2$
($\sim g_s^2 T$). The coefficient $\delta_{\rm G}$
denotes a dimensionless number which is not
perturbatively calculable but is expected to be
$|\delta_{\rm G}| \sim 1$ (note that $8 \alpha_{\rm G}
\approx 1.6$). We will allow the following, rather generous,
variation of $\delta_{\rm G}$:
\begin{equation}
- 5 \ < \ \delta_{\rm G} \ < \ + 5 \ .
\label{dGvar}
\end{equation}
We note that, according to the
procedure mentioned before, the result (\ref{pG2})
is obtained from the corresponding result for
$p_{\rm G}(T)/(T {\widetilde \mu}^{-2 \epsilon})$
of Ref.~\cite{Kajantie:2002wa} by removing
the term $1/\epsilon$, replacing in the
logarithm ${\overline \mu}$ by the UV cutoff ${\Lambda}_{\rm M}$
of MQCD, and then taking $\epsilon \to 0$.

The same procedure gives for $p_{\rm M}$ the following expression:
\begin{eqnarray}
p_{\rm M}(T) &=& T m_{\rm E}^3 \left( \frac{2}{3 \pi} \right)
{\Bigg \{} 1 + \frac{1}{4 \pi} 3^2 
\left( \frac{ g_{\rm E}^2 }{m_{\rm E}} \right) 
\left[ - \frac{3}{4} - 
\ln \left( \frac{ {\Lambda}_{\rm E} }{2 m_{\rm E} } \right) \right]
\nonumber\\
&& + \frac{1}{ (4 \pi)^2} 3^3 
\left( \frac{ g_{\rm E}^2 }{m_{\rm E}} \right)^2
\left[ - \frac{89}{24} - \frac{ \pi^2}{6} +
\frac{11}{6} \ln 2 \right]
\nonumber\\
&& +  \frac{1}{ (4 \pi)^3} 3^4 
\left( \frac{ g_{\rm E}^2 }{m_{\rm E}} \right)^3
\left[ 8 (\alpha_{\rm M}+\alpha_{\rm G}) 
\ln \left( \frac{ {\Lambda}_{\rm E} }{2 m_{\rm E} } \right)
- 8~\alpha_{\rm G}
\ln \left( \frac{ {\Lambda}_{\rm M} }{2 m_{\rm E} } \right)
+ \beta_{\rm M} \right]
\nonumber\\
&& + \frac{3}{4 \pi} \frac{(-5)}{2} 
\frac{ \lambda^{(1)}_{\rm E} }{ m_{\rm E} }
+ \ldots {\Bigg \}} \ ,
\label{pM2}
\end{eqnarray}
The expression in the brackets containing $\alpha_{\rm M}$
and $\alpha_{\rm G}$ is obtained from the term
$8~\alpha_{\rm M} \ln( {\overline \mu}/(2 m_{\rm E}) )$
in the expansion for $p_{\rm M}/(T {\widetilde \mu}^{-2 \epsilon})$
of Ref.~\cite{Kajantie:2002wa}, by replacing 
${\overline \mu}$ in part of the logarithm by the
corresponding UV cutoff $\Lambda_{\rm E}$ of EQCD
and in the rest of the logarithm by the
IR cutoff $\Lambda_{\rm M}$ in such a way
as to guarantee that the sum $p_{\rm G}+p_{\rm M}$
is independent of $\Lambda_{\rm M}$. For this,
relation (\ref{gM}) for $g_{\rm M}$
must be inserted into expression (\ref{pG2})
\begin{equation}
p_{\rm G} = T m_{\rm E}^3 \left( \frac{2}{3 \pi} \right)
\left( \frac{3^4}{ (4 \pi)^3 } \right)
\left( \frac{ g_{\rm E}^2 }{m_{\rm E}} \right)^3
\left[ 8~\alpha_{\rm G} 
\ln \left( \frac{ \Lambda_{\rm M} }{ 6 g_{\rm E}^2 } \right)
+ \delta_{\rm G} \right] \ .
\label{pG3}
\end{equation}
In expression (\ref{pM2}) for $p_{\rm M}$, the values of constants
$\alpha_{\rm M}$ and $\beta_{\rm M}$ have been obtained
in Refs.~\cite{Kajantie:2002wa} and \cite{Kajantie:2003ax},
respectively
\begin{eqnarray}
\alpha_{\rm M} &=& \frac{43}{32} - \frac{491}{6144} \pi^2
\approx 0.555017 \ ,
\label{alM}
\\
\beta_{\rm M} & \approx & -1.391512 \ .
\label{beM}
\end{eqnarray}
The last term in expansion (\ref{pM2}) is written in
the convention $\lambda_{\rm E}^{(2)} = 0$ were $\lambda_{\rm E}^{(1)}$
is equal to
\cite{Nadkarni:1988fh}
\begin{equation}
\lambda_{\rm E}^{(1)} = T g_s^4({\overline \mu}) \frac{1}{24 \pi^2}
(9 - n_f) + {\cal O}(g_s^6) \ .
\label{l1}
\end{equation}
Here, $n_f$ is the number of active quark flavors.
In ref.~\cite{Landsman:1989be} a different convention is adopted yielding
$\lambda_{\rm E}^{(1)} = T g_s^4 n_f/(432 \pi^2) + {\cal O}(g_s^6)$
and $\lambda_{\rm E}^{(2)} = T g_s^4 (1 - n_f/3)/(24 \pi^2)
+ {\cal O}(g_s^6)$, but this then leads to the same result
for $p_{\rm M}$. 

Adding expressions (\ref{pM2}) and (\ref{pG3}), we obtain
the sum $p_{\rm M+G}$ as expansion in powers of the
EQCD parameters $g_{\rm E}$ and $m_{\rm E}$
\begin{eqnarray}
{\widetilde p}_{\rm M+G} & \equiv &
\frac{3 \pi}{2} \frac{1}{T m_{\rm E}^3} ( p_{\rm M} + p_{\rm G} )
\nonumber\\
&=& 
{\Bigg \{} 1 + \frac{1}{4 \pi} 3^2 
\left( \frac{ g_{\rm E}^2 }{m_{\rm E}} \right) 
\left[ - \frac{3}{4} - 
\ln \left( \frac{ {\Lambda}_{\rm E} }{2 m_{\rm E} } \right) \right]
\nonumber\\
&& + \frac{1}{ (4 \pi)^2} 3^3 
\left( \frac{ g_{\rm E}^2 }{m_{\rm E}} \right)^2
\left[ - \frac{89}{24} - \frac{ \pi^2}{6} +
\frac{11}{6} \ln 2 \right]
\nonumber\\
&& +  \frac{1}{ (4 \pi)^3} 3^4 
\left( \frac{ g_{\rm E}^2 }{m_{\rm E}} \right)^3
\left[ 8 \alpha_{\rm M}
\ln \left( \frac{ {\Lambda}_{\rm E} }{2 m_{\rm E} } \right)
+ 8~\alpha_{\rm G}
\ln \left( \frac{ {\Lambda}_{\rm E} }{6 g_{\rm E}^2 } \right)
+ \beta_{\rm M} + \delta_{\rm G} \right]
+ \frac{3}{4 \pi} \frac{(-5)}{2} 
\frac{ \lambda^{(1)}_{\rm E} }{ m_{\rm E} }
+ \ldots {\Bigg \}} \ .
\label{pMGeff}
\end{eqnarray}

The other matching coefficients $m_{\rm E}$ and $g_{\rm E}$
can be expanded in powers of QCD coupling 
$g_s \equiv g_s({\overline \mu})$ 
(cf.~\cite{Kajantie:2002wa}, with $\epsilon \mapsto 0$)
\begin{eqnarray}
m_{\rm E}^2 & = &
T^2 (1\!+\!n_f/6) g_s^2
\left\{ 1 + \left( \frac{g_s}{2 \pi} \right)^2
\left[ P_m(n_f) +
2 \beta_0 \ln \left( \frac{{\overline \mu}}{2 \pi T} \right) \right]
+ {\cal O}(g_s^4) \right\} \ ,
\label{mE}
\\
g_{\rm E}^2 & = &
T g_s^2 \left\{ 1 + \left( \frac{g_s}{2 \pi} \right)^2
\left[ P_g(n_f) +
2 \beta_0 \ln \left( \frac{{\overline \mu}}{2 \pi T} \right) \right]
+ {\cal O}(g_s^4) \right\} \ ,
\label{gE}
\end{eqnarray}
where $\beta_0 = (11/4) (1 - 2 n_f/33)$ is the first beta coefficient, 
and 
\begin{eqnarray}
P_m(n_f) &=& \frac{ ( 0.612377 - 0.488058 \; n_f - 0.0427979 \; n_f^2 )}
{(1 + n_f/6)} \ ,
\label{Pm}
\\
P_g(n_f) & = & ( - 0.387623 - 0.423454 \; n_f ) \ .
\label{Pg}
\end{eqnarray}
The $\lambda_{\rm E}^{(1)}/m_{\rm E}$-term in expansion
(\ref{pMGeff}) can be written with the help of the
leading order relation (\ref{l1}) and relations
(\ref{mE})-(\ref{gE}) in several ways, for example
in the following two ways:
\begin{eqnarray}
 \frac{3}{4 \pi} \frac{(-5)}{2} 
\frac{ \lambda^{(1)}_{\rm E} }{ m_{\rm E} } & = &
- \frac{5}{(4 \pi)^3} \frac{(9 - n_f)}{(1 + n_f/6)^{1/2}}
g_s^3({\overline \mu}) \left[ 1 + {\cal O}(g_s^2) \right] \ ,
\label{l1a}
\\
& = & 
- \frac{5}{(4 \pi)^3} (9 - n_f) (1 + n_f/6)
\left( \frac{g_{\rm E}^2}{m_{\rm E}} \right)^3
\left[ 1 + {\cal O} \left(
\left( \frac{g_{\rm E}^2}{m_{\rm E}} \right)^2 \right) \right] \ .
\label{l1b}
\end{eqnarray}
The more conservative approach in the resummations based on 
EQCD expansion (\ref{pMGeff}) should be to resum separately 
the expansion of terms which are powers of $g_{\rm E}^2/m_{\rm E}$ 
and the expansion of terms
which are powers of $\lambda_{\rm E}^{(1)}/m_{\rm E}$,
because the two expansions represent probably two
topologically different families of diagrams.
The problem with the latter expansion is that we
know only the leading term there, which is written
in terms of the QCD coupling parameter $g_s({\overline \mu})$
in Eq.~(\ref{l1a}), and in terms of the first EQCD
coupling parameter $g_{\rm E}^2/m_{\rm E}$ in Eq.~(\ref{l1b}).
The latter equation, in comparison to the former,
represents some kind of (EQCD-)resummation of the
$\lambda_{\rm E}^{(1)}/m_{\rm E}$-family of terms.
Since the parameters $g_{\rm E}^2/m_{\rm E}$ and
$\lambda_{\rm E}^{(1)}/m_{\rm E}$ are both
of the effective EQCD theory, the expression
(\ref{l1b}) may be considered as better,
but less conservative. While
$\lambda_{\rm E}^{(1)}/m_{\rm E}$ should be
independent of the renormalization scale
${\overline \mu} = \mu_{\rm M}$ ($\sim m_{\rm E}$),
we will later see that the
leading order EQCD expression (\ref{l1b}) is significantly less
$\mu_{\rm M}$-dependent than
the leading order QCD expression (\ref{l1a}),
especially when the Pad\'e resummations ${\rm P[1/1]}(a(\mu_{\rm M}))$
are applied to expansions (\ref{mE})-(\ref{gE})
for $m_{\rm E}^2$ and $g_{\rm E}^2$.

The coupling parameter $g_{\rm E}$ and the
Debye screening mass $m_{\rm E}$ are physical quantities,
and thus they are independent of the
renormalization scale (${\overline \mu}$).
This independence is reflected in the coefficients of
expansions (\ref{mE})-(\ref{gE}) in the terms 
proportional to $\beta_0 \ln {\overline \mu}$.  
%The expansion of the dimensionless
%EQCD coupling parameter $(g_{\rm E}^2/m_{\rm E})$ is
%\begin{equation}
%\frac{g_{\rm E}^2}{m_{\rm E}} = \frac{1}{(1 + n_f/6)^{1/2}}
%g_s \left\{ 1 + \left( \frac{g_s}{2 \pi} \right)^2
%\left[ P_{g}(n_f) - \frac{1}{2} P_{m}(n_f)+
%\beta_0 \ln \left( \frac{\mu}{2 \pi T} \right) \right]
%+ {\cal O}(g_s^4) \right\} \ .
%\label{gE2mE}
%\end{equation}
Inserting expansions (\ref{mE}) and (\ref{gE}) 
into series (\ref{pMGeff}), and using relation (\ref{l1}),
leads to an expansion for the
sum ${\widetilde p}_{\rm M+G}$ in terms of the
QED coupling $g_s \equiv g_s({\overline \mu})$
\begin{eqnarray}
{\widetilde p}_{\rm M+G} & \equiv &
\frac{3 \pi}{2} \frac{1}{T m_{\rm E}^3} ( p_{\rm M} + p_{\rm G} )
\nonumber\\
&=& 1 + g_s \frac{3^2}{4 \pi} \frac{1}{(1 + n_f/6)^{1/2}} 
\left[ - \frac{3}{4} - 
\ln \left( \frac{ {\Lambda}_{\rm E} }{2 m_{\rm E} } \right) \right]
\nonumber\\
&& + g_s^2 \frac{3^3}{(4 \pi)^2} \frac{1}{(1 + n_f/6)} 
\left[ - \frac{89}{24} - \frac{ \pi^2}{6} +
\frac{11}{6} \ln 2 \right]
\nonumber\\
&& + g_s^3 \frac{3}{(4 \pi)^3} \frac{1}{(1 + n_f/6)^{1/2}} 
{\Bigg \{} K_3 \left[ - \frac{3}{4} - 
\ln \left( \frac{ {\Lambda}_{\rm E} }{2 m_{\rm E} } \right) \right] 
\nonumber\\
&&+ \frac{3^3}{(1 + n_f/6)}
\left[ 8 (\alpha_{\rm M}+\alpha_{\rm G}) 
\ln \left( \frac{ {\Lambda}_{\rm E} }{2 m_{\rm E} } \right)
+ 8~\alpha_{\rm G}
\ln \left( \frac{ m_{\rm E} }{3 g_{\rm E}^2 } \right)
+ \beta_{\rm M} + \delta_{\rm G} \right]
\nonumber\\
&& - \frac{5}{3} (9 - n_f) + 
12 \beta_0 \ln \left( \frac{{\overline \mu}}{ 2 \pi T} \right)
\left[ - \frac{3}{4} - 
\ln \left( \frac{ {\Lambda}_{\rm E} }{2 m_{\rm E} } \right) \right]
{\Bigg \}} + {\cal O}(g_s^4) \ ,
\label{pMG}
\end{eqnarray}
where the constant $K_3$ is
\begin{equation}
K_3 = \frac{1}{12} \frac{1}{(1 + n_f/6)}
(-99.9089 - 35.1402 \; n_f - 7.08145 \; n_f^2) \ .
\label{K3}
\end{equation}
The dependence on the factorization scale
${\Lambda}_M$ in the expansion (\ref{pMG}) disappeared as it should.
Further, the quantity $p_{\rm M+G}$ is independent of the
renormalization scale ${\overline \mu}$, and this is reflected
by the term proportional to $\beta_0 \ln {\overline \mu}$
in the coefficient at $g_s^3$ in Eq.~(\ref{pMG}).

We note that the only unknown dimensionless coefficient in (\ref{pMG})
is $\delta_{\rm G}$ which, as argued before, is expected to
be $|\delta_{\rm G}| \sim 1$.

The knowledge of the expansion of the short-distance pressure $p_{\rm E}$
is less complete -- this is an expansion in powers of $g_s^2$,
and it is known only up to $\sim g_s^4$ (cf.~Ref.~\cite{Braaten:1996ju}).
However, important parts of the coefficient at $g_s^6$
in $p_{\rm E}$ can be deduced from the requirement of
${\overline \mu}$-independence of $p_{\rm E}$
and of the ${\Lambda}_{\rm E}$-independence of
$p_{\rm E+M+G}$ (${\Lambda}_{\rm E}$ is the factorization
scale for the sum $p_{\rm E} + p_{\rm M+G}$). After
some (tedious) algebra, we end up with the
following expansion of $p_{\rm E}$ up to $\sim g_s^6$:
\begin{equation}
p_{\rm E}(T) = p_{\rm ideal}(T) \left[ 1 -
\frac{15}{4} \frac{ (1 + 5 n_f/12) }{ (1 + 21 n_f/32) }
R_{\rm E}^{\rm can}(T) \right] \ ,
\label{pEcandef}
\end{equation}
where
\begin{eqnarray}
p_{\rm ideal}(T) & = & \frac{8 \pi^2}{45} T^4 \left(
1 + \frac{21}{32} n_f \right) \ ,
\label{pideal}
\\
R_{\rm E}^{\rm can}(T) & = & \left( \frac{g_s}{ 2 \pi} \right)^2
{\Bigg \{} 1 + \left( \frac{g_s}{ 2 \pi} \right)^2
\left[ 2 \beta_0 \ln \left( \frac{ {\overline \mu} }{ 2 \pi T}
\right) - 36 \frac{(1 + n_f/6)}{(1 + 5 n_f/12)} 
\ln \left( \frac{ \Lambda_{\rm E} }{ \kappa T } \right) \right]
\nonumber\\
&& + \left( \frac{g_s}{ 2 \pi} \right)^4
{\bigg [} 4 \beta_0^2 \ln^2 \left( \frac{ {\overline \mu} }{ 2 \pi T}
\right) + 2 \ln \left( \frac{ {\overline \mu} }{ 2 \pi T} \right)
\left( \beta_1 - 72 \beta_0 \frac{(1 + n_f/6)}{(1 + 5 n_f/12)} 
\ln \left( \frac{ \Lambda_{\rm E} }{ \kappa T } \right) \right)
\nonumber\\
&& + \frac{36}{(1 + 5 n_f/12)} 
\ln \left( \frac{ \Lambda_{\rm E} }{ \kappa T } \right)
\left( - \frac{1}{2} (1 + n_f/6) (3 P_m(n_f) + K_3/6)
+ 18 (\alpha_{\rm M}+\alpha_{\rm G}) \right) + \delta_{\rm E} 
{\bigg ]} + {\cal O}(g_s^6) {\Bigg \}} \ ,
\label{pE2}
\end{eqnarray}
and the parameter $\kappa$ was introduced such that the NLO coefficient 
is made up
of only two logarithmic terms proportional to
$\ln({\overline \mu}/ 2 \pi T)$ and 
$\ln ( \Lambda_{\rm E}/ \kappa T)$ as shown above
\begin{equation}
\ln \left( \frac{2 \pi}{\kappa} \right) =  \frac{1}{135}
\frac{1}{(1 + n_f/6)} (244.898 + 17.2419 \; n_f -
0.415029 \; n_f^2) \ .
\label{kappa}
\end{equation}
For example, $\kappa \approx 1.0241, 1.47922$ for
$n_f = 0, 3$, respectively.
At order $g_s^6$ in expansion (\ref{pE2}), only the dimensionless 
number $\delta_{\rm E}$ is unknown -- it is
independent of the energy scales and of their ratios, just like
$\delta_{\rm G}$. We organized the coefficient at $g_s^6$
in Eq.~(\ref{pE2}) in the following way:
the ${\overline \mu}$-dependent terms are written
in powers of $\ln ({\overline \mu}/2 \pi T)$;
the IR cutoff ($\Lambda_{\rm E}$)-dependent
terms are written in terms of $\ln ( \Lambda_{\rm E}/ \kappa T )$,
because this combination absorbs all the 
$\ln({\overline \mu}/ 2 \pi T)$-independent
terms in the coefficient at
$g_s^4$. It is reasonable to expect that the
$\ln ({\overline \mu}/2 \pi T)$-independent terms at $\sim g_s^6$
would also be absorbed to a large degree by a quantity
proportional to the combination $\ln ( \Lambda_{\rm E}/ \kappa T )$. 
That's why we expect the number $\delta_{\rm E}$ to be small.
A conservative expectation would be that $|\delta_{\rm E}|$
is not larger than the $\ln ( \Lambda_{\rm E}/ \kappa T )$-term
there:
\begin{equation}
- | k_2^{(0)}(\Lambda_{\rm E}) | \ < \ \delta_{\rm E} \ < \
+ | k_2^{(0)}(\Lambda_{\rm E}) | \ ,
\label{dEest}
\end{equation}
where 
\begin{equation}
k_2^{(0)}(\Lambda_{\rm E}) \equiv 
\frac{36}{(1 + 5 n_f/12)} 
\ln \left( \frac{ \Lambda_{\rm E} }{ \kappa T } \right)
\left( - \frac{1}{12} K_3 (1 + n_f/6) + 
18 (\alpha_{\rm M}+\alpha_{\rm G}) - \frac{3}{2}
(1 + n_f/6) P_m(n_f) \right) \ .
\label{R3}
\end{equation} 
While $\delta_{\rm E}$ is independent of any scale,
$k_2^{(0)}$ will have only slight dependence on temperature
$T$ when we will take $\Lambda_{\rm E} =
\sqrt{ 2 \pi T m_{\rm E}(T) }$ ($\sim g_s^{1/2} T$).
This choice of $\Lambda_{\rm E}$ was taken also
in Ref.~\cite{Cvetic:2002ju}.

\section{Numerical analysis of unphysical dependence on scales}
\label{sec:numscales}

Expansions (\ref{pMG}) or (\ref{pMGeff}) for $p_{\rm M+G}$,
and (\ref{pEcandef})-(\ref{pE2}) for $p_{\rm E}$,
in conjunction with expansions (\ref{mE}) and (\ref{gE})
for $m_{\rm E}^2$ and $g_{\rm E}^2$,
will form the basis of our numerical analysis. 
The only unknown parameters are $\delta_{\rm G}$ and
$\delta_{\rm E}$ which will be allowed to vary
in the intervals (\ref{dGvar}) and (\ref{dEest}).
The numerical analysis will be performed in analogy with that
of our previous work \cite{Cvetic:2002ju}. There, 
expansion corresponding to present Eq.~(\ref{pMG}) 
for ${\widetilde p}_{\rm M+G}$ was applied but
contained only terms up to $\sim g_s^2$; expansion for 
$R_{\rm E}^{\rm can}$ of present Eq.~(\ref{pE2}) 
contained only terms up to $\sim g_s^4$; and expansion (\ref{gE}) 
only the leading term. The resummations in Ref.~\cite{Cvetic:2002ju}
were performed with Pad\'e approximants, or with
simple evaluation of the truncated perturbation series (TPS). 
In the present work, the resummations will be performed
with Pad\'e, Borel-Pad\'e (cf.~Appendix \ref{app:PBP}), or TPS.

As argued in the Introduction and in Ref.~\cite{Cvetic:2002ju}, the 
${\overline {\rm MS}}$ renormalization scale
${\overline \mu}$ should be chosen accordingly in different regimes
for the resummation of the different quantities
(\ref{pMG}), (\ref{pE2}), and (\ref{mE})-(\ref{gE}).
The scale regime for ${\overline \mu}$ should be of the order of
a typical physical scale that corresponds to the
quantity to be resummed.
Therefore, the renormalization scale ${\overline \mu} \equiv \mu_{\rm E}$
in the short-distance quantity (\ref{pE2}) is 
$\mu_{\rm E} \sim 2 \pi T$. For the long-distance EQCD quantities
(\ref{mE})-(\ref{gE}), the relevant scale is
${\overline \mu} \equiv \mu_m$ such that $\mu_m \sim m_{\rm E}$
($\sim g_s T$). For the long-distance quantity $p_{\rm M+G}$,
Eqs.~(\ref{pMG}) and (\ref{pMGeff}), the relevant scale
 ${\overline \mu} \equiv \mu_{\rm M+G}$ should be somewhere
between $\sim g_s T$ and $g_s^2 T$; we will take it
$\sim g_s T$, i.e., ${\overline \mu} = \mu_{\rm M} \sim g_s T$.
Unless otherwise stated, we will take 
$\mu_{\rm M} = \mu_m = m_{\rm E}(\mu_m)$.
For the factorization scale $\Lambda_{\rm E}$ we take,
unless otherwise stated, the geometric mean between the
hard scale $2 \pi T$ and the EQCD scale $m_{\rm E}$:
$\Lambda_{\rm E} = [2 \pi T m_{\rm E}]^{1/2}$ 
($\sim g_s^{1/2} T$). 
For $m_{\rm E}^2$ and $g_{\rm E}^2$ we will take, unless otherwise stated,
the ${\rm P}[1/1](a)$ Pad\'e approximant (PA) with respect to
$a$ (cf.~Appendix \ref{app:PBP}), where 
$a \equiv a({\overline \mu}) \equiv ( g_s({\overline \mu})/ 2 \pi )^2$ 
and the scale ${\overline \mu}$ ($\equiv \mu_m$) is adjusted so that
${\overline \mu} = m_{\rm E}$. We will see below that
${\rm P}[1/1](a)$ is a very reasonable resummation for
$m_{\rm E}^2$.

Further, if not stated otherwise, we will take for the
number of active (and massless) quark flavors $n_f=3$.
For the QCD coupling parameter 
$\alpha_s({\overline \mu}) \equiv 
g_s^2({\overline \mu})/(4 \pi)$ we take the reference value 
$\alpha_s(m_{\tau}^2,{\overline {\rm MS}}) = 0.334$
which is approximately the value extracted
from the hadronic $\tau$ decay data 
\cite{Geshkenbein:2001mn,Cvetic:2001sn}.
We work in the ${\overline {\rm MS}}$ scheme and use
for the $\beta$ function ${\rm P}[2/3](a)$ Pad\'e approximant (PA)
($a=\alpha_s/\pi$), unless otherwise stated. 
This approximant, as shown in 
Refs.~\cite{Cvetic:2000mh}, represents a reasonable 
(quasi)analytic continuation of the TPS of 
$\beta_{\overline {\rm MS}}(a)$ into the regime of large $a$
[with ${\overline \mu}$ down to $\alpha_s(\mu) \approx 1.0$]
where the TPS is practically inapplicable.

\begin{figure}[htb]
\begin{minipage}[b]{.49\linewidth}
 \centering\epsfig{file=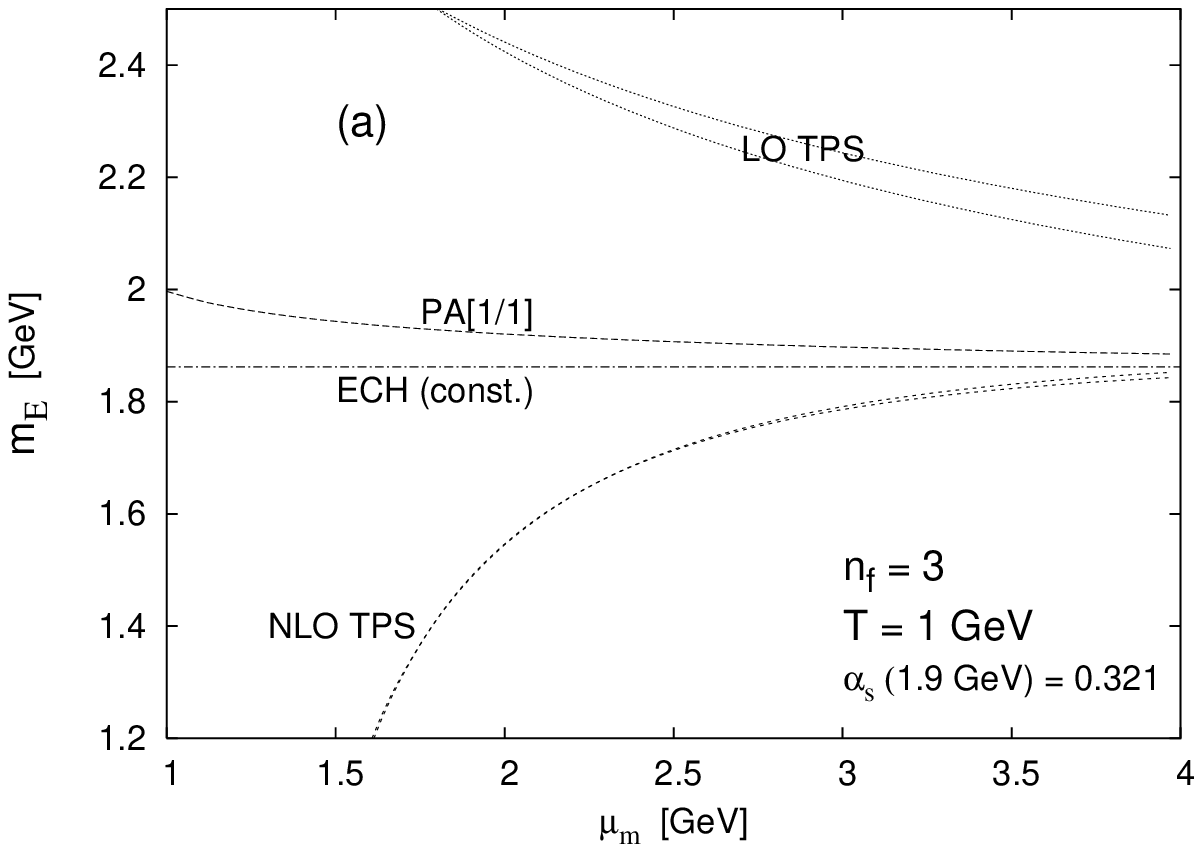,width=\linewidth}
\end{minipage}
\begin{minipage}[b]{.49\linewidth}
 \centering\epsfig{file=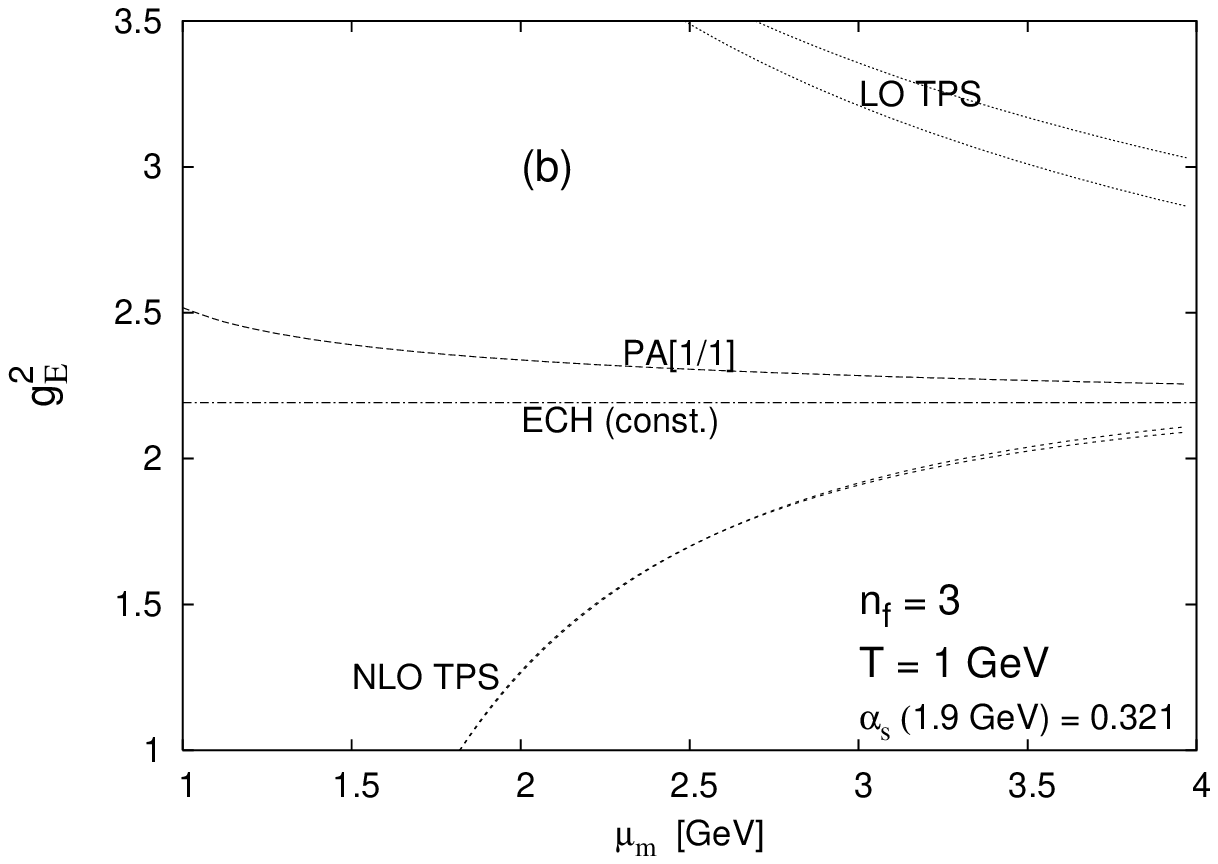,width=\linewidth}
\end{minipage}
%\vspace{0.2cm}
\caption{\footnotesize (a) The Debye screening mass $m_{\rm E}$, 
and (b) the EQCD coupling parameter $g_{\rm E}^2$, as functions
of the renormalization scale $\mu_m$, when $T\!=\!1$ GeV.
The upper of the LO TPS (NLO TPS) twin curves
has $a(\mu_m^2)$ evolved by the one-loop (two-loop)
RGE from $a(m_{\tau}^2)$. The other curves have
$a(\mu_m^2)$ evolved by the four-loop PA $[2/3]$ beta function.} 
\label{figmEgE2}
\end{figure} 
First, the resummation of the expansion of the 
squared Debye screening mass
$m_{\rm E}^2$  (\ref{mE}) and of the EQCD coupling parameter
$g_{\rm E}^2$ (\ref{gE}), respectively, are performed. 
Both expansions are next-to-leading order
(NLO) in $a(\mu_m) = [g_s(\mu_m)/ 2 \pi]^2$. 
Therefore, the diagonal ${\rm P}[1/1](a)$ can be
constructed and should be a good candidate.
Figs.~\ref{figmEgE2}(a) and \ref{figmEgE2}(b) show the results,
as a function of the renormalization scale $\mu_m$ ($\sim m_{\rm E}$).
The values of $m_{\rm E}$ in Fig.~\ref{figmEgE2}(a)
are obtained by the corresponding resummation of expansion
(\ref{mE}) for $m_{\rm E}^2$ and then taking the square root.
In addition to ${\rm P}[1/1](a(\mu_m))$, also the effective charge
method result (ECH) \cite{ECH,KKP,Gupta}, as well as
the TPS results, are presented. The ECH result is
the NLO TPS at a specific value of the renormalization scale,
so it is automatically independent of $\mu_m$.
We see from Figs.~\ref{figmEgE2}(a)
and \ref{figmEgE2}(b)
that PA's ${\rm P}[1/1](a)$ for $m_{\rm E}^2$ and $g_{\rm E}^2$
are  very good approximations as they almost eliminate
$\mu_m$-dependence contrary to the TPS-expressions which show (both for LO and
NLO) a very strong unphysical $\mu$-dependence.
This is in accordance with Ref.~\cite{Gardi:1996iq}
where it is argued that any diagonal ${\rm P}[n/n](a(\mu))$
of a QCD observable has significantly reduced $\mu$-dependence,
i.e., it is $\mu$-independent in the large-$\beta_0$
limit.\footnote{
There exists an extension of the diagonal PA's such that
it is completely $\mu$-independent \cite{Cvetic:1997ca}, or 
$\mu$- and scheme-independent \cite{Cvetic:2000mh}.} 
As mentioned before, we will choose the mass $m_{\rm E}$
by the condition $(m_{\rm E}^2)^{{\rm P}[1/1]} = \mu_m^2$,
and we will denote the square root of this value 
as $m_{\rm E}^{(0)}(T)$
\begin{equation}
\left( m_{\rm E}^2(T) \right)^{{\rm P}[1/1]} = 
\mu_m^2 \  \equiv \  \left( m_{\rm E}^{(0)} (T) \right)^2 \ .
\label{m0T}
\end{equation}
At $T=1$ GeV (and $n_f=3$), this value is 
$m_{\rm E}^{(0)} \approx 1.923$ GeV.\footnote{
Fig.~\ref{figmEgE2}(a) was presented in our previous
work \cite{Cvetic:2002ju}, but there the curves are
slightly lower (by about $0.02$ GeV) due to inadvertent
omission of the factor $(1 + n_f/6)^{-1}$ appearing
in our Eq.~(\ref{Pm}) for $P_m$.} 

\begin{figure}[htb]
\begin{minipage}[b]{.49\linewidth}
 \centering\epsfig{file=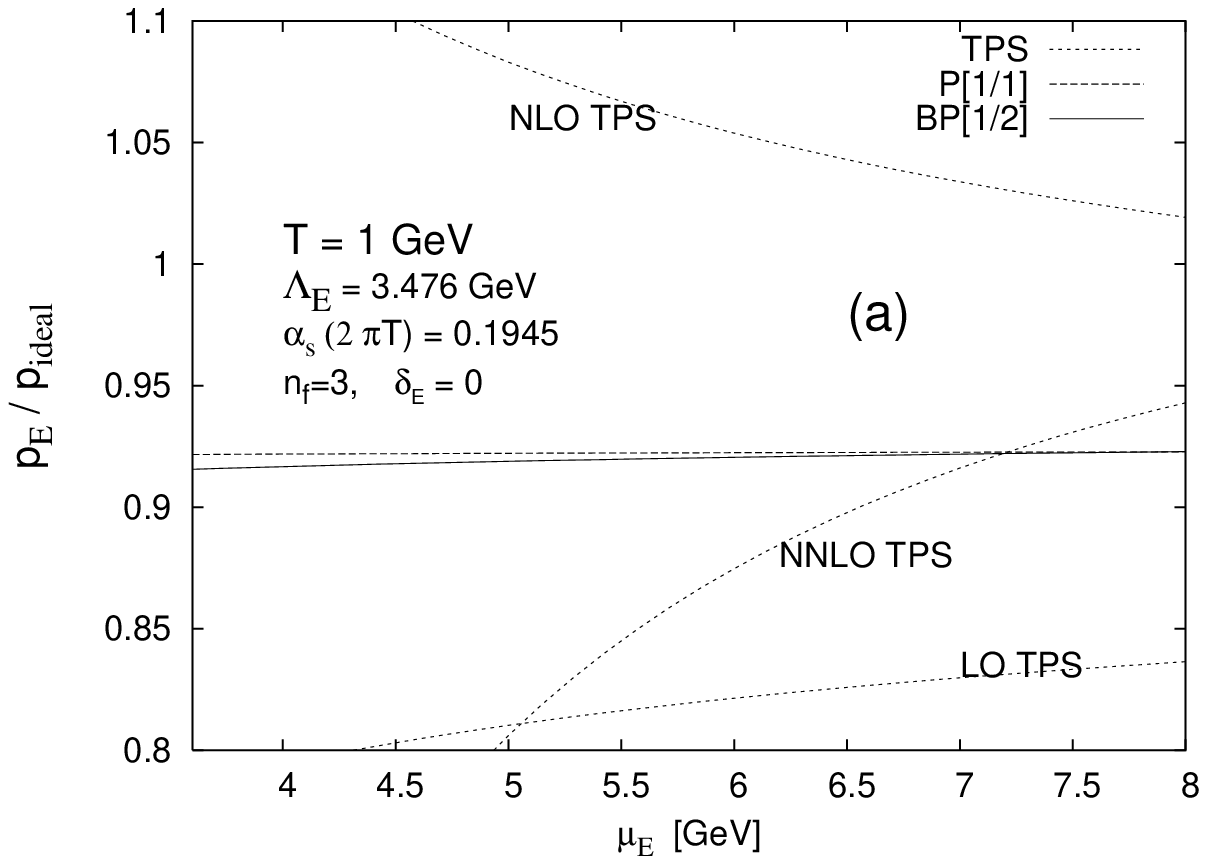,width=\linewidth}
\end{minipage}
\begin{minipage}[b]{.49\linewidth}
 \centering\epsfig{file=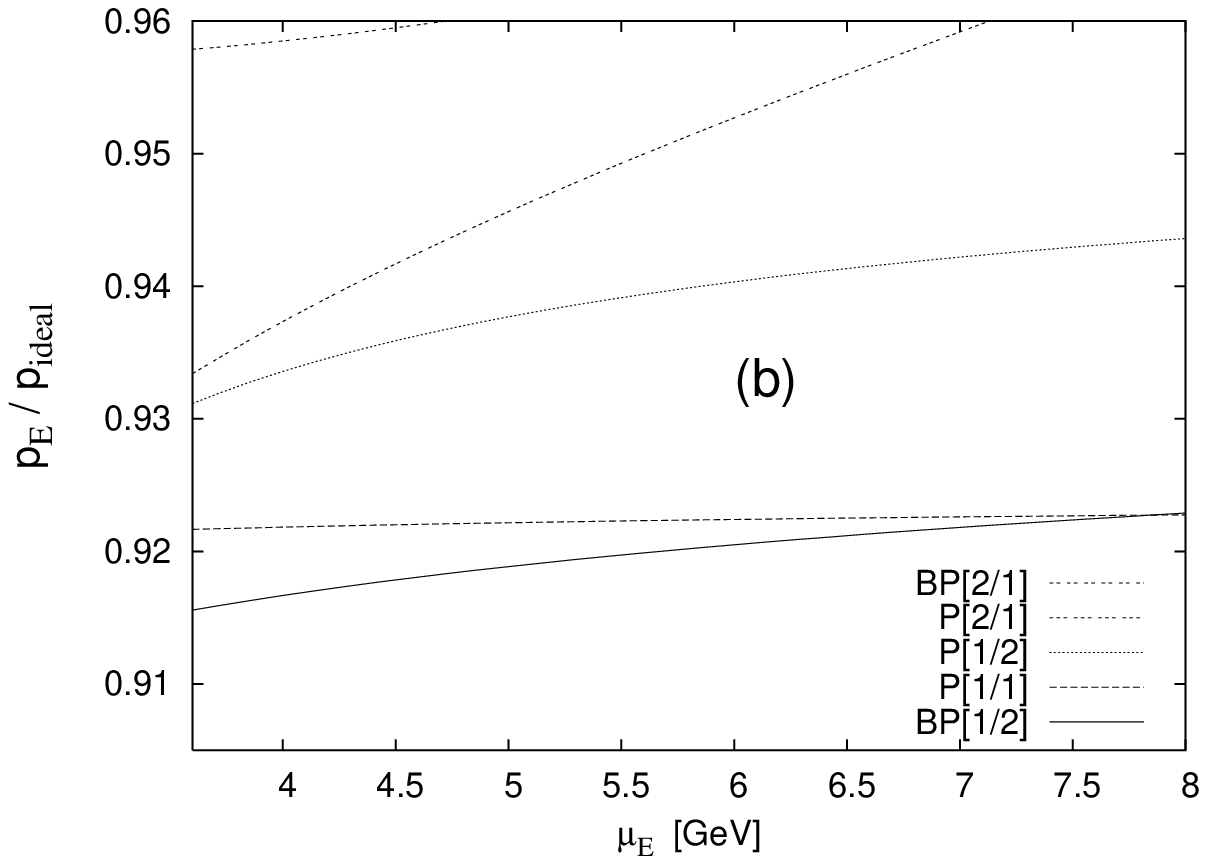,width=\linewidth}
\end{minipage}
%\vspace{0.2cm}
\caption{\footnotesize The short-distance pressure
$p_{\rm E}$, at $T=1$ GeV, as a function of the corresponding 
renormalization scale $\mu_{\rm E}$ -- various approximants are applied
to the perturbation expansion (\ref{pE2}) in powers of
$a(\mu_{\rm E}) = \left[g_s(\mu_{\rm E})/ 2 \pi \right]^2$,
and $\delta_{\rm E}=0$ was taken.
Figure (b) has a finer vertical scale and includes some
approximants not shown in Fig.~(a).}
\label{pEvsmu}
\end{figure} 
In Figs.~\ref{pEvsmu}
we present the short-distance 
pressure $p_{\rm E}$ as a function of the respective
renormalization scale $\mu_{\rm E}$ -- various
resummed expressions based on the perturbation
series (\ref{pEcandef})-(\ref{pE2}) in powers of
$a(\mu_{\rm E}) = [g_s(\mu_{\rm E})/ 2 \pi ]^2$,
for $T=1$ GeV and $\delta_{\rm E}=0$. The naming
of each resummation refers to the use of the corresponding
approximant for $R_{\rm E}^{\rm can}$ in
$p_{\rm E}$ of Eq.~(\ref{pEcandef}) as a function of 
$a(\mu_{\rm E}) = [g_s(\mu_{\rm E})/ 2 \pi ]^2$.
For example, P[2/1] means that we apply to the
perturbation series (\ref{pE2}) in powers of
$a(\mu_{\rm E})$ the Pad\'e approximant 
${\rm P}[2/1](a(\mu_{\rm E}))$;
BP[1/2] means that we apply P[1/2] to the Borel transform of
expansion (\ref{pE2}) -- cf.~Appendix \ref{app:PBP}.
The range of $\mu_{\rm E}$ is around $2 \pi T$,
i.e., the order of the relevant physical modes
contributing to $p_{\rm E}$. 
Further, the IR cutoff $\Lambda_{\rm E}$ is taken fixed:
$\Lambda_{\rm E} = [ 2 \pi T m_{\rm E}^{(0)}(T) ]^{1/2}$.
The results are normalized by 
$p_{\rm ideal} = ( 8 \pi^2/45) (1 + 21 n_f/32)$,
i.e., the expression for the pressure when $T \to  \infty$ 
[cf.~Eq.~(\ref{pEcandef})].
{}From Fig.~\ref{pEvsmu}(a) we see that the TPS's
are not acceptable, due to too strong
${\mu}_{\rm E}$-dependence. For the same reason,
as seen from Fig.~\ref{pEvsmu}(b), the
approximants P[1/2], P[2/1], and Borel-Pad\'e
BP[2/1] are not acceptable. The only
acceptable candidates are P[1/1], and BP[1/2].
We note that P[1/1] is based explicitly
only on the NLO series (up to $\sim a^2$) of Eq.~(\ref{pE2}).
Nonetheless, we will include later P[1/1] of
$R_{\rm E}^{\rm can}$ in some of our results,
because it is close to BP[1/2] and is very stable under
variation of $\mu_{\rm E}$ (NL ECH also gives values
close to P[1/1] and BP[1/2]).
Although the results are presented only for the case
$T=1$ GeV and $\delta_{\rm E}=0$, the behaviour and the
conclusions remain 
the same for other temperatures, and for other
values of $\delta_{\rm E}$ in the interval (\ref{dEest}).

\begin{figure}[htb]
\begin{minipage}[b]{.49\linewidth}
 \centering\epsfig{file=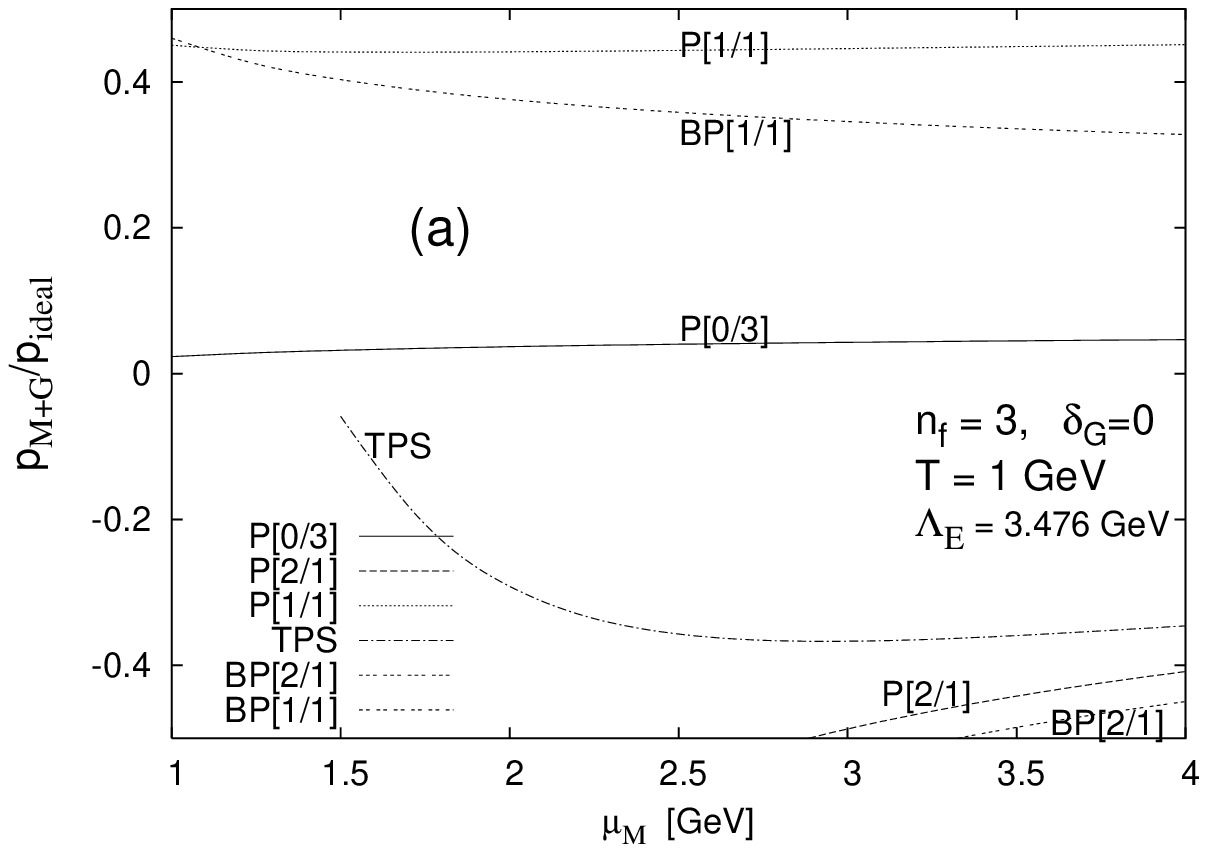,width=\linewidth}
\end{minipage}
\begin{minipage}[b]{.49\linewidth}
 \centering\epsfig{file=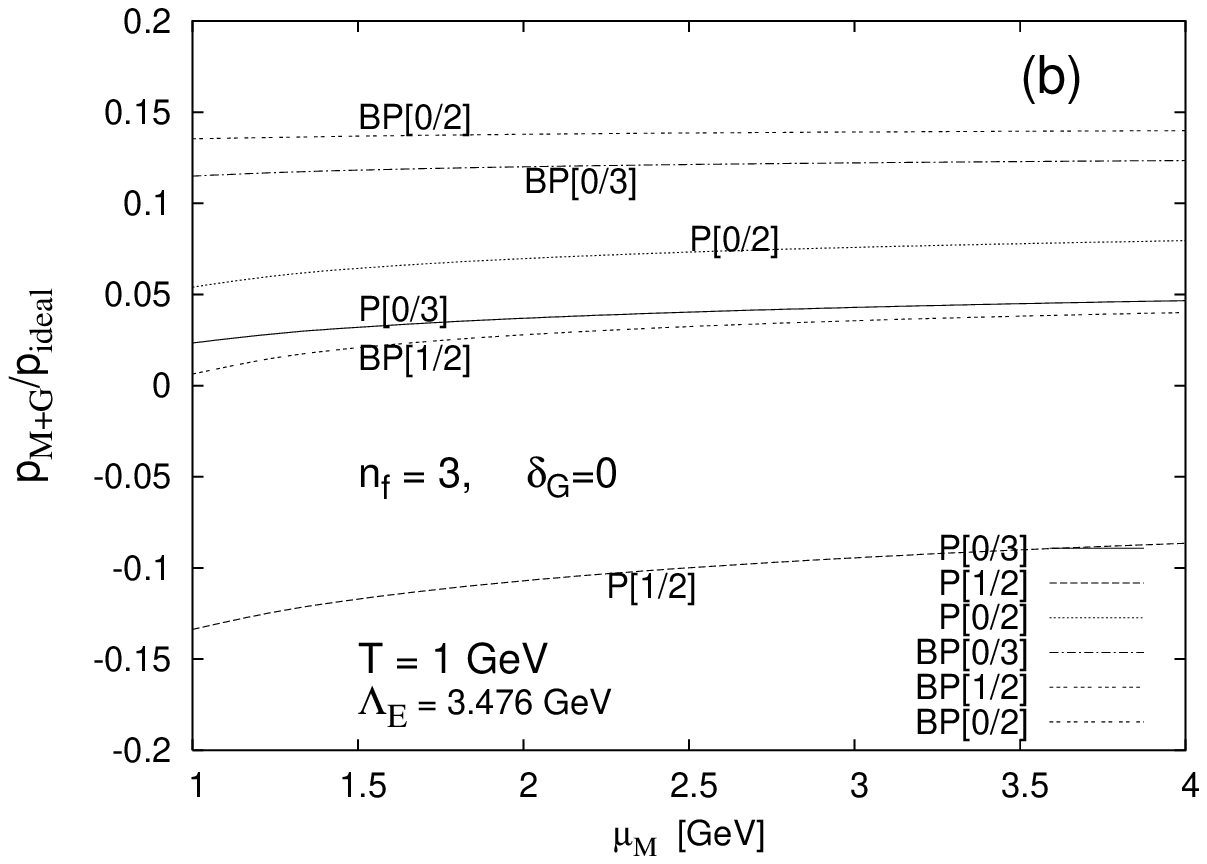,width=\linewidth}
\end{minipage}
%\vspace{0.2cm}
\caption{\footnotesize The long-distance pressure
$p_{\rm M+G}$, at $T=1$ GeV, as a function of the corresponding 
renormalization  scale $\mu_{\rm M}$ -- various approximants are applied
to the perturbation expansion (\ref{pMG}) in powers of
$g_s(\mu_{\rm M})$, and $\delta_{\rm G}=0$ was taken.
Figure (b) has a finer vertical scale and includes some
approximants not shown in Fig.~(a).}
\label{pMGvsmu}
\end{figure} 
\begin{figure}[htb]
\begin{minipage}[b]{.49\linewidth}
 \centering\epsfig{file=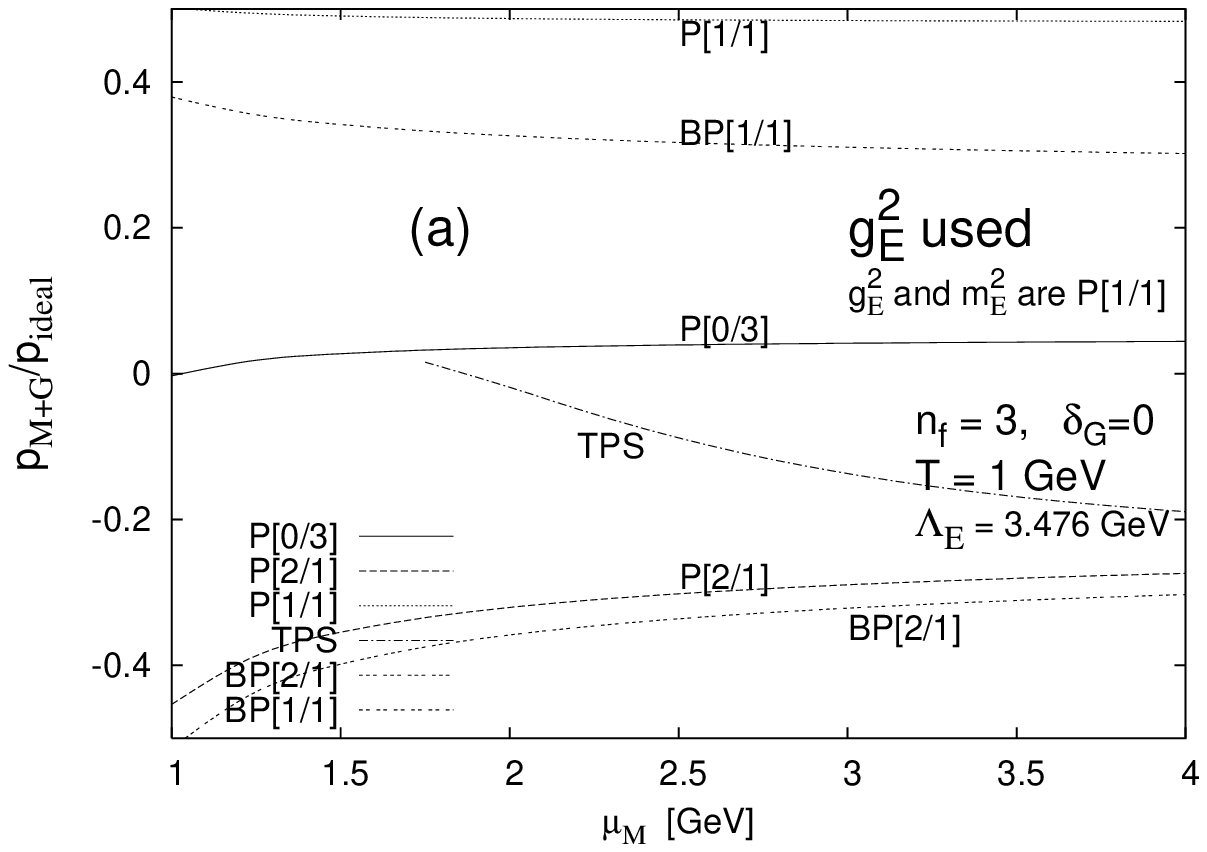,width=\linewidth}
\end{minipage}
\begin{minipage}[b]{.49\linewidth}
 \centering\epsfig{file=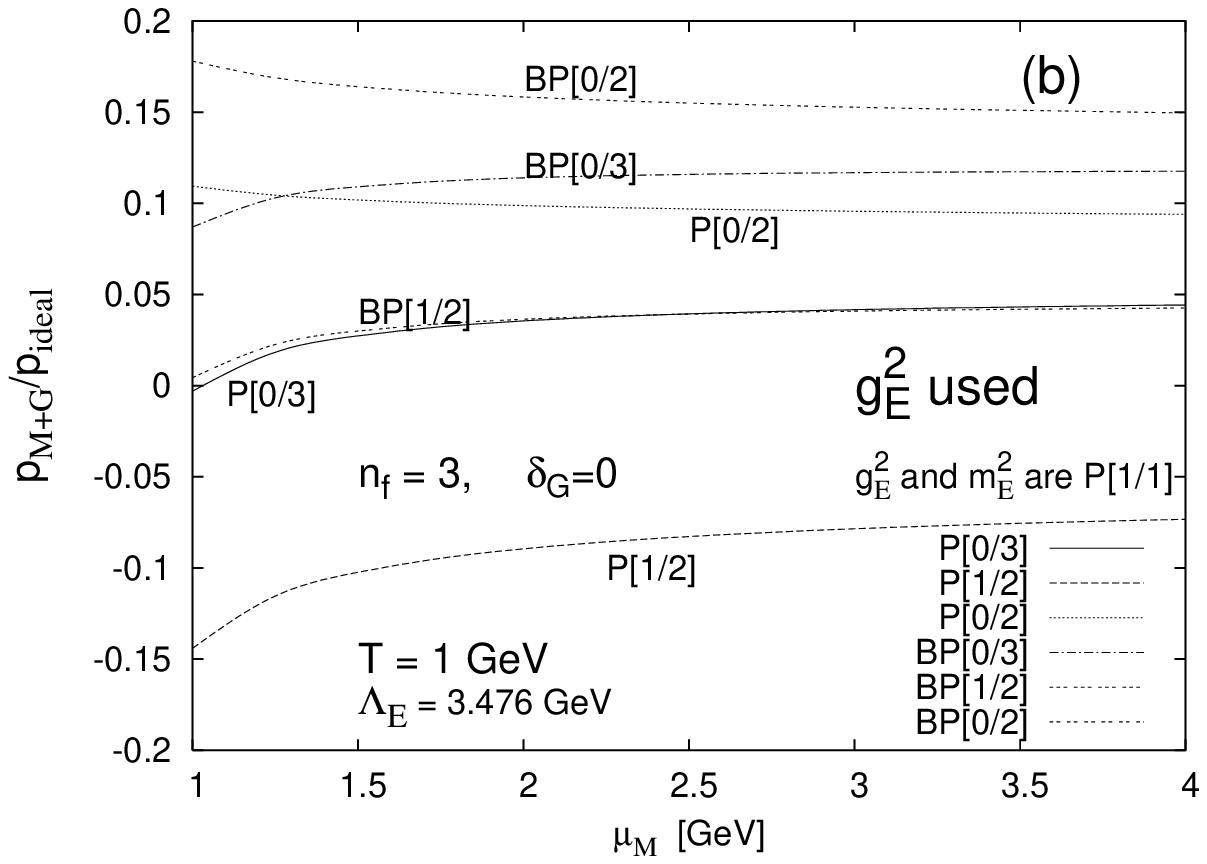,width=\linewidth}
\end{minipage}
%\vspace{0.2cm}
\caption{\footnotesize Same as in Figs.~\ref{pMGvsmu}, but now
${\widetilde p}_{\rm M+G}$ is based on expansion (\ref{pMGeff})
in $g_{\rm E}^2/m_{\rm E}$. 
Details are given in the text.}
\label{pMGeffvsmu}
\end{figure} 
Similar analysis is now performed for the
long-distance pressure $p_{\rm M+G}$. Pad\'e
and Borel-Pad\'e resummations are first applied to
expansion (\ref{pMG}) for ${\widetilde p}_{\rm M+G}$
($\propto p_{\rm M+G}/m_{\rm E}^3$) which starts with one and
is in powers of $g_s(\mu_{\rm M})$ -- cf.~Appendix
\ref{app:PBP} for more details. 
In Figs.~\ref{pMGvsmu} we present
the results of various resummations
as a function of the respective
renormalization scale $\mu_{\rm M}$,
for $T=1$ GeV and $\delta_{\rm G}=0$.
The factor $m_{\rm E}^3$ in the first line of Eq.~(\ref{pMG})
was taken with P[1/1] for $m_{\rm E}^2(\mu_m)$,
with the renormalization scale $\mu_m = \mu_{\rm M}$.
The UV cutoff $\Lambda_{\rm E}$ was taken fixed according to 
the formula: $\Lambda_{\rm E} = ( 2 \pi T m_{\rm E}^{(0)} )^{1/2}$.
The TPS and some of the other resummations show significant
$\mu_{\rm M}$-dependence. 
Further, BP[0/2] and P[0/2] 
are of lower order (NNLO) and do not take explicitly into account the 
numerically important $\sim g_s^3$ terms of expansion (\ref{pMG})
of ${\widetilde p}_{\rm M+G}$ ($\sim g_s^6$ in $p_{\rm M+G}$). 
We decide at this stage to consider only the ${\rm N}^3{\rm LO}$ 
resummations: P[1/2], P[0/3], BP[1/2], and BP[0/3].
The approximant ${\rm P[0/2]}(g_s)$ for ${\widetilde p}_{\rm M(+G)}$
was considered in our previous work
\cite{Cvetic:2002ju}, and we include it later in 
Figs.~\ref{pEMGvsT1} and \ref{pEMGvsTB03}
in the presentation of $p(T)$.
The conclusions of this paragraph survive when other values of $T$ and 
$\delta_{\rm G}$ are used.

We alternatively apply our resummation procedure to the EQCD 
expansion (\ref{pMGeff}) instead of (\ref{pMG})
for ${\widetilde p}_{\rm M+G}$. 
The results are presented in Figs.~\ref{pMGeffvsmu}.
The values of $g_{\rm E}^2$
and $m_{\rm E}^2$ in (\ref{pMGeff}) are chosen as
${\rm P}[1/1](a(\mu_m))$,
due to their ${\mu}_m$-stability 
(${\mu}_m = \mu_{\rm M}$ taken) as seen in Figs.~\ref{figmEgE2},
and then Pad\'e or Borel-Pad\'e are applied to expansion (\ref{pMGeff})
in powers of the EQCD parameter $g_{\rm E}^2/m_{\rm E}$ ($\sim g_s$),
without the $\lambda_{\rm E}^{(1)}/m_{\rm E}$-term which is 
then added separately as the leading order QCD term 
$\propto g_s^3(\mu_{\rm M})$, Eq.~(\ref{l1a}).
When Pad\'e or Borel-Pad\'e based on TPS terms of order
lower than $\sim g_s^3$ for ${\widetilde p}_{\rm M+G}$
of Eq.~(\ref{pMGeff}) are applied -- such as P[1/1], BP[1/1], P[0/2] or 
BP[0/2] to TPS (\ref{pMGeff}) of ${\widetilde p}_{\rm M+G}$
-- the $\lambda_{\rm E}^{(1)}/m_{\rm E}$-term is not included
and not added as it represents a term $\sim g_s^3$ of
${\widetilde p}_{\rm M+G}$ ($\sim g_s^6$ to $p_{\rm M+G}$).
The TPS curve in Fig.~\ref{pMGeffvsmu}(a) was obtained
by evaluating expansion
(\ref{pMGeff}) of ${\widetilde p}_{\rm M+G}$ directly
as EQCD TPS, with the values for $g_{\rm E}^2$ and $m_{\rm E}^2$ taken as
simple (NLO) TPS's (\ref{mE})-(\ref{gE}) at 
${\overline \mu} = {\mu}_{\rm M}$. 
The conclusions from Figs.~\ref{pMGeffvsmu} are the same as in the
procedure leading to Figs.~\ref{pMGvsmu}: 
the ${\rm N}^3{\rm LO}$ 
resummations P[1/2], P[0/3], BP[1/2], and BP[0/3] 
all remain acceptable candidates at this stage,
and this conclusion turns out to be
independent of $T$ and $\delta_{\rm G}$.

\begin{figure}[htb]
\begin{minipage}[b]{.49\linewidth}
 \centering\epsfig{file=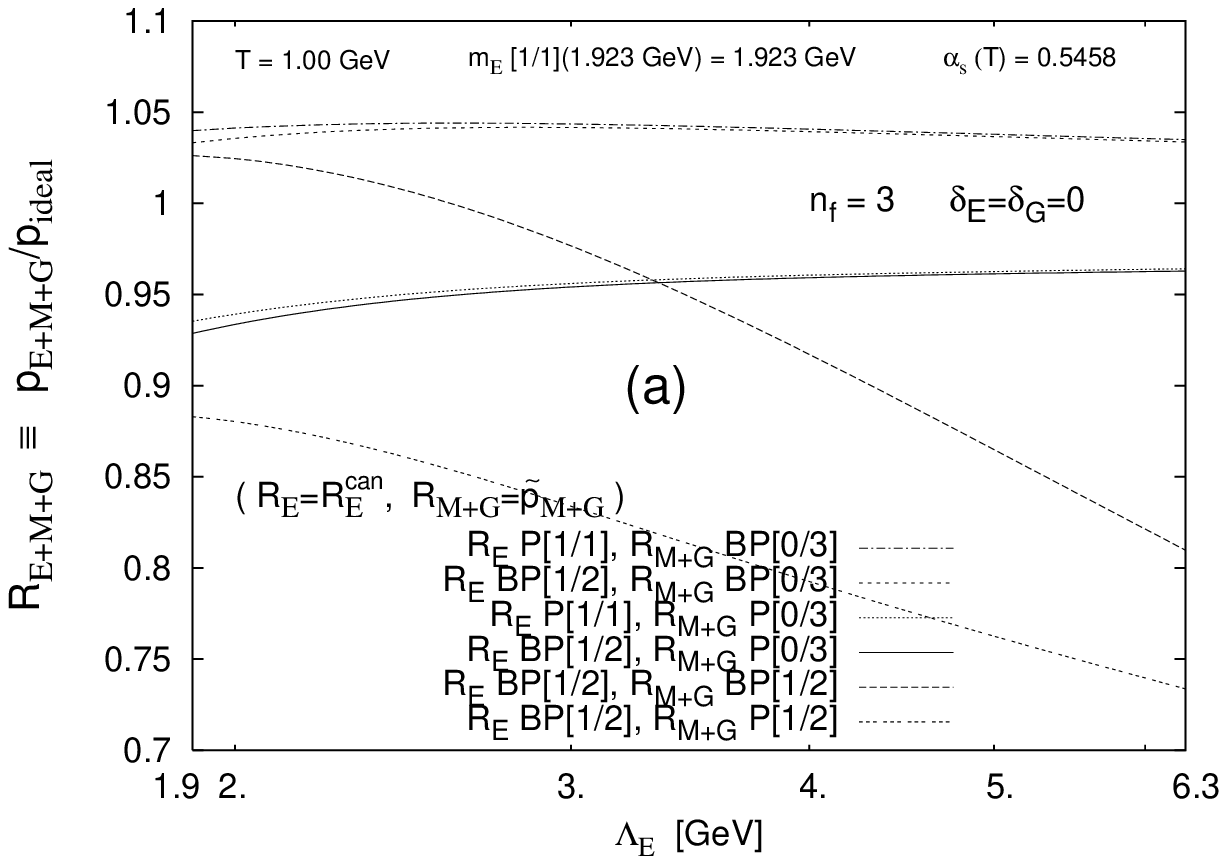,width=\linewidth}
\end{minipage}
\begin{minipage}[b]{.49\linewidth}
 \centering\epsfig{file=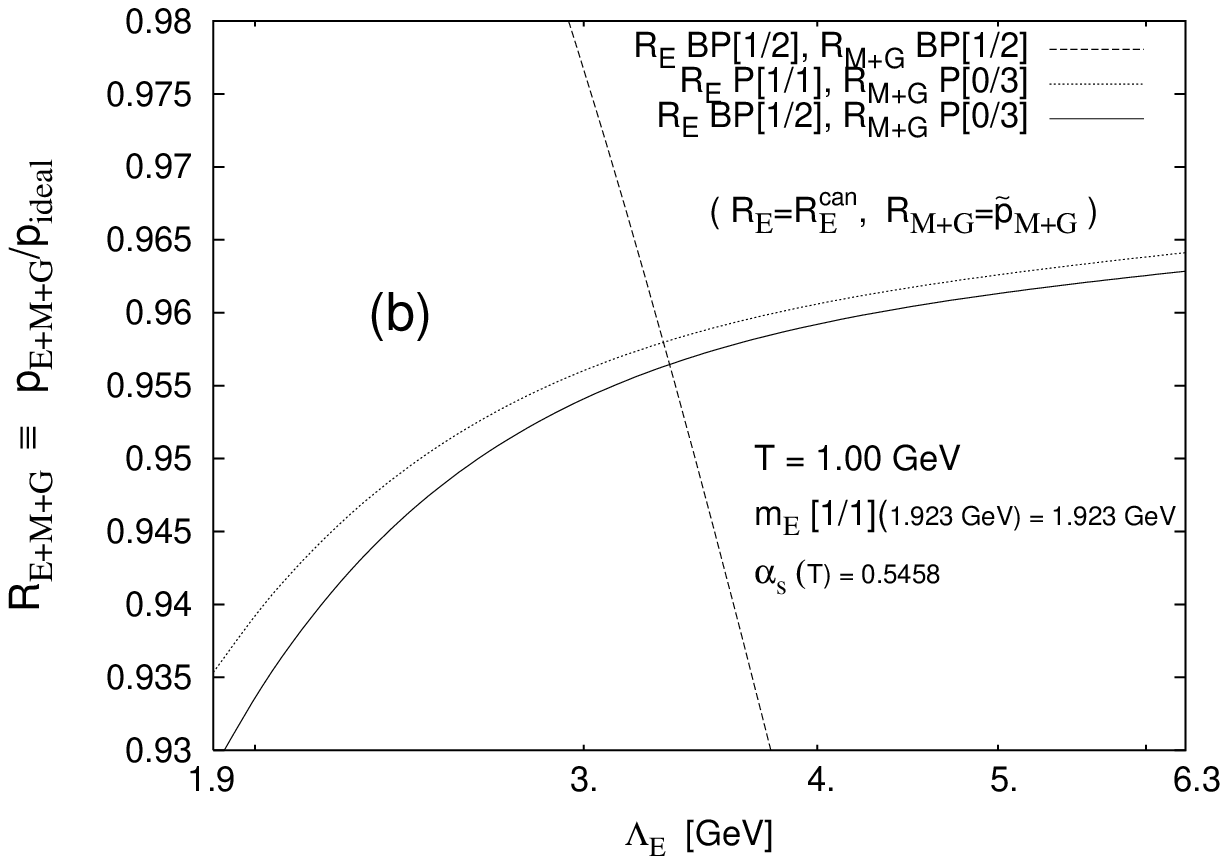,width=\linewidth}
\end{minipage}
%\vspace{0.2cm}
\caption{\footnotesize The total pressure 
$p_{\rm QCD} = p_{\rm E} + p_{\rm M+G}$,
at $T=1$ GeV, as a function of the corresponding 
factorization scale $\Lambda_{\rm E}$ -- various approximants are 
applied separately to the perturbation expansions
$R_{\rm E}^{\rm can}(a(\mu_{\rm E}))$ (\ref{pE2}) and 
${\widetilde p}_{\rm M+G}(g_s(\mu_{\rm M}))$ (\ref{pMG}). 
Details are given in the
text. Figure (b) has a finer vertical scale.}
\label{pEMGvsLE}
\end{figure} 
\begin{figure}[htb]
\begin{minipage}[b]{.49\linewidth}
 \centering\epsfig{file=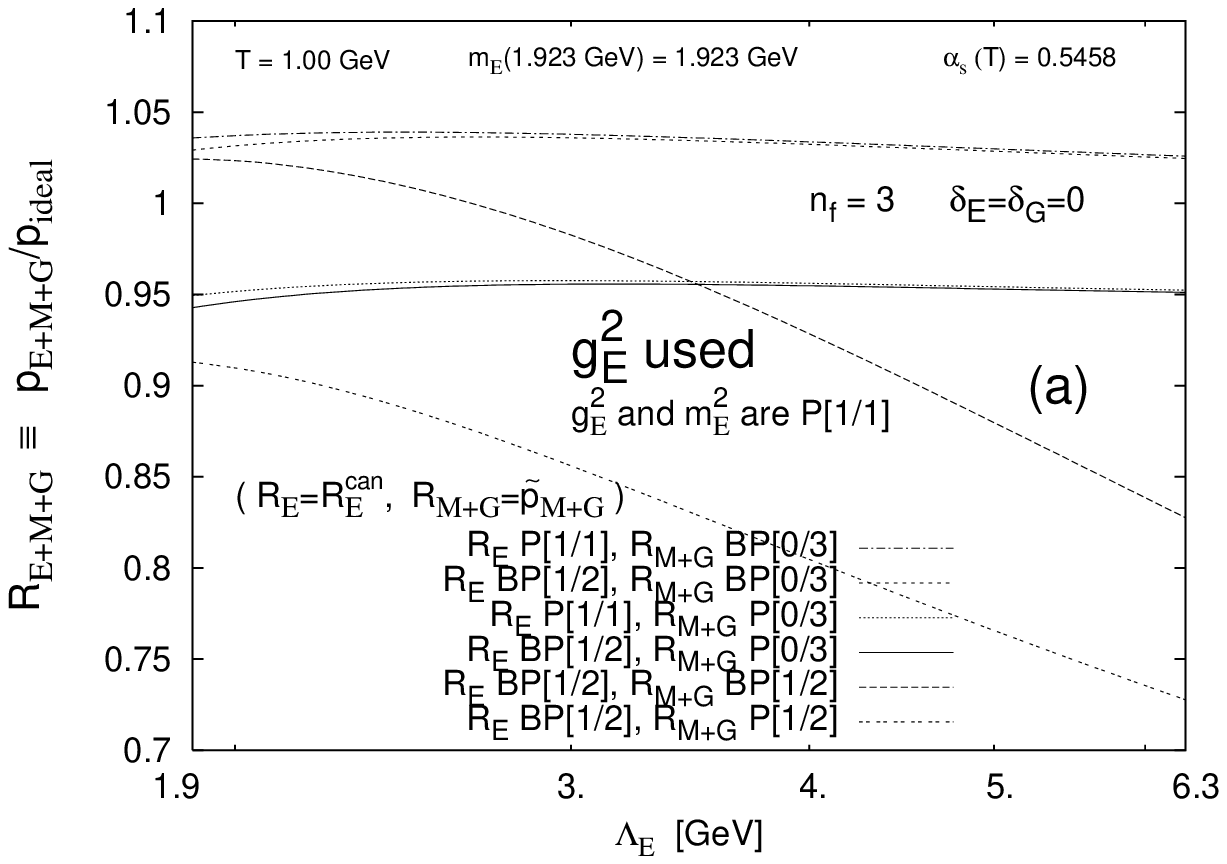,width=\linewidth}
\end{minipage}
\begin{minipage}[b]{.49\linewidth}
 \centering\epsfig{file=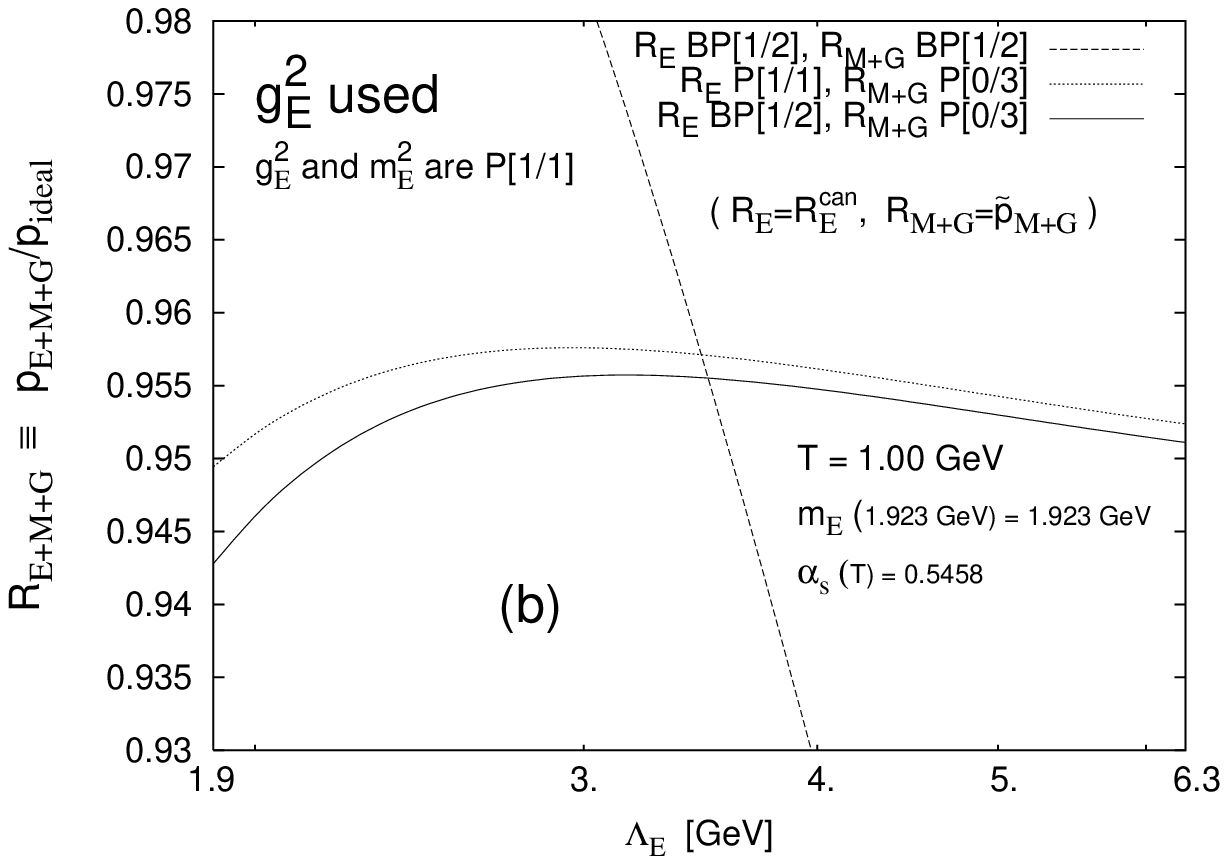,width=\linewidth}
\end{minipage}
%\vspace{0.2cm}
\caption{\footnotesize Same as in Figs.~\ref{pEMGvsLE},
but now ${\widetilde p}_{\rm M+G}$ is based on expansion (\ref{pMGeff})
in $g_{\rm E}^2/m_{\rm E}$.
Further details are given in the text.}
\label{pEMGeffvsLE}
\end{figure} 
Another necessary condition for an acceptable
resummation is that the spurious $\Lambda_{\rm E}$-dependence
in the total sum $p_{\rm E} + p_{\rm M+G}$ be weak.
In Figs.~\ref{pEMGvsLE} we present the 
results on $\Lambda_{\rm E}$-dependence,
at $T=1$ GeV, combining various aforementioned resummations 
for $p_{\rm E}$ and $p_{\rm M+G}$ that were found acceptable so far.
Here, $m_{\rm E} = m_{\rm E}^{(0)}$ 
was fixed in the aforementioned way (\ref{m0T}),
$\mu_{\rm E} = 2 \pi T$, and $\mu_{\rm M} = \mu_m = m_{\rm E}^{(0)}$.
Now, $\Lambda_{\rm E} \not= \sqrt{ \mu_{\rm E} \mu_{\rm M} }$,
but varies between $\mu_{\rm M}$ ($\approx 1.9$ GeV) and 
$\mu_{\rm E}$ ($\approx 6.3$ GeV). The two unknown parameters
were chosen to be $\delta_{\rm E} = \delta_{\rm G} = 0$. 
We see that the curves with P[0/3] and B[0/3] used for
${\widetilde p}_{\rm M+G}$ have acceptably weak $\Lambda_{\rm E}$-dependence,
while those with P[1/2] and BP[1/2] used for
${\widetilde p}_{\rm M+G}$ have unacceptably strong
$\Lambda_{\rm E}$-dependence.
Thus, we are left with just two types of resummations which
fulfill the necessary conditions of weak $\mu_{\rm E}$,
$\mu_{\rm M}$ and $\Lambda_{\rm E}$ dependence:
\begin{enumerate}
\item
${\rm P}[0/3](g_s(\mu_{\rm M}))$ for ${\widetilde p}_{\rm M+G}$;
and either ${\rm P}[1/1](a(\mu_{\rm E}))$ or 
${\rm BP}[1/2](a(\mu_{\rm E}))$ for 
$R_{\rm E}^{\rm can}(a(\mu_{\rm E}))$;

\item
${\rm BP}[0/3](g_s(\mu_{\rm M}))$ for ${\widetilde p}_{\rm M+G}$;
and either ${\rm P}[1/1](a(\mu_{\rm E}))$ or 
${\rm BP}[1/2](a(\mu_{\rm E}))$ for 
$R_{\rm E}^{\rm can}(a(\mu_{\rm E}))$.\footnote{
NL ECH could be used instead of P[1/1] or BP[1/2] for
$R_{\rm E}^{\rm can}$, but the results are similar
in all three cases, at any temperature.}
\end{enumerate}
These conclusions do not change when the values of $T$,
$\delta_{\rm E}$ and $\delta_{\rm G}$ are changed.
We note that ${\rm P}[1/1](a(\mu_{\rm E}))$ is of lower order
and therefore does not use the term $\sim a^3$ in expansion
(\ref{pE2}).

We can repeat the same analysis, but using expansion
(\ref{pMGeff}) for ${\widetilde p}_{\rm M+G}$
in powers of $(g_{\rm E}^2/m_{\rm E})$
instead of expansion (\ref{pMG}) in powers of $g_s(\mu_{\rm M})$.
The values of $g_{\rm E}^2$ and $m_{\rm E}^2$ are taken
as ${\rm P}[1/1](a(\mu_{\rm M}))$, with
$\mu_{\rm M} = \mu_m = m_{\rm E}^{(0)}$.
The results are given in Figs.~\ref{pEMGeffvsLE},
they are similar to those of Figs.~\ref{pEMGvsLE}, and the
conclusions are the same.

\section{Numerical results as a function of temperature}
\label{sec:numres}

\begin{figure}[htb]
\begin{minipage}[b]{.49\linewidth}
 \centering\epsfig{file=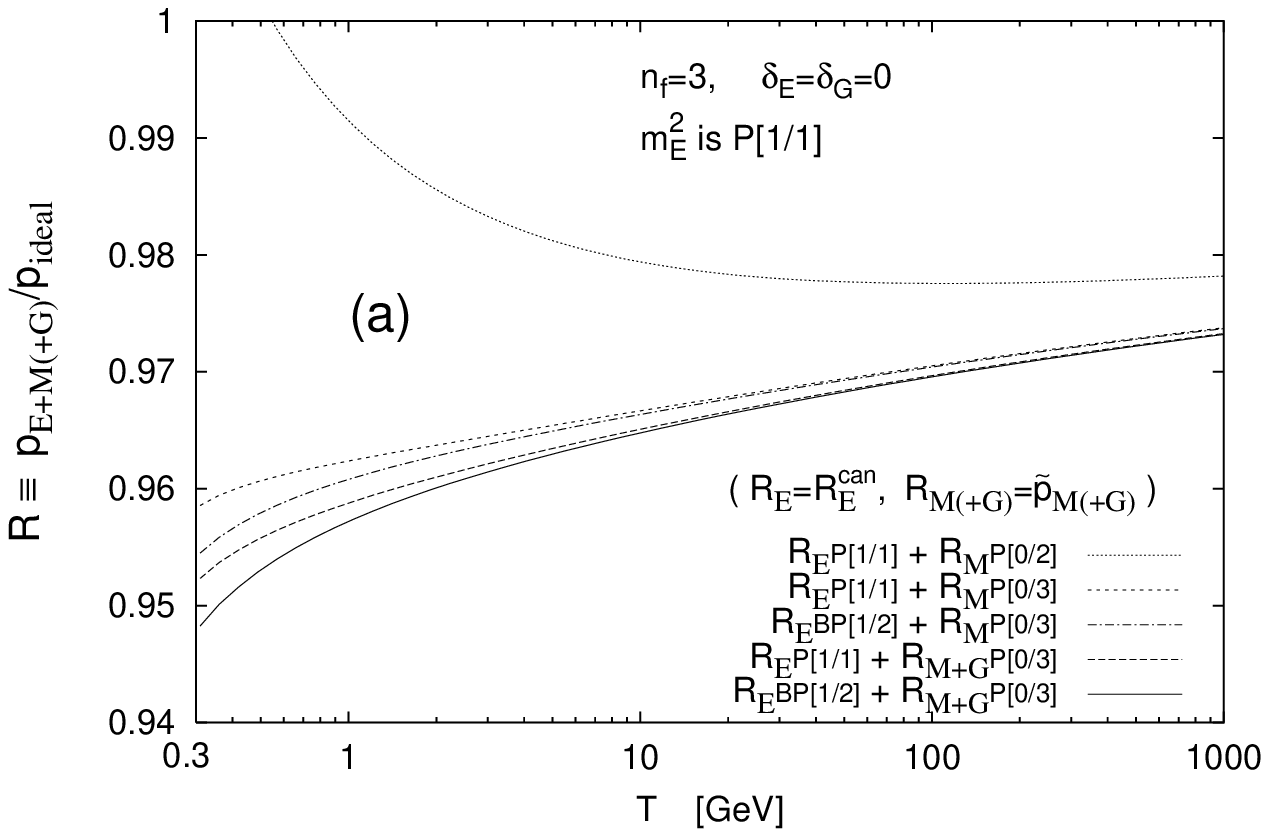,width=\linewidth}
\end{minipage}
\begin{minipage}[b]{.49\linewidth}
 \centering\epsfig{file=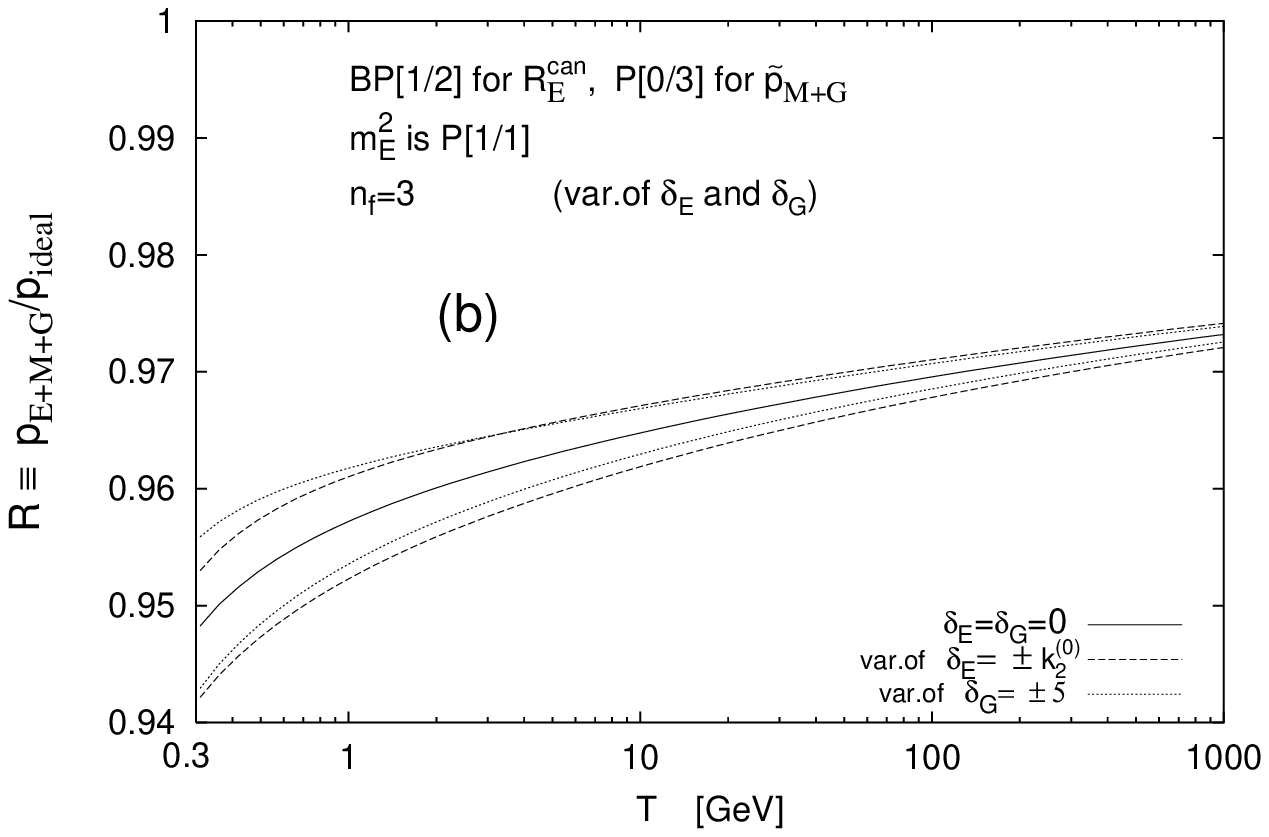,width=\linewidth}
\end{minipage}
%\vspace{0.2cm}
\caption{\footnotesize 
(a) The total pressure $p$ (normalized by $p_{\rm ideal}$)
as a function of temperature $T$, 
with $\delta_{\rm E} = \delta_{\rm G} = 0$ --
various resummations are applied: 
for $R_{\rm E}^{\rm can}(a(\mu_{\rm E}))$
the approximants ${\rm P}[1/1](a(\mu_{\rm E}))$ 
and ${\rm BP}[1/2](a(\mu_{\rm E}))$; 
for ${\widetilde p}_{\rm M+G}(g_s(\mu_{\rm M}))$ the 
approximant ${\rm P}[0/3](g_s(\mu_{\rm M}))$. 
Shown are also the analogous results when $p_{\rm G}$ is
excluded. In addition, a resummation which does
not account for the $\sim g_s^6$ terms in $p_{\rm M+G}$
is included (dotted line). (b) Variation of a specific 
Pad\'e/Borel-Pad\'e
resummation when the unknown parameters $\delta_{\rm G}$
and $\delta_{\rm E}$ are varied. The full curves in (a) and (b)
are the same curves. Further explanations are
given in the text.}
\label{pEMGvsT1}
\end{figure}

We will first concentrate on the first family of
resummations, i.e., those with P[0/3] for ${\widetilde p}_{\rm M+G}$.
We present the results for these resummations
as a function of temperature $T$ in Fig.~\ref{pEMGvsT1}(a),
for $\delta_{\rm E} = \delta_{\rm G} = 0$, and 
${\widetilde p}_{\rm M+G}$ is based on expansion (\ref{pMG})
in powers of $g_s(\mu_{\rm M})$.
In addition to these resummations, we include also
the same type of resummations where the $p_{\rm G}$
part is excluded ($p_{\rm E} + p_{\rm M}$), where we use
in $p_{\rm M}$ for the IR cutoff: $\Lambda_{\rm M} =
(m_{\rm E}^{(0)})^2/\Lambda_{\rm E}$. We use
$\mu_{\rm E} = 2 \pi T$; 
$\mu_{\rm M} = \mu_m = m_{\rm E} = m_{\rm E}^{(0)}(T)$;
$\Lambda_{\rm E} = \sqrt{ \mu_{\rm E} \mu_{\rm M} }$.
\begin{figure}[htb]
\begin{minipage}[b]{.49\linewidth}
 \centering\epsfig{file=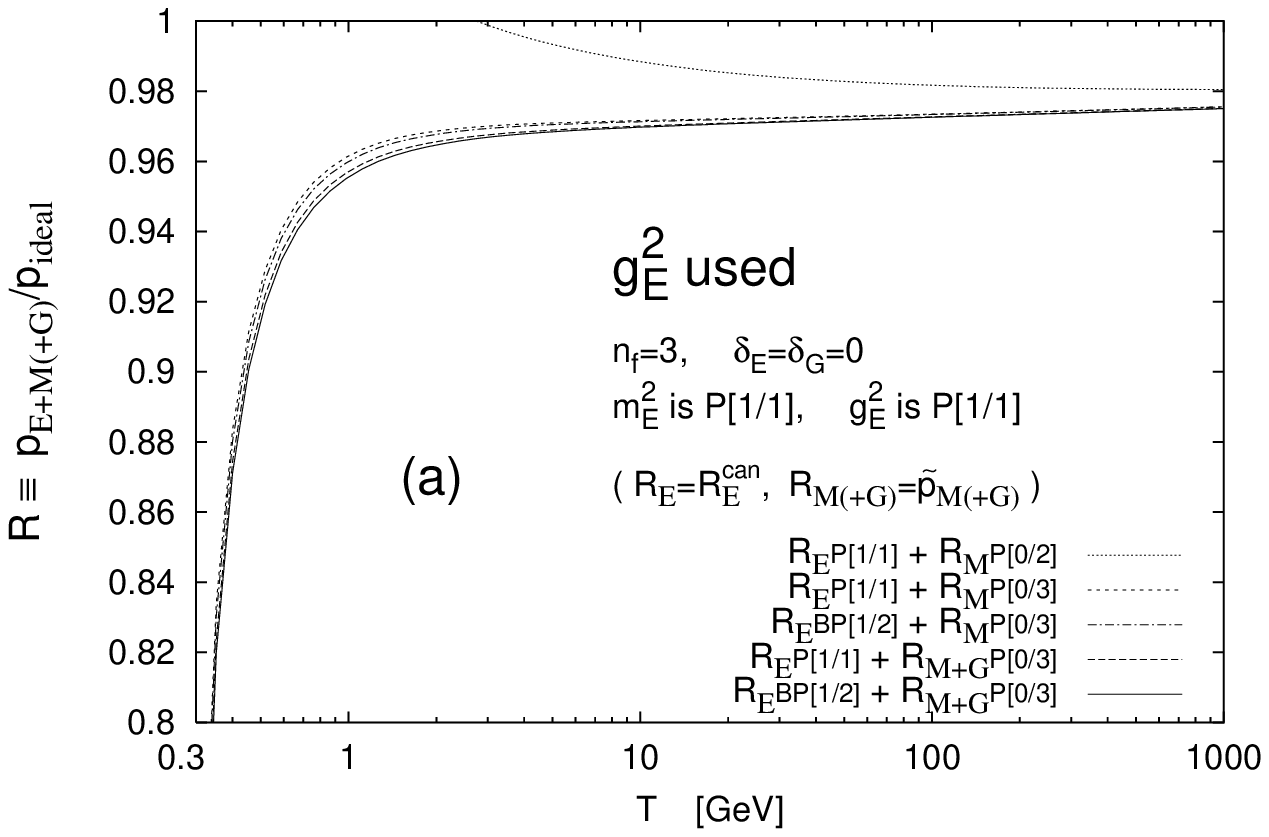,width=\linewidth}
\end{minipage}
\begin{minipage}[b]{.49\linewidth}
 \centering\epsfig{file=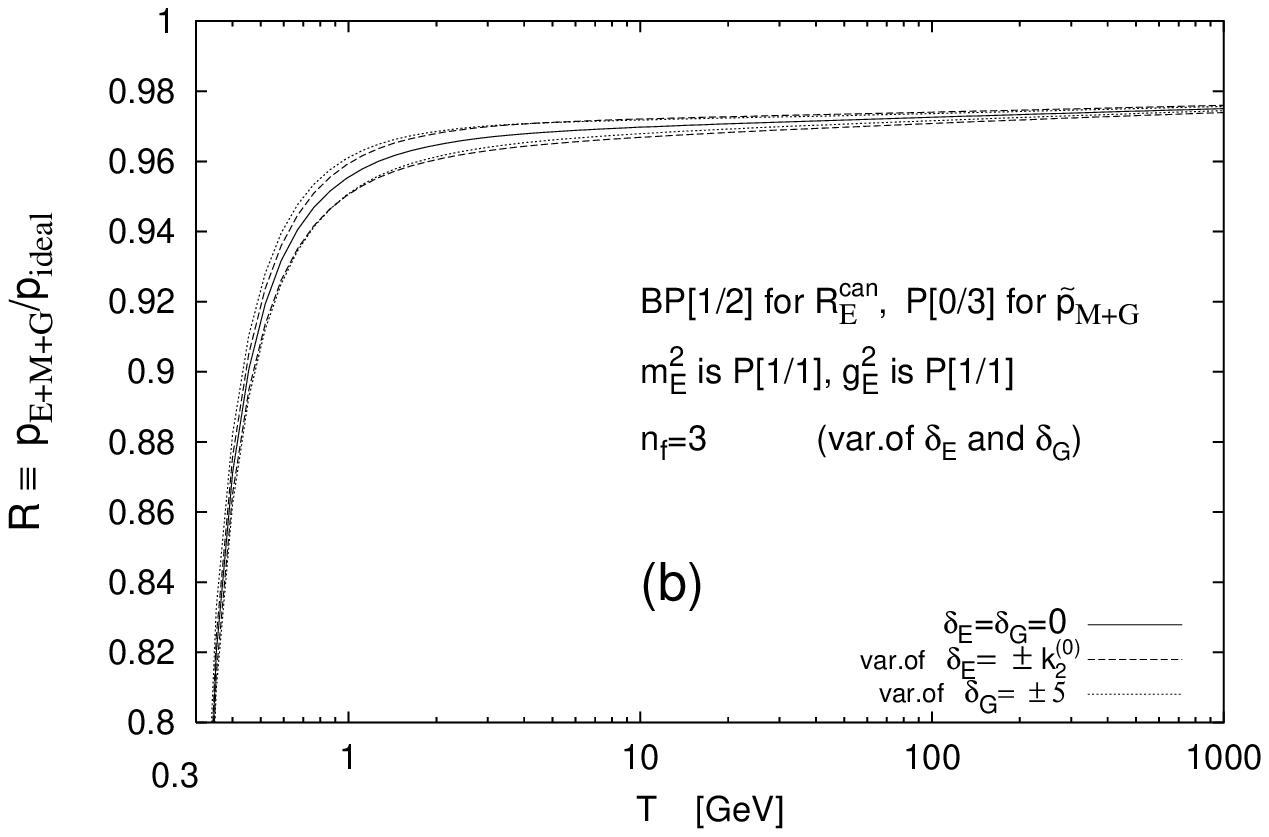,width=\linewidth}
\end{minipage}
%\vspace{0.2cm}
\caption{\footnotesize Same as in Fig.~\ref{pEMGvsT1},
but now ${\widetilde p}_{\rm M+G}$ is based on expansion (\ref{pMGeff})
in $g_{\rm E}^2/m_{\rm E}$ -- using the resummed $g_{\rm E}^2$ and 
$m^2_{\rm E}$ as ${\rm P}[1/1](a(\mu_{\rm M}))$.} 
\label{pEMGvsT1eff}
\end{figure} 
We can see in Fig.~\ref{pEMGvsT1}(a) that the presence of $p_{\rm G}$,
at least for $\delta_{\rm G} = 0$, decreases the value of
the total pressure somewhat. For comparison, we also include
the result of resummation of P[1/1] for $R_{\rm E}^{\rm can}$
and P[0/2] for ${\widetilde p}_{\rm M}$ (cf.~Ref.~\cite{Cvetic:2002ju}),
i.e., the case where the terms $\sim g_s^3$ in ${\widetilde p}_{\rm M+G}$
(terms $\sim g_s^6$ in $p_{\rm M+G}$) are not explicitly
accounted for (and neither are the terms $\sim g_s^6$ in $p_{\rm E}$).
Fig.~\ref{pEMGvsT1}(a) shows one interesting
feature: when the terms $\sim g_s^6$ in $p_{\rm M+G}$
are explicitly accounted for in the resummation, the sign of the 
curvature becomes negative
in the entire temperature interval -- i.e., at least there where
the resummation is applicable: $T > 0.3$ GeV. Thus the curvature
at low temperatures has now the same sign as suggested by the
known relation $p/p_{\rm ideal} << 1$ at $T \sim T_c \approx 0.2$ GeV
(see later in this Section).
For $T < 0.3$ GeV our resummations cannot be
applied any more, because in that case $\mu_m (=\mu_{\rm M}) < 0.81$ GeV,
but the ${\rm P}[2/3](a)$ beta function does not allow running
below such values [$a(\mu_m)$ blows up]. 
When we vary the values of $\delta_{\rm G}$ and $\delta_{\rm E}$
in the intervals (\ref{dGvar}) and (\ref{dEest}), respectively,
the predictions do not change much. This is presented
in Fig.~\ref{pEMGvsT1}(b). 

When we do not base the resummations of ${\widetilde p}_{\rm M+G}$ 
on expansion (\ref{pMG}) in powers of $g_s(\mu_{\rm M})$,
but rather on expansion (\ref{pMGeff})
in powers of the EQCD parameter $g_{\rm E}^2/m_{\rm E}$ 
[using the resummed $g_{\rm E}^2$ and 
$m^2_{\rm E}$ as ${\rm P}[1/1](a(\mu_{\rm M}))$,
and adding the $\lambda_{\rm E}^{(1)}/m_{\rm E}$-term 
separately as the leading order QCD term 
$\propto g_s^3(\mu_{\rm M})$, Eq.~(\ref{l1a})],
remarkable changes occur for the results of
$p/p_{\rm ideal}$ as a function of temperature.
The obtained results, analogous to those of Figs.~\ref{pEMGvsT1}(a)
and \ref{pEMGvsT1}(b), 
are presented in Figs.~\ref{pEMGvsT1eff}(a) and
\ref{pEMGvsT1eff}(b), respectively. 
The ``low order'' dotted curve in Fig.~\ref{pEMGvsT1eff}(a)
(the uppermost), which 
does not use information on $\sim g_s^6$ terms (and thus does not
include the $\lambda_{\rm E}^{(1)}$-term), still does not change the
curvature in the low temperature regime.
On the other hand, ``higher order'' Pad\'e resummations
of ${\widetilde p}_{\rm M+G}$, which use information on
$\sim g_s^6$ terms in $p_{\rm M+G}$ ($\sim g_s^3$ in
${\widetilde p}_{\rm M+G}$) and include, added separately,
the $\lambda_{\rm E}^{(1)}/m_{\rm E}$-term 
in ${\widetilde p}_{\rm M+G}$ as the leading order QCD term 
$\propto g_s^3(\mu_{\rm M})$ [Eq.~(\ref{l1a})],
result in a pronounced negative curvature and
a rapid fall of $p/p_{\rm ideal}$ when the temperature
falls down toward the critical values $T_c \approx 0.2$ GeV.
This behavior is qualitatively correct
because we know that it must be
$p/p_{\rm ideal} << 1$ at $T \approx T_c$ (see later).
In that respect, this phenomenon indicates
that the resummations of ${\widetilde p}_{\rm M+G}$
based on expansion (\ref{pMGeff})
in powers of the effective EQCD theory parameter
$g_{\rm E}^2/m_{\rm E}$ are more reliable 
than those based on expansion (\ref{pMG}) in powers of $g_s$.
The low-$T$ results of Figs.~\ref{pEMGvsT1eff} turn out to be closer 
to those of lattice calculations, the feature which
will be discussed and presented in more detail in Section \ref{sec:comp}.
Another very positive feature can be read off from
Fig.~\ref{pEMGvsT1eff}(b): variation of the
Pad\'e and Borel-Pad\'e resummed predictions, when the 
unknown parameters $\delta_{\rm G}$ and $\delta_{\rm E}$ are varied
in the generously wide intervals (\ref{dGvar}) and
(\ref{dEest}), is weak.

\begin{figure}[htb]
\begin{minipage}[b]{.49\linewidth}
 \centering\epsfig{file=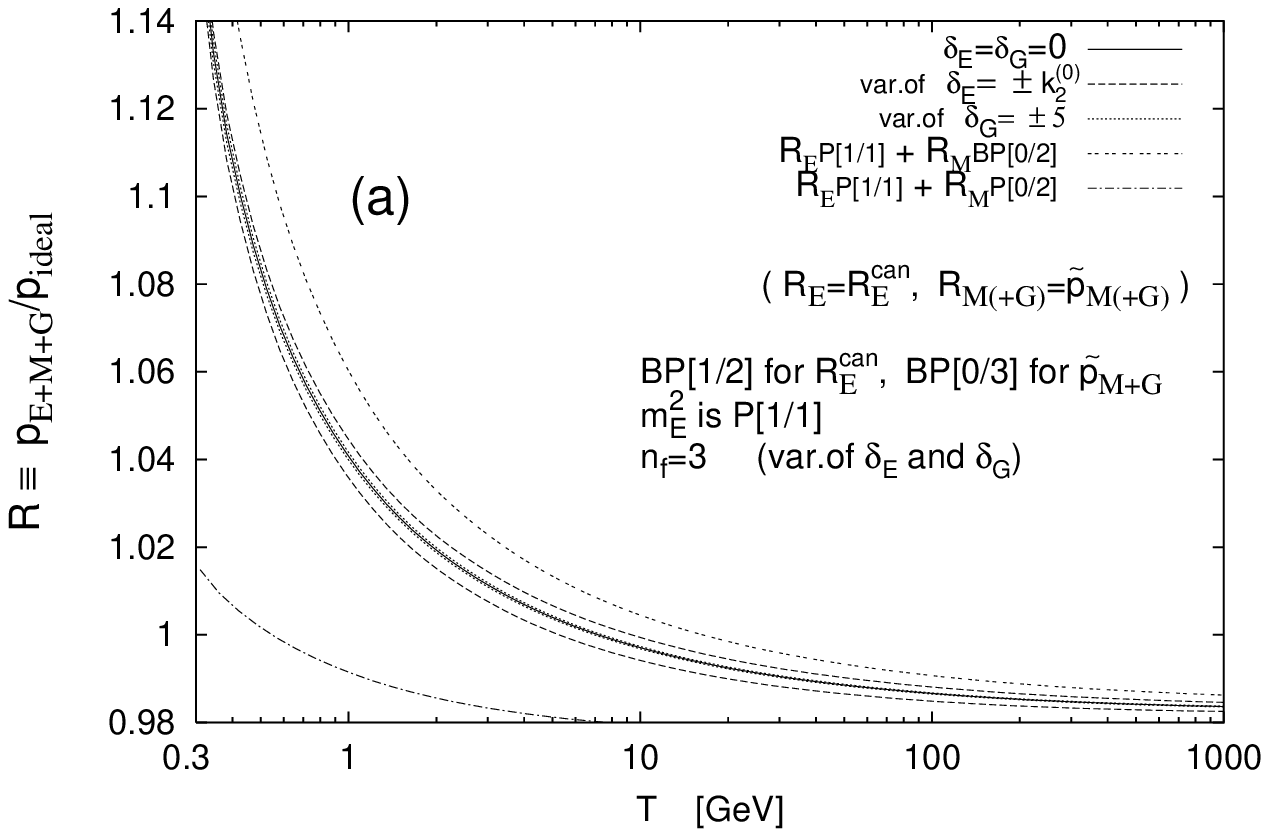,width=\linewidth}
\end{minipage}
\begin{minipage}[b]{.49\linewidth}
 \centering\epsfig{file=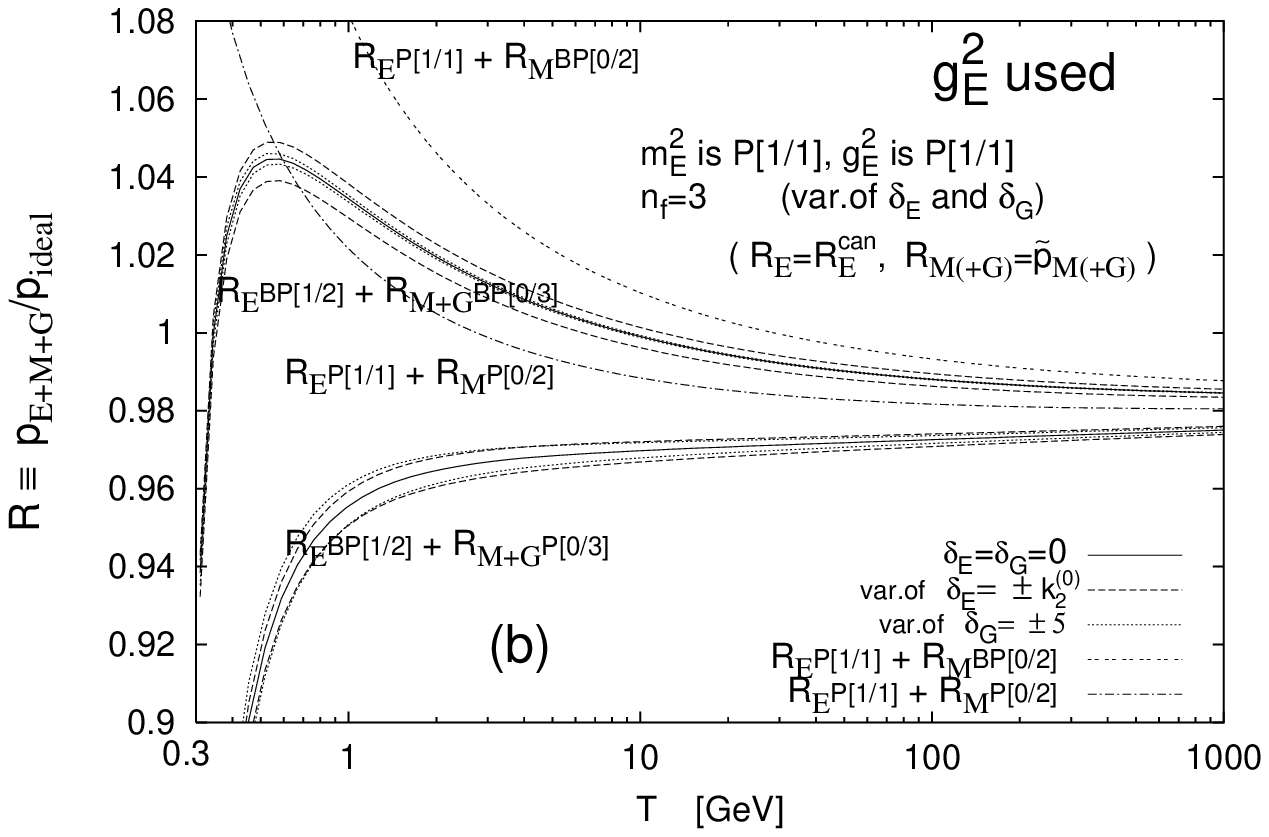,width=\linewidth}
\end{minipage}
%\vspace{0.2cm}
\caption{\footnotesize (a) The total pressure $p$ (normalized by
$p_{\rm ideal}$) as a function of temperature $T$, for various
values of $\delta_{\rm E}$ and $\delta_{\rm G}$, when
BP[0/3] is applied to expansion (\ref{pMG}) for
${\widetilde p}_{\rm M+G}$ (and BP[1/2] to $R_{\rm E}^{\rm can}$);
(b) Same as in (a), but BP[0/3] for ${\widetilde p}_{\rm M+G}$ is
based on expansion (\ref{pMGeff}) in $g_{\rm E}^2/m_{\rm E}$,
using the resummed $g_{\rm E}^2$ and 
$m^2_{\rm E}$ as ${\rm P}[1/1](a(\mu_{\rm M}))$.}
\label{pEMGvsTB03}
\end{figure}

Now we turn to the second family of resummations, i.e.,
those with BP[0/3] for ${\widetilde p}_{\rm M+G}$
instead of P[0/3].
We present these results, as a function of temperature $T$,
with values of $\delta_{\rm E}$ and $\delta_{\rm G}$
varied in the intervals (\ref{dEest}) and (\ref{dGvar}),
in Figs.~\ref{pEMGvsTB03}(a) and \ref{pEMGvsTB03}(b).
In Fig.~\ref{pEMGvsTB03}(a), ${\rm BP}[0/3]$ was applied to expansion
(\ref{pMG}) for ${\widetilde p}_{\rm M+G}$; 
in Fig.~\ref{pEMGvsTB03}(b) to expansion (\ref{pMGeff}) 
for ${\widetilde p}_{\rm M+G}$
without the $\lambda_{\rm E}^{(1)}/m_{\rm E}$-term, the latter
was taken as the leading order $\propto g_s^3(\mu_{\rm M})$
of QCD expansion (\ref{l1a}) 
and added separately
(only in the ``higher order'' P[0/3] and BP[0/3] cases).
Further, the other parameters and procedures
are the same as in Figs.~\ref{pEMGvsT1}-\ref{pEMGvsT1eff}.
For comparison, the lower order counterpart, i.e., with
BP[0/2] for ${\widetilde p}_{\rm M+G}$ (and P[1/1] for
$R_{\rm E}^{\rm can}$) is included in these Figures.
In addition, for comparison, the corresponding
curves when P[0/3] is applied for ${\widetilde p}_{\rm M+G}$,
and at lower order P[0/2], are included whenever visible\footnote{
It is natural to regard P[0/2] and P[0/3] (for ${\widetilde p}_{\rm M+G}$)
as part of a sequence of approximations, and BP[0/2] and BP[0/3]
as part of another sequence of approximations -- e.g., Ref.~\cite{Baker}
and Figs.~\ref{pEMGvsTB03}(a) and \ref{pEMGvsTB03}(b).} --
see also Figs.~\ref{pEMGvsT1}-\ref{pEMGvsT1eff}.
We observe from Figs.~\ref{pEMGvsTB03} and \ref{pEMGvsT1}-\ref{pEMGvsT1eff}
that the choice BP[0/3] for $p_{\rm M+G}$ gives
$p/p_{\rm ideal} > 1$ for most of the low temperatures $T$,
and the choice P[0/3] for $p_{\rm M+G}$ gives $p/p_{\rm ideal} < 1$.

We recall that BP[0/3] for ${\widetilde p}_{\rm M+G}$
was equally acceptable as P[0/3], 
when only the demand for weak renormalization and factorization 
scale dependence is used as a criterion, 
Figs.~\ref{pMGvsmu}-\ref{pEMGeffvsLE}.
However, there
are several indications (not using the approximate knowledge of the
low-temperature curves of $p/p_{\rm ideal}$ from lattice calculations)
that the choice BP[0/3] is less acceptable than P[0/3].
Some of them are due to general physical considerations, others are
connected with the specific numerical approximation 
techniques applied here.

Physical considerations provide at least two arguments for expecting
$p/p_{ideal} < 1$ and, consequently, for favoring the choice P[0/3]:
\begin{enumerate}
\item
{}From the physical point of view one would expect that the
pressure of a (relativistic) quark-gluon gas gets lowered relative
to the free particle case once the interaction is switched on, simply
because we expect that the behavior of interacting massless quarks
and gluons is approximately described by (almost free) massive 
quasi-particles, the mass stemming from Debye screening. Such a 
behavior is manifest not only within a non-relativistic electromagnetic
plasma (calculated according to Debye-H\"uckel), but shows up also in
specific model calculations for a relativistic plasma \cite{Biro,Blaizot}.
Therefore, we expect that $p/p_{\rm ideal}< 1$ for $T$ close to critical
temperatures $T_c$ ($\approx 0.15-0.25$ GeV).
\item
The same qualitative behavior is inferred from a  thermodynamic
consideration \cite{Satz:2000bn}: We know that the pressure 
(considered as a function of $T$) remains continuous at the 
phase transiton point $T = T_c$. But
below $T_c$ the system is a hadron (mostly pion) gas, which to a good
approximation can be described by an ideal pion gas. The corresponding
pressure is 
$$
p_{\pi} = 3 \frac{\pi^2}{90} T^4 = \frac{3}{16} \frac{8 \pi^2}{45} T^4
\qquad (T < T_c)
$$ 
which is much
smaller than the pressure for an ideal gas of quarks (with $n_f$ flavours)
and gluons, Eq.~(\ref{pideal}).
Consequently, the true pressure of the interacting plasma at temperature
$T \agt T_c$ (but close to $T_c$)
must be much smaller than the ideal gas value at the same temperature, and
we again obtain $p/p_{\rm ideal} < 1$ for $T \agt T_c$.
\end{enumerate}
{}From Fig.~\ref{pEMGvsTB03}(a) it is seen that 
$p/p_{\rm ideal} > 1$ above $T_c$ and even seems to increase when
$T  \to T_c$ if the approximant BP[0/3] is applied to expansion 
(\ref{pMG}) of $\widetilde{p}_{\rm M+G}$ 
($g_{\rm E}^2$ not being used/resummed), 
whereas P[0/3] is in accordance with the aforementioned expectation that 
$p/p_{\rm ideal} < 1$ [Fig.~\ref{pEMGvsT1}(a)]. On the other hand,
when expansion (\ref{pMGeff}) is used as the basis for resummation
(with $g_E^2$ resummed), $p/p_{\rm ideal}$ falls below one
for temperatures very close to $T_c$ even in the case BP[0/3]
(but $g_{\rm E}^2$ resummed), but it is still larger than $1$ for a
considerable temperature region above but not far from $T_c$ 
[Fig.~\ref{pEMGvsTB03}(b)]. 
This indicates that using expansion (\ref{pMGeff}) instead of 
(\ref{pMG}) gives in general more realistic results.

Numerical considerations provide at least two other arguments in favor
of the choice P[0/3] for $p_{\rm M+G}$:
\begin{enumerate}
\item 
If BP[0/3] for ${\widetilde p}_{\rm M+G}$ gave results
closer to the true values of $p_{\rm M+G}$ than P[0/3],
then Fig.~\ref{pEMGvsTB03}(b) would suggest that
the lower order counterpart P[0/2] (to P[0/3])
for ${\widetilde p}_{\rm M+G}$ gives results which
lie closer to the true values of $p_{\rm M+G}$
than those from P[0/3] for $T > 0.5$ GeV -- a situation that has
to be regarded as unlikely.\footnote{
Note that the lower order ``counterpart'' P[1/1] (to BP[1/2])
for $R_{\rm E}^{\rm can}$ gives, at any temperature, 
values of $p_{\rm E}$ very similar to those
of BP[1/2] -- cf.~Figs.~\ref{pEMGvsT1}(a) and
\ref{pEMGvsT1eff}(a).}
\item
The ${\cal O}(g_s^3)$ terms of ${\widetilde p}_{\rm  M+G}$
predicted by re-expansion of BP[0/2] are clearly
worse than those predicted by P[0/2], cf. Tables 1 and 2.
This suggests for the respective higher
order approximants BP[0/3] and P[0/3] the
corresponding hierarchy of reliability.
\end{enumerate}
\begin{table}
\caption{\label{table1}
The exact and predicted coefficients $r_j$ of the
expansion (\ref{pMG}) for ${\widetilde p}_{\rm  M+G}$
in powers of $g_s(\mu_{\rm M})$.
The scales are fixed as usual:
$\mu_{\rm M} = \mu_m = m_{\rm E}^{(0)}$ of
Eq.~(\ref{m0T}), $\Lambda_{\rm E} = ( 2 \pi T m_{\rm E}^{(0)})^{1/2}$.
The temperatures are $T = 1$ GeV and $0.5$ GeV; $n_f = 3$. The results
$r_3(\delta_{\rm G})$ are given for the values 
$\delta_{\rm G} = 0 \pm 5$ Eq.~(\ref{dGvar}).}
\begin{ruledtabular}
\begin{tabular}{cccccc}
 $T$ [GeV] & $r_1$(exact) & $r_2$(exact) & $r_3(\delta_{\rm G})$ (exact) & 
$r_3$ (pred.~P[0/2]) & $r_3$ (pred.~BP[0/2]) \\
\hline
1.0 & -0.3794 & -0.4654 & $-0.0643 \pm 0.1111$ & 0.4077 & 1.387
\\
0.5 & -0.3374 & -0.4654 & $-0.0852 \pm 0.1111$  & 0.3524 & 1.172
\\
\end{tabular}
\end{ruledtabular}
\end{table}
\begin{table}
\caption{\label{table2}
The exact and predicted coefficients ${r}_j^{\rm eff}$ of the
expansion (\ref{pMGeff}) for ${\widetilde p}_{\rm  M+G}$
in powers of the EQCD parameter $g_{\rm E}^2/m_{\rm E}$, without the
$\lambda_{\rm E}^{(1)}$-term. All the other parameters
as in Table \ref{table1}.}
\begin{ruledtabular}
\begin{tabular}{cccccc}
 $T$ [GeV] & $r_1^{\rm eff}$(exact) & $r_2^{\rm eff}$(exact) & 
$r_3^{\rm eff}(\delta_{\rm G})$ (exact) & 
$r_3^{\rm eff}$ (pred.~P[0/2]) & $r_3^{\rm eff}$ (pred.~BP[0/2]) \\
\hline
1.0 & -0.4647 & -0.6980 & $-0.1645 \pm 0.2041$ & 0.7490 & 2.548
\\
0.5 & -0.4132 & -0.6980 & $-0.1902 \pm 0.2041$ & 0.6474 & 2.154
\\
\end{tabular}
\end{ruledtabular}
\end{table}

For all these reasons, we will regard as the acceptable
resummation at the ${\cal O}(g_s^6)$ level
to be the one using P[0/3] for ${\widetilde p}_{\rm  M+G}$,
and BP[1/2] for $R_{\rm E}^{\rm can}$. Further,
as mentioned before,
the resummation of ${\widetilde p}_{\rm  M+G}$
appears to be more consistent with the physical
expectation of $p/p_{\rm ideal} < 1$ at $T \to T_c$
when it is based on expansion (\ref{pMGeff})
of ${\widetilde p}_{\rm  M+G}$ in powers of the EQCD parameter
$g_{\rm E}^2/m_{\rm E}$ 
(with the $\lambda_{\rm E}^{(1)}/m_{\rm E}$-term
added separately). 

\begin{figure}[htb]
\begin{minipage}[b]{.49\linewidth}
 \centering\epsfig{file=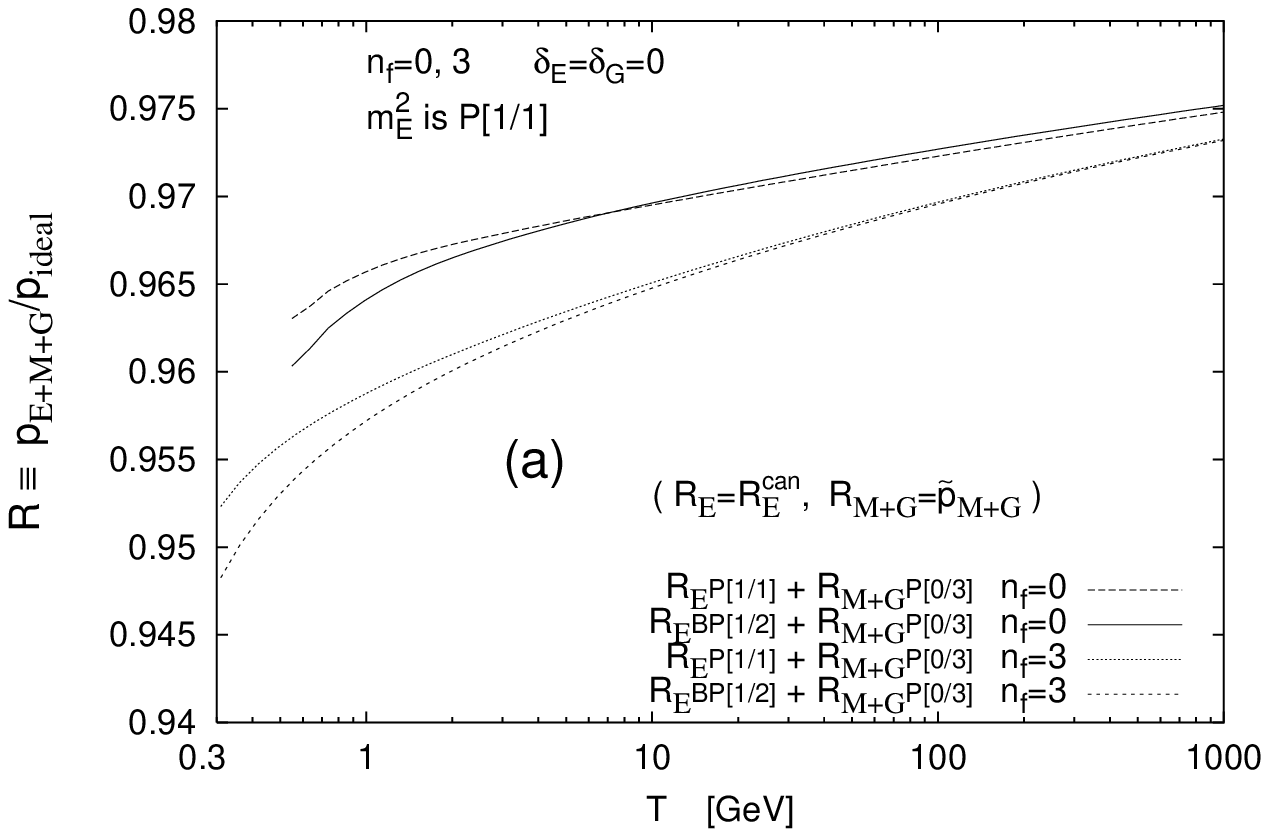,width=\linewidth}
\end{minipage}
\begin{minipage}[b]{.49\linewidth}
 \centering\epsfig{file=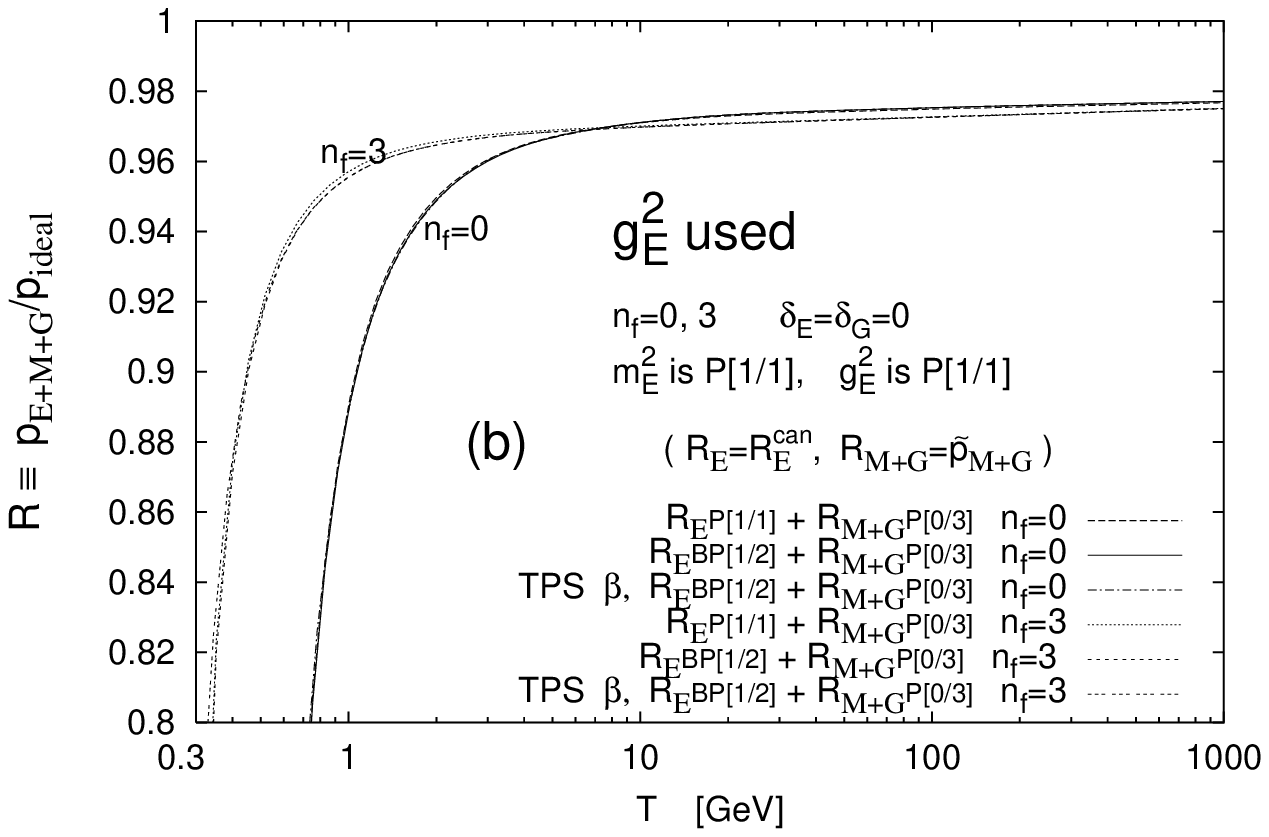,width=\linewidth}
\end{minipage}
%\vspace{0.2cm}
\caption{\footnotesize (a)
Analogous to Fig.~\ref{pEMGvsT1}(a), but with
$n_f=0$ for comparison; (b) as in Fig.~(a), 
but now ${\widetilde p}_{\rm M+G}$ is based on expansion (\ref{pMGeff})
in $g_{\rm E}^2/m_{\rm E}$ -- using the resummed $g_{\rm E}^2$ and 
$m^2_{\rm E}$ as ${\rm P}[1/1](a(\mu_{\rm M}))$,
i.e., analogous to Fig.~\ref{pEMGvsT1eff}(a) but
with $n_f=0$ for comparison.}
\label{pEMGvsTnf0}
\end{figure}

Up until now, all the results presented were for $n_f=3$.
In Figs.~\ref{pEMGvsTnf0} we present 
comparison of  (Borel-)Pad\'e predictions in the cases 
$n_f=0$ and $n_f=3$: in Fig.~\ref{pEMGvsTnf0}(a) when
${\widetilde p}_{\rm M+G}$ is based on expansion (\ref{pMG})
in powers of $g_s$; and in Fig.~\ref{pEMGvsTnf0}(b) when
${\widetilde p}_{\rm M+G}$ is based on expansion (\ref{pMGeff})
in powers of $g_{\rm E}^2/m_{\rm E}$, with the aforementioned
treatment of the $\lambda_{\rm E}^{(1)}/m_{\rm E}$-term. 
In the latter Figure we also included two curves for the case when
the ${\overline {\rm MS}}$ $\beta$ function is taken
as TPS instead of ${\rm P}[2/3](a(\mu))$. 
We see that the negative curvature at low $T$ survives
also in the $n_f=0$ case. Further, the results for
$n_f =0, 3$ depend only very little on the type of the
resummation used for the $\beta$-function, even at low $T$.

\begin{figure}[htb]
\begin{minipage}[b]{.49\linewidth}
 \centering\epsfig{file=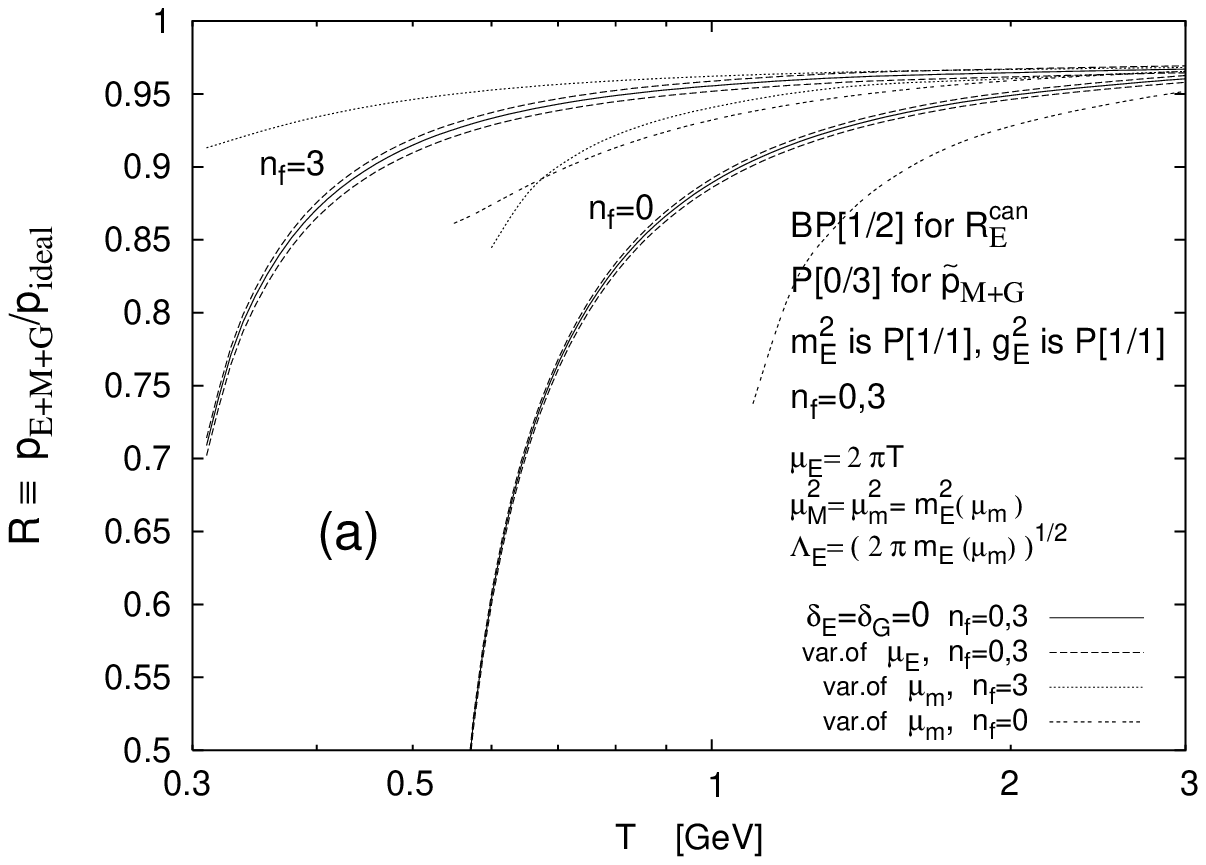,width=\linewidth}
\end{minipage}
\begin{minipage}[b]{.49\linewidth}
 \centering\epsfig{file=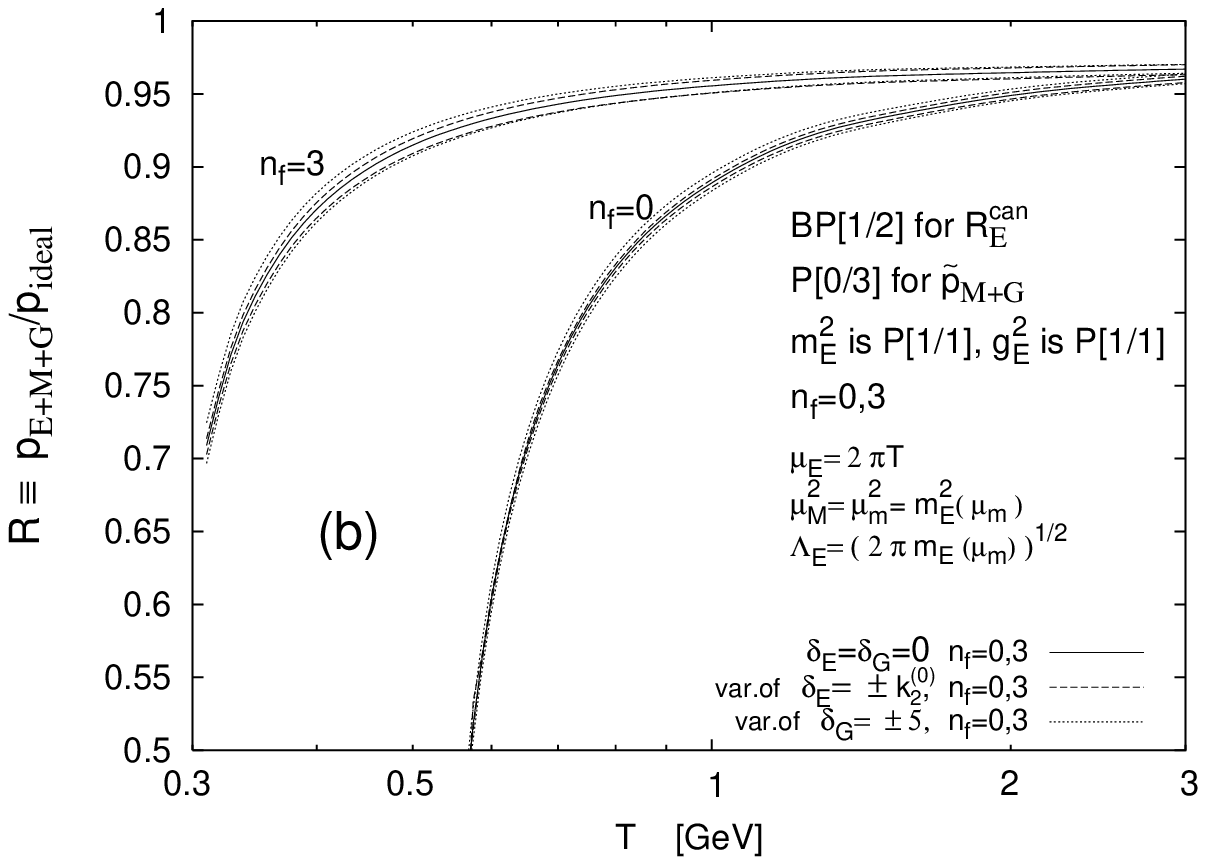,width=\linewidth}
\end{minipage}
%\vspace{0.2cm}
\caption{\footnotesize (a) The pressure $p$, normalized by
$p_{\rm ideal}$, as a function of temperature $T$,
for the most acceptable resummation approach: P[0/3]
for ${\widetilde p}_{\rm  M+G}$ of expansion (\ref{pMGeff}),
and BP[1/2] for $R_{\rm E}^{\rm can}$, for $\delta_{\rm E} =
\delta_{\rm G} = 0$, and $n_f=3, 0$.
The variation of the
renormalization scales $\mu_{\rm E}$ and $\mu_m=\mu_{\rm M}$
is by factor $1.5$ and $1/1.5$ around the
values $2 \pi T$ and $m_{\rm E}^{(0)}(T)$ of Eq.~(\ref{m0T}), 
respectively.
(b) Same as in Fig.~(a), but now the unknown paremeters
$\delta_{\rm E}$ and $\delta_{\rm G}$ are varied, according to
Eqs.~(\ref{dEest}) and (\ref{dGvar}), respectively.}
\label{pEMGeffvsTvsc}
\end{figure}

We present the main results for $p/p_{\rm ideal}$
as a function of temperature, which
are given also in Figs.~\ref{pEMGvsT1eff}(b) and \ref{pEMGvsTnf0}(b),
in a detailed form in the low-temperature regime
in Figs.~\ref{pEMGeffvsTvsc}.
In the latter Figures, we present further the variation of the
curves when the renormalization scales
$\mu_{\rm E}$, and $\mu_{\rm M}=\mu_m$ are varied
around their central values $2 \pi T$ and $m_{\rm E}^{(0)}(T)$
of Eq.~(\ref{m0T}) by factors $1.5$ and $1/1.5$.\footnote{
The factorization scale is 
$\Lambda_{\rm E} = (2 \pi T m_{\rm E})^{1/2}$
where $m_{\rm E}$ is the square root of P[1/1] of
$m_{\rm E}^2(\mu_m)$. $\Lambda_{\rm E}$ changes
only little with the variation of the scale $\mu_{\rm M}=\mu_m$.}
The variation of $\mu_{\rm E}$
changes the curves insignificantly. 

However, the variation of the lower scale $\mu_m=\mu_{\rm M}$
influences significantly the results for ${\widetilde p}_{\rm M+G}$
and thus $p/p_{\rm ideal}$, at $T \alt 1$ GeV, as seen in
Figs.~\ref{pEMGeffvsTvsc}.
This strong variation is a reflection of 
at least two aspects present at $T \alt 1$ GeV: 
(a) the scale $\mu_{\rm M}=\mu_m  \sim m_{\rm E}^{(0)}(T)$
falls down to $\alt 1$ GeV,
a region where the perturbative approach and RGE running
eventually break down; 
(b) the hierarchy of scales $(2 \pi T) \gg  m_{\rm E}^{(0)}(T)$ 
gets very narrowed down.
The first aspect is reflected in the strong instability
of the leading order QCD term $\propto g_s^3(\mu_m)$
[Eq.~(\ref{l1a})] for the $\lambda_{\rm E}^{(1)}/m_{\rm E}$-term
in expansion (\ref{pMGeff}) under the variation of $\mu_m$ ($=\mu_{\rm M}$)
at low $T$'s. This term is not included in the resummation,
as emphasized earlier. The variation of this term is a major
source of the appreciable variation of the
$p/p_{\rm ideal}$-curve at low $T$'s in Fig.~\ref{pEMGeffvsTvsc}(a).
The variation of $\mu_{\rm M}=\mu_m$ downwards to 
$m_{\rm E}^{(0)}/1.5 = 0.73$ GeV is not allowed
because the coupling parameter $g_s(\mu_m)$ blows up
at such low scales by the renormalization group equation
when the beta function is continued into the
strong coupling region by the Pad\'e [2/3] (as is the case
in our numerical results). 
The renormalization
scales $\mu_{\rm M}=\mu_m = m_{\rm E}^{(0)}/1.5$
correspond to the lower curves in Fig.~\ref{pEMGeffvsTvsc},
and they were drawn down to such temperatures where
the corresponding coupling constant $a(\mu_m)$
blew up. For the other renormalization scales, 
the curves were drawn down to approximately
such temperatures where the physically motivated
condition (\ref{m0T}) cannot be fulfilled any more.
\begin{figure}[htb]
 \centering\epsfig{file=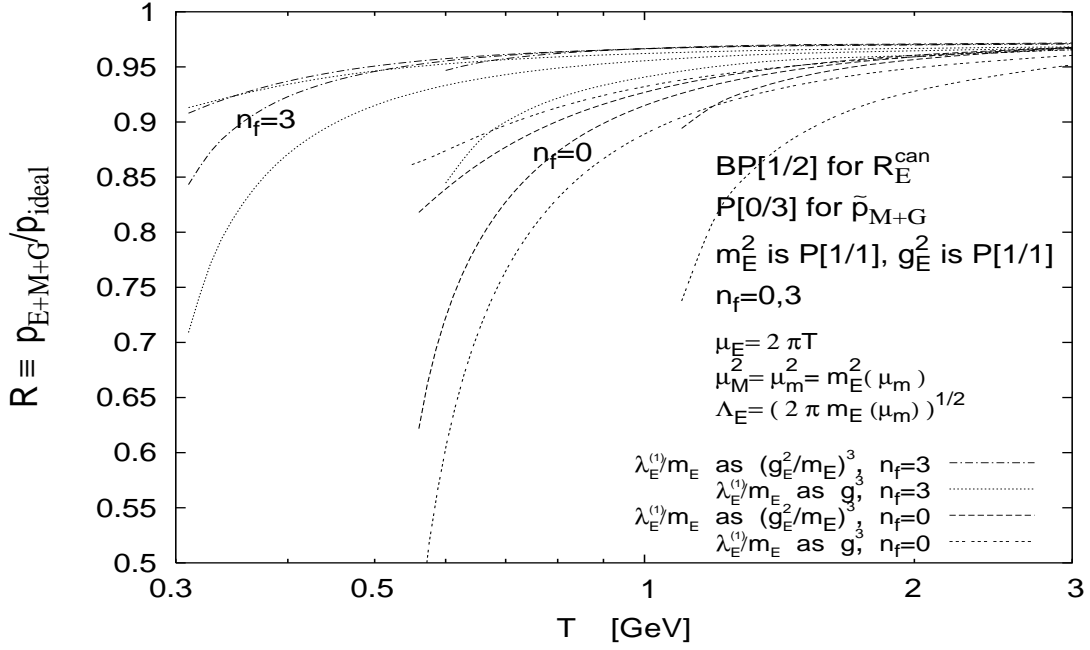,width=15.cm,height=8.8cm}
\caption{\footnotesize Analogous to Fig.~\ref{pEMGeffvsTvsc}(a),
but now the results are included when the 
$\lambda_{\rm E}^{(1)}/m_{\rm E}$-term in expansion (\ref{pMGeff})
is written as proportional to $(g_{\rm E}^2/m_{\rm E})^3$
[Eq.~(\ref{l1b})], and not $g_s^3$ [Eq.~(\ref{l1a})]. 
The variation under the aforementioned
changes of the scale $\mu_{\rm M}=\mu_m$ are included;
$\delta_{\rm E} = \delta_{\rm G} = 0$.}
\label{pEMGeffmvsTvsc}
\end{figure}

We can apply for the $\lambda_{\rm E}^{(1)}/m_{\rm E}$-term 
in expansion (\ref{pMGeff}) of ${\widetilde p}_{\rm M+G}$
the leading term of expansion (\ref{l1b})
in powers of the first EQCD parameter $g_{\rm E}^2/m_{\rm E}$
(with P[1/1] for $g_{\rm E}^2$ and P[1/1] for $m_{\rm E}^2$,
at renormalization scale $\mu_{\rm M}$),
instead of the leading term of expansion (\ref{l1a})
in powers of the QCD parameter $g_s(\mu_{\rm M})$.
The $\lambda_{\rm E}^{(1)}/m_{\rm E}$-term is again not
included in the [0/3]-Pad\'e resummation, but is added separately.
These results are shown in Fig.~\ref{pEMGeffmvsTvsc},
along with the results of Fig.~\ref{pEMGeffvsTvsc}.
Only the variation of the curves under the
aforementioned variation of the $\mu_{\rm M}=\mu_m$ scales
is presented. The variation of the two types of curves, at any given
temperature $T$, under the changes of $\mu_{\rm E}$,
$\delta_{\rm E}$ and $\delta_{\rm G}$, is identical,
thus very weak (shown in Figs.~\ref{pEMGeffvsTvsc}). 
We see from Fig.~\ref{pEMGeffmvsTvsc} that the variation of the
curves under the changes of $\mu_{\rm M}=\mu_m$ is
now significantly weaker. This has primarily to do with
the significantly weaker $\mu_{\rm M}$-dependence of the 
EQCD term $(g_{\rm E}^2/m_{\rm E})^3$ of Eq.~(\ref{l1b}),
in comparison to the $\mu_{\rm M}$-dependence of
the the QCD term $g_s^3(\mu_{\rm M})$ of Eq.~(\ref{l1a}).

\section{Evaluations of truncated perturbation series}
\label{sec:TPS}

\begin{figure}[htb]
\begin{minipage}[b]{.49\linewidth}
 \centering\epsfig{file=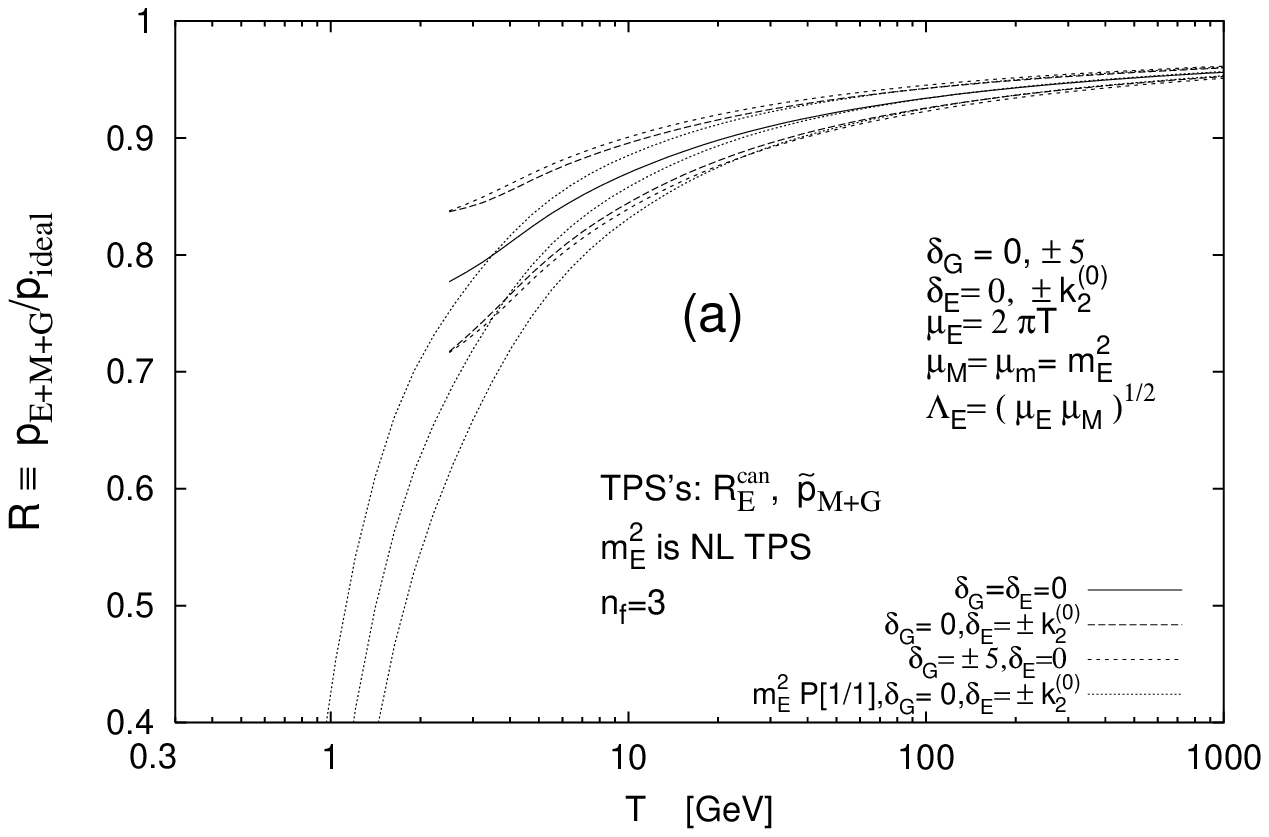,width=\linewidth}
\end{minipage}
\begin{minipage}[b]{.49\linewidth}
 \centering\epsfig{file=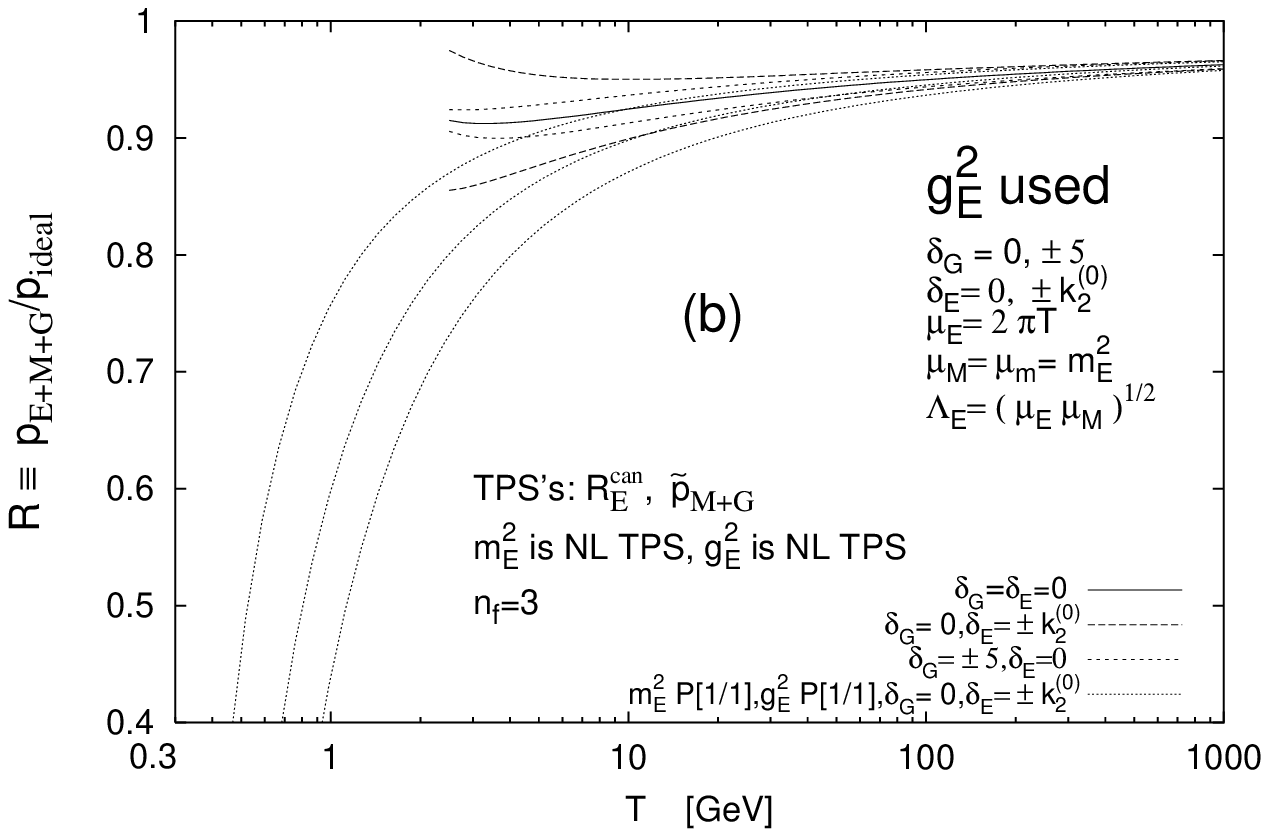,width=\linewidth}
\end{minipage}
%\vspace{0.2cm}
\caption{\footnotesize 
(a) The total pressure as a function of
temperature $T$ when TPS evaluation is employed for $R_{\rm E}^{\rm can}$
and ${\widetilde p}_{\rm M+G}$, instead of the
Pad\'e and Borel-Pad\'e of Fig.~\ref{pEMGvsT1};
(b) Same as in (a), 
but now ${\widetilde p}_{\rm M+G}$ is evaluated as expansion (\ref{pMGeff})
in $g_{\rm E}^2/m_{\rm E}$.
Further explanations are given in the text.}
\label{pEMGvsTTPS1}
\end{figure}

\begin{figure}[htb]
\begin{minipage}[b]{.49\linewidth}
 \centering\epsfig{file=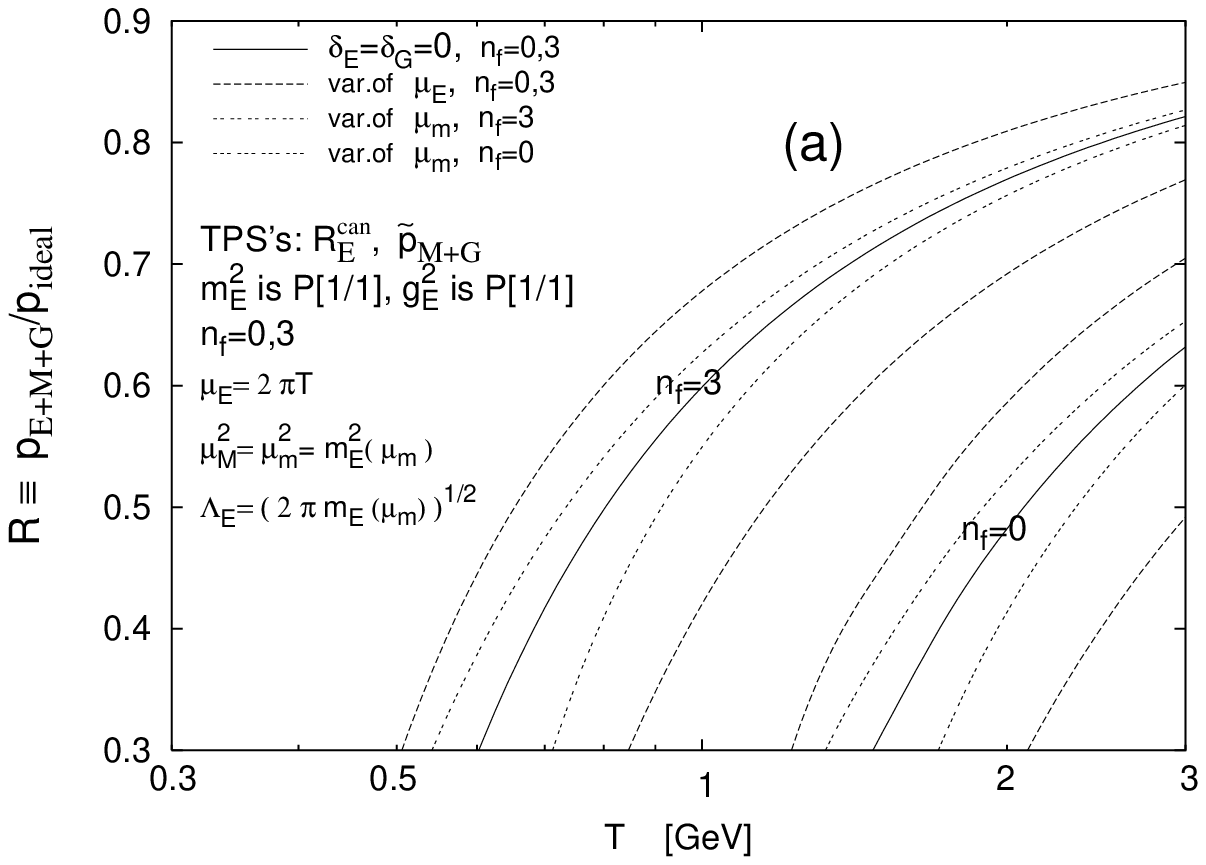,width=\linewidth}
\end{minipage}
\begin{minipage}[b]{.49\linewidth}
 \centering\epsfig{file=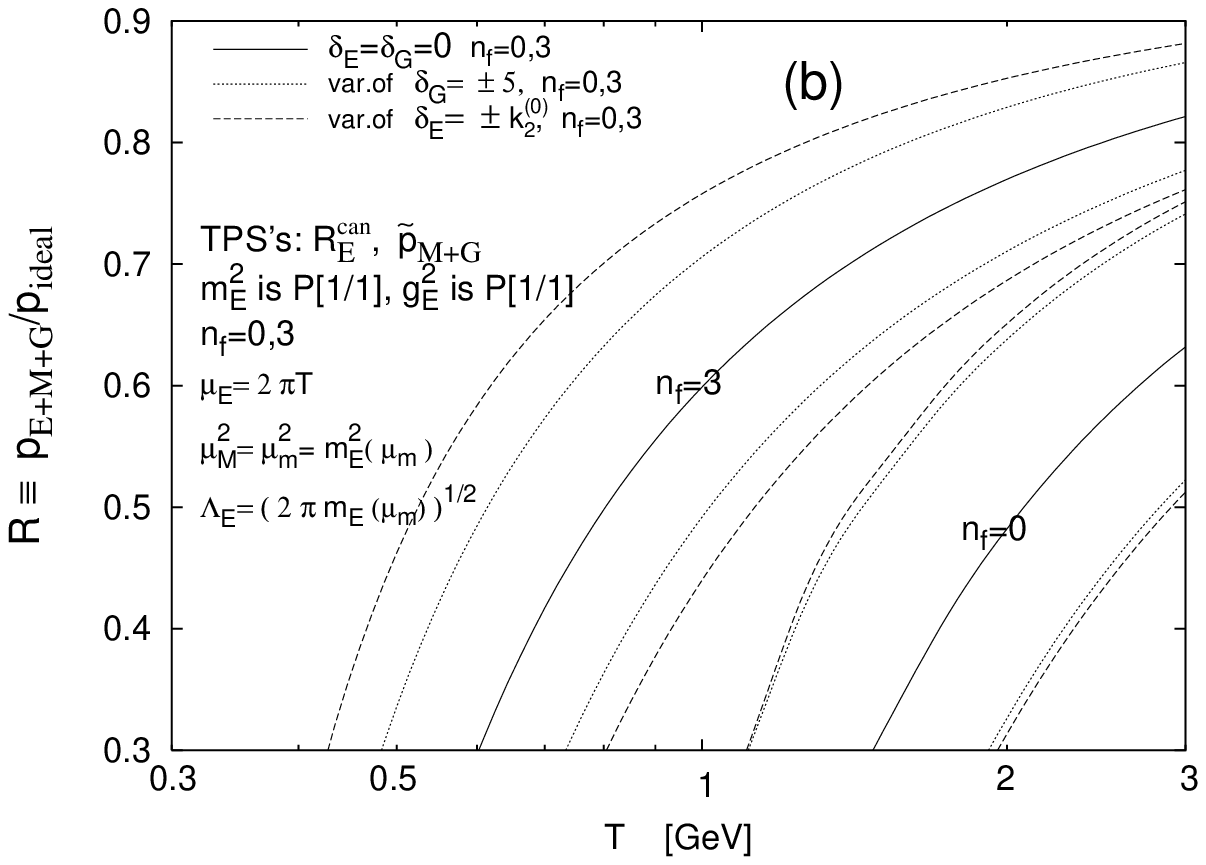,width=\linewidth}
\end{minipage}
%\vspace{0.2cm}
\caption{\footnotesize 
(a) The low-temperature total pressure (for $n_f=3,0$) 
when TPS evaluation is employed for $R_{\rm E}^{\rm can}$
and ${\widetilde p}_{\rm M+G}$, but for $m_{\rm E}^2$
and $g_{\rm E}^2$ Pad\'e P[1/1]. The variation of the
renormalization scales $\mu_{\rm E}$ and $\mu_m=\mu_{\rm M}$
is by factor $1.5$ and $1/1.5$ around the
values $2 \pi T$ and $m_{\rm E}^{(0)}(T)$ of Eq.~(\ref{m0T}), 
respectively.
(b) Same as in Fig.~(a), but now the unknown paremeters
$\delta_{\rm E}$ and $\delta_{\rm G}$ are varied, according to
Eqs.~(\ref{dEest}) and (\ref{dGvar}), respectively.}
\label{pEMGvsTTPS1b}
\end{figure}

Now we investigate how the results change if, instead of
Pad\'e or Borel-Pad\'e, 
simple TPS evaluations are applied to all expansions:
to expansion (\ref{pE2}) for $R_{\rm E}^{\rm can}$,
to (\ref{mE}) and (\ref{gE}) for $m_{\rm E}^2$
and $g_{\rm E}^2$, and to (\ref{pMG}) or
(\ref{pMGeff}) for ${\widetilde p}_{\rm M+G}$.
The results for the pressure, as a function of temperature,
for $n_f=3$, are presented in Figs.~\ref{pEMGvsTTPS1}(a)
and \ref{pEMGvsTTPS1}(b), when expansions (\ref{pMG})
and (\ref{pMGeff}) are used for  ${\widetilde p}_{\rm M+G}$,
respectively.
We enforce here: $\mu_{\rm M} = \mu_m$; and either 
$\mu_m = \left[(m_{\rm E}^2)^{(\rm TPS)}(\mu_m) \right]^{1/2}$, or
$\mu_m = \left[(m_{\rm E}^2)^{\rm P[1/1]}(\mu_m) \right]^{1/2}$
[the latter is $m_{\rm E}^{(0)}(T)$ of Eq.~(\ref{m0T})].
The curves with (NL)TPS $m_{\rm E}^2$ are continued down to such
temperatures where the aforementioned condition
$\mu_m = m_{\rm E}$ cannot be enforced any more,
indicating that this (perturbative) TPS
method becomes inapplicable below such temperatures
$T \approx 2$ GeV. The curves with P[1/1] $m_{\rm E}^2$
in Figs.~\ref{pEMGvsTTPS1}
can in principle be shown, as earlier, for temperatures
down to $T \approx 0.3$ GeV; however, their
values (for $|\delta_{\rm E}| \leq |k_2^{(0)}|$)
fall drastically: $p/p_{\rm ideal} < 0.45$
already at $T \approx 1$ GeV (which, incidentally, is far lower
than the lattice results),
and $p < 0$ already at $T \approx 0.5$ GeV. 
These curves, with P[1/1]-resummed $m_{\rm E}^2$
and $g_{\rm E}^2$ are presented in Figs.~\ref{pEMGvsTTPS1b} in more
detail for the low temperatures,
where now $n_f=3$ and $n_f=0$. The $\lambda_{\rm E}^{(1)}/m_{\rm E}$-term
in expansion (\ref{pMGeff}) is added as the
term $\propto g_s^3(\mu_{\rm M})$ [Eq.~(\ref{l1a})].
In Fig.~\ref{pEMGvsTTPS1b}(a), the variation of these TPS curves
under the changes of the two renormalization scales $\mu_{\rm E}$ and
$\mu_{\rm M} = \mu_m$, as explained for Fig.~\ref{pEMGeffvsTvsc}(a),
is presented. In Fig.~\ref{pEMGvsTTPS1b}(b), the variation under 
the changes of the parameters $\delta_{\rm E}$ and $\delta_{\rm G}$
is presented. In Fig.~\ref{pEMGvsTTPS1b}(a) we see that the
TPS curves vary more strongly under the changes of the
high-energy scale $\mu_{\rm E}$ than
under those of the low-energy scale $\mu_{\rm M} = \mu_m$.
This result, at first sight paradoxical, occurs mainly
because we used for $p_{\rm M+G}$ the EQCD
expansion (\ref{pMGeff}), with the dominant part
coming from powers of the EQCD parameter $g_{\rm E}^2/m_{\rm E}$,
where the $\mu_m$-dependence of this parameter is
rather weak because we used ${\rm P[1/1]}(a(\mu_m))$ for
$g_{\rm E}^2$ and for $m_{\rm E}^2$.
Comparing with the corresponding
Pad\'e--Borel-Pad\'e (P+BP) resummed curves of Figs.~\ref{pEMGeffvsTvsc},
we see that the latter have much weaker dependence
on the parameters $\delta_{\rm E}$ and $\delta_{\rm G}$
and on the high-energy renormalization scale $\mu_{\rm E}$,
while the dependence on the low energy renormalization
scale $\mu_{\rm M} = \mu_m$ ($\sim m_{\rm E} \sim g_s T$)
is reduced by P+BP resummation only by a factor of 2-3.
For example, at $T=0.7$ GeV and $n_f=3$, this variation is about
$0.18$ and $0.06$ GeV in the TPS and P+BP cases, respectively.
This has largely to do with the $\lambda_{\rm E}^{(1)}/m_{\rm E}$-term
which is not resummed in the P+BP case, but is also taken as
$\propto g_s^3(\mu_{\rm M})$ [Eq.~(\ref{l1a})] and added separately. 
On the other hand,
taking that term as $\propto (g_{\rm E}^2/m_{\rm E})^3$
[Eq.~(\ref{l1b})], and adding it after the resummation,
significantly weakened the
$\mu_{\rm M}$-dependence at low temperatures,
as was seen in Fig.~\ref{pEMGeffmvsTvsc}. 

On all these grounds, the Pad\'e and Borel-Pad\'e
results of Figs.~\ref{pEMGeffvsTvsc} and \ref{pEMGeffmvsTvsc}
are likely to give more realistic results at low $T \sim 1$ GeV
than the corresponding TPS results.

\section{Comparisons with other approaches, and conclusions}
\label{sec:comp}

\begin{figure}[htb]
\begin{minipage}[b]{.49\linewidth}
 \centering\epsfig{file=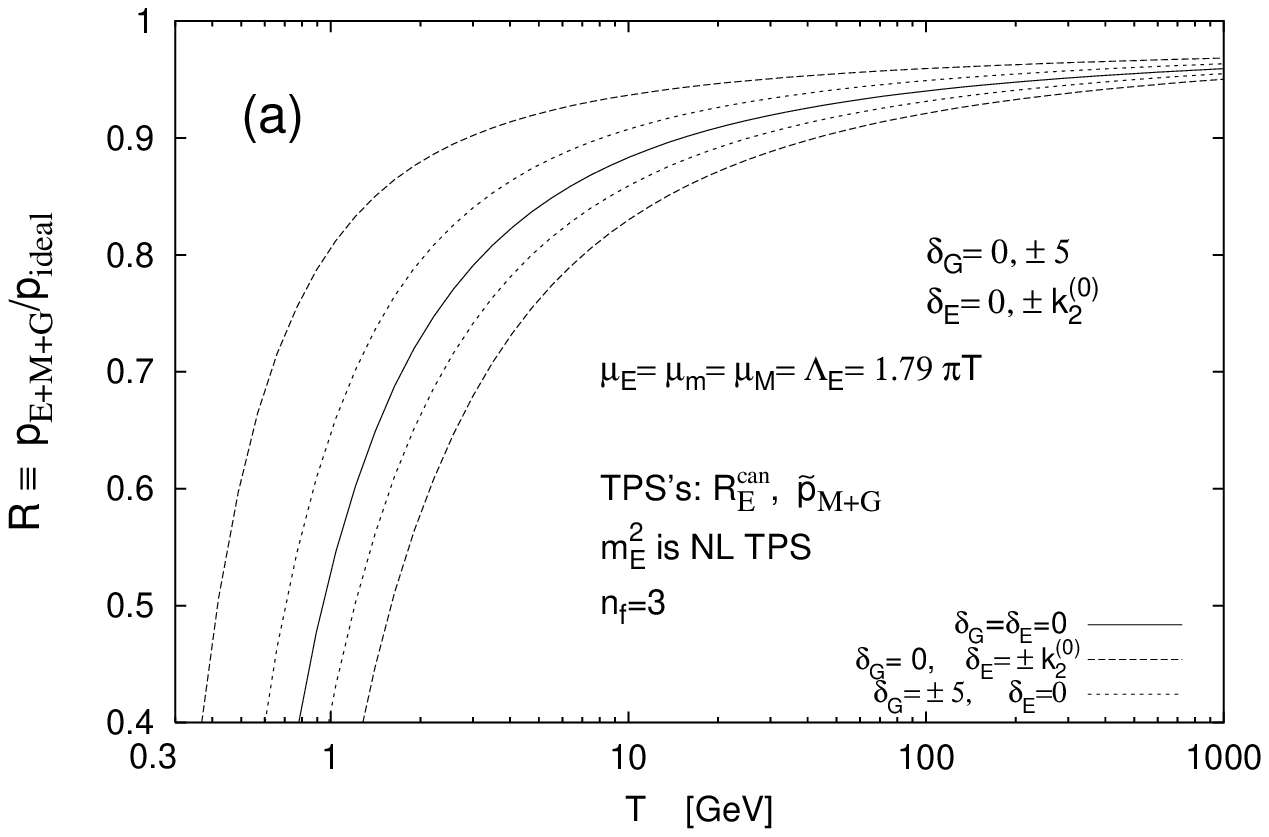,width=\linewidth}
\end{minipage}
\begin{minipage}[b]{.49\linewidth}
 \centering\epsfig{file=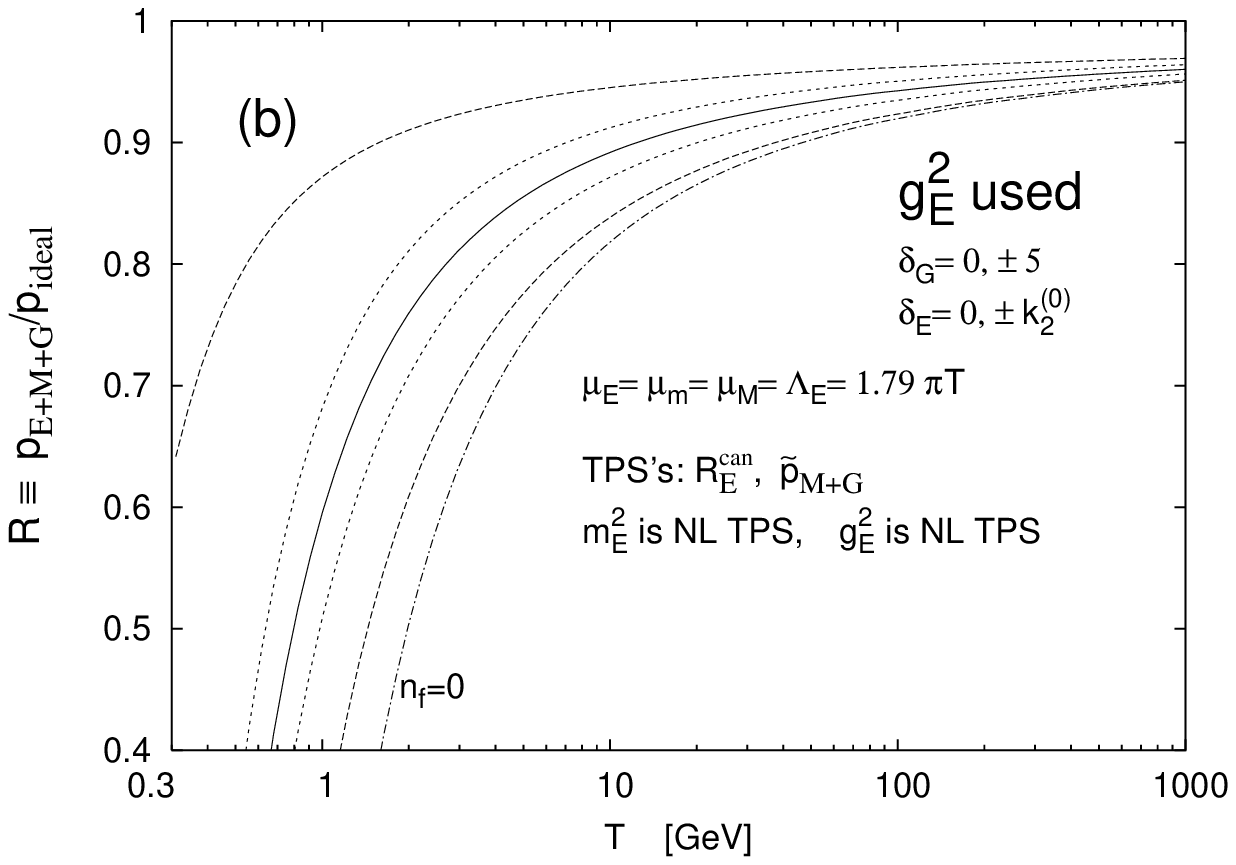,width=\linewidth}
\end{minipage}
%\vspace{0.2cm}
\caption{\footnotesize 
(a) The total pressure (with $n_f=3$) as a function of the
temperature $T$ when TPS evaluation is employed for $R_{\rm E}^{\rm can}$,
$m_{\rm E}^2$, and ${\widetilde p}_{\rm M+G}$ of (\ref{pMG}), 
but with common scales used: 
$\mu_{\rm E} = \mu_{\rm m} = \mu_{\rm M} = \Lambda_{\rm E}
= 1.79 \pi T$. The unknown parameters $\delta_{\rm G}$
and $\delta_{\rm E}$ are varied in the interval (\ref{dGvar})
and (\ref{dEest}), respectively, for $n_f=3$.
(b) same as Fig.~(a), but using for $p_{\rm M+G}$ expansion
(\ref{pMGeff}) in powers of $g_{\rm E}^2/m_{\rm E}$.
Further explanations are given in the text.}
\label{pEMGvsTTPS2}
\end{figure}
\begin{figure}[htb]
\begin{minipage}[b]{.49\linewidth}
 \centering\epsfig{file=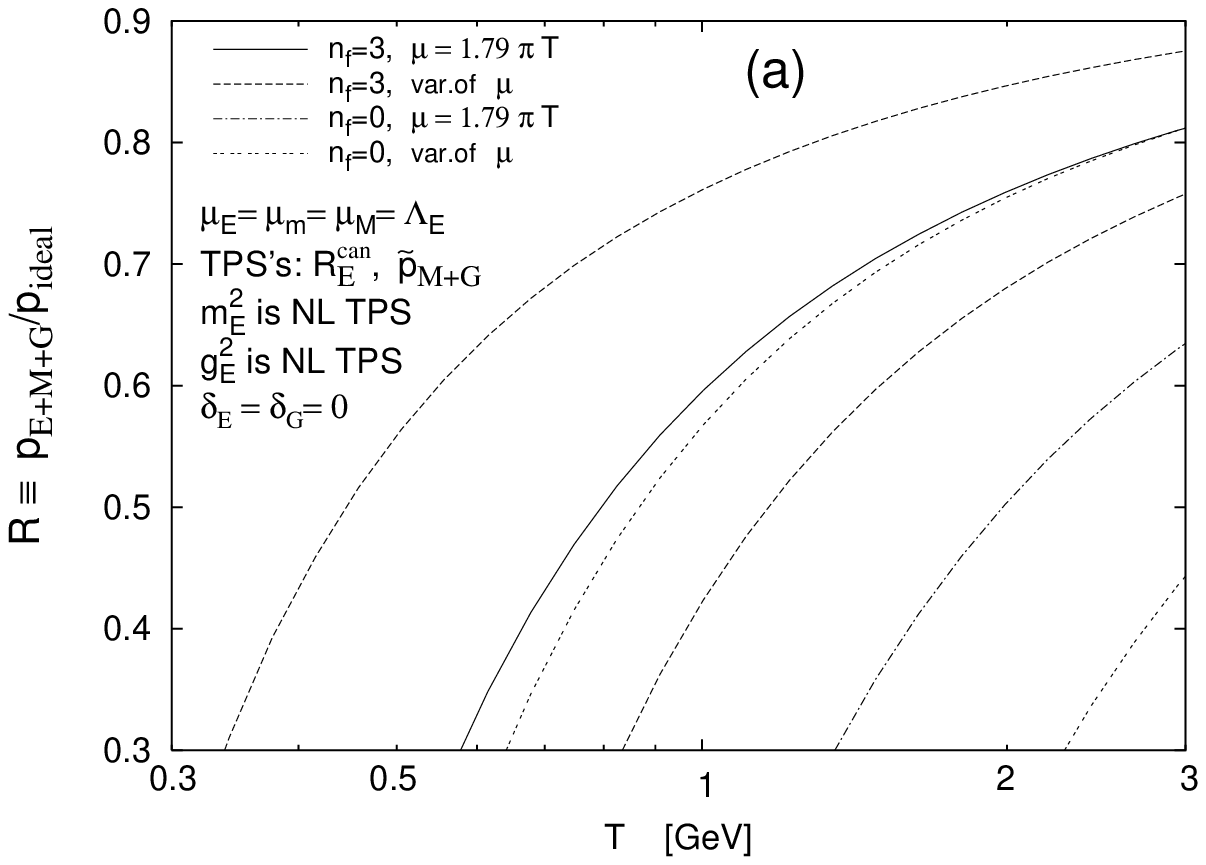,width=\linewidth}
\end{minipage}
\begin{minipage}[b]{.49\linewidth}
 \centering\epsfig{file=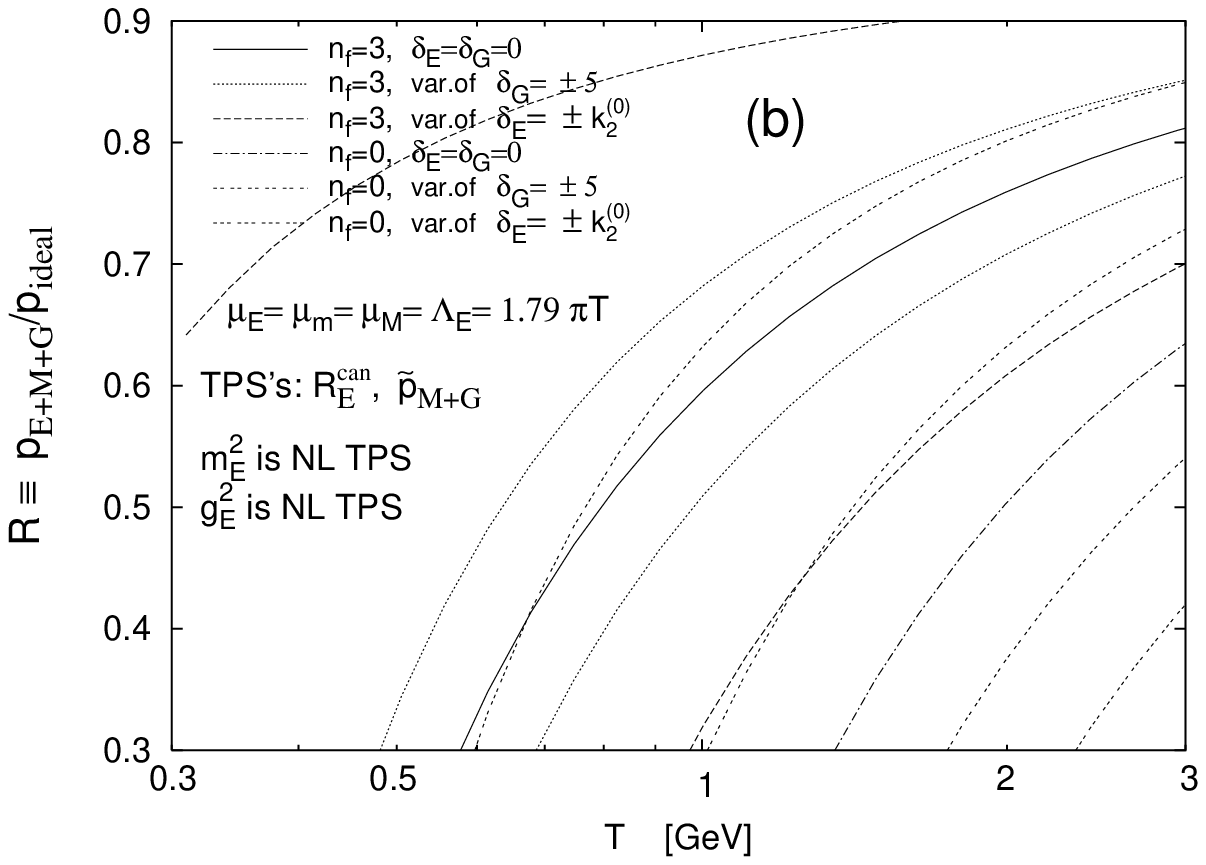,width=\linewidth}
\end{minipage}
%\vspace{0.2cm}
\caption{\footnotesize 
(a) The low-temperature total pressure (for $n_f=3,0$) 
when TPS evaluation is employed for $R_{\rm E}^{\rm can}$,
$m_{\rm E}^2$, $g_{\rm E}^2$, and ${\widetilde p}_{\rm M+G}$ 
of Eq.~(\ref{pMGeff}), but with common scales used: 
$\mu_{\rm E} = \mu_{\rm m} = \mu_{\rm M} = \Lambda_{\rm E}
= 1.79 \pi T$. Variation of the scale is by factor $1.5$:
$\mu_{\rm max} = 1.5 1.79 \pi T$, $\mu_{\rm min} = (1/1.5) 1.79 \pi T$. 
(b) The unknown parameters $\delta_{\rm G}$
and $\delta_{\rm E}$ are varied in the interval (\ref{dGvar})
and (\ref{dEest}), respectively, for $n_f=3,0$.}
\label{pEMGvsTTPS3}
\end{figure}
An approach different from our resummation
is to set all (${\overline {\rm MS}}$) renormalization
scales equal: ${\overline \mu} = \mu_{\rm E} = \mu_{\rm m} = \mu_{\rm M} =
\Lambda_{\rm E}$, and then evaluate $m_{\rm E}$, $p_{\rm E}$
and ${\widetilde p}_{\rm M+G}$ ($\Rightarrow \ p_{\rm M+G}$). 
This was the approach of Ref.~\cite{Blaizot:2003iq}, and the evaluation 
in Ref.~\cite{Kajantie:2002wa} was similar as well. 
In both of these references, truncated perturbation series (TPS) 
evaluations were applied.
In Figs.~\ref{pEMGvsTTPS2}(a) and \ref{pEMGvsTTPS2}(b), we present the results
of such type of resummation as a function of temperature, for $n_f=3$,
when ${\widetilde p}_{\rm M+G}$ is evaluated as TPS (\ref{pMG})
and TPS (\ref{pMGeff}), respectively. 
The common scale was chosen to be ${\overline \mu} = 1.79 \pi T$
and the expansions (\ref{pE2}), (\ref{mE}), (\ref{gE}), (\ref{pMG}) 
[or (\ref{pMGeff})] were evaluated as TPS.  
This should correspond roughly to the method leading
to the the dash-dotted curve in Fig.~5 of Ref.~\cite{Blaizot:2003iq}
for $n_f = 0$. We can see from Figs.~\ref{pEMGvsTTPS2} that
this approach gives, at low $T < 10$ GeV, results similar
to the dotted TPS curves of Figs.~\ref{pEMGvsTTPS1} where
${\rm P}[1/1]$ was used for $m_{\rm E}^2$ and
different renormalization scales were used for TPS $p_{\rm E}$
and TPS ${\widetilde p}_{\rm M+G}$. 
However, in contrast to Ref.~\cite{Blaizot:2003iq},
most of our TPS curves in Figs.~\ref{pEMGvsTTPS1} and \ref{pEMGvsTTPS2}
fall down faster then theirs when temperature decreases to
$T< 1$ GeV ($T/T_c \alt 5$), and fall down faster than
the lattice curves. Only the upper $n_f=3$ curve in Fig.~\ref{pEMGvsTTPS2}(b),
corresponding to the choice $\delta_{\rm E} = - |k_2^{(0)}|$
(and $\delta_{\rm G}=0$), is marginally compatible 
with the lattice results.

In Figs.~\ref{pEMGvsTTPS3}(a) and (b) we present, at low
temperatures,  and for $n_f = 3$ and $0$,
the variation of these curves when the common renormalizatio
scale $\mu$ changes around $1.79 \pi T$ (from $1.79 \pi T/1.5$
to $1.5 \times 1.79 \pi T$),
and when the parameters $\delta_{\rm E}$ and $\delta_{\rm G}$
change, respectively (in analogy with Figs.~\ref{pEMGvsTTPS1b}
and \ref{pEMGeffvsTvsc}). We see that the variation of the
curves is then similarly strong as for the TPS curves of
Figs.~\ref{pEMGvsTTPS1b}.

\begin{figure}[htb]
 \centering\epsfig{file=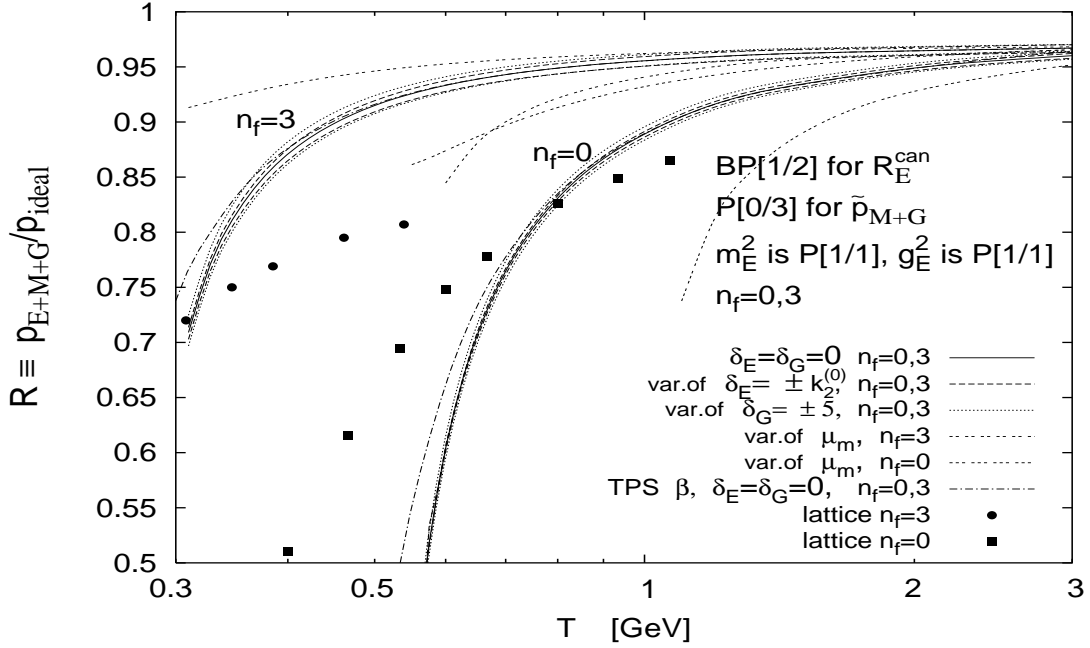,width=15.cm,height=8.8cm}
\caption{\footnotesize Summary of our results for $n_f=3$ and
$n_f=0$, when the unknown parameters $\delta_{\rm E}$ and
$\delta_{\rm G}$ are varied. The resummation of $p_{\rm M+G}$ 
was performed on the basis of expansion (\ref{pMGeff})
for ${\widetilde p}_{\rm M+G}$ in powers of $g_{\rm E}^2/m_{\rm E}$.
Other details given in the text.}
\label{pEMGeffvsTco}
\end{figure}
On the other hand, Pad\'e and Borel-Pad\'e resummations,
Figs.~\ref{pEMGvsT1eff}-\ref{pEMGeffvsTvsc} (falling down
as well  at low decreasing $T$)  are much less dependent on the unknown
parameters $\delta_{\rm G}$ and $\delta_{\rm E}$,
as can be seen by comparing with the TPS curves
Figs.~\ref{pEMGvsTTPS1} and \ref{pEMGvsTTPS2}. 

\begin{figure}[htb]
 \centering\epsfig{file=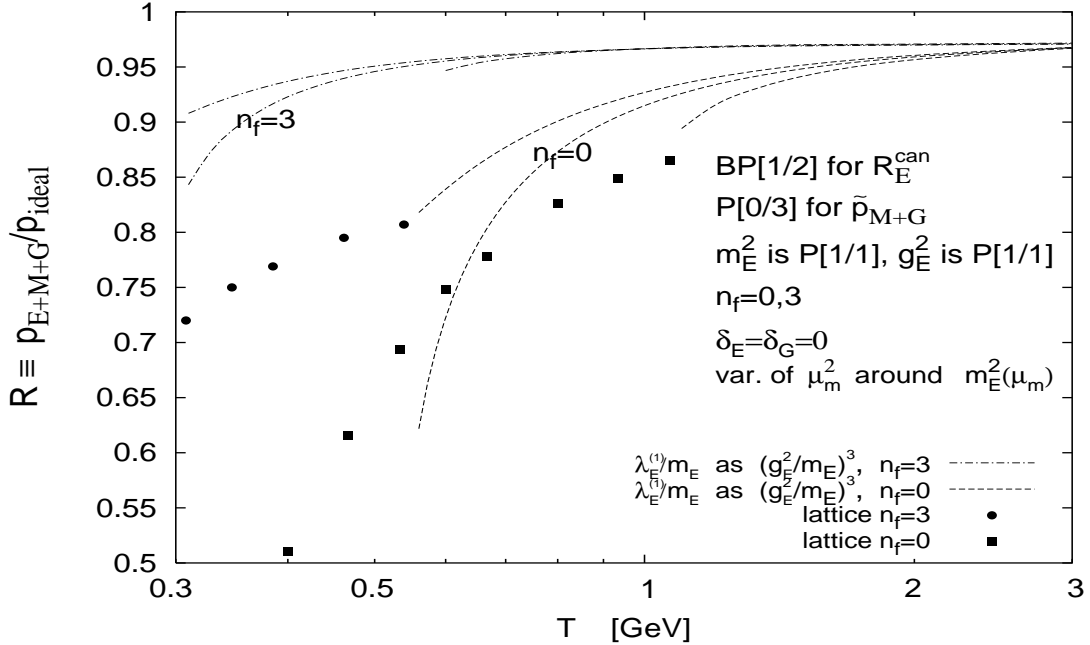,width=15.cm,height=8.8cm}
\caption{\footnotesize Analogous to Fig.~\ref{pEMGeffvsTco}, but
the $\lambda_{\rm E}^{(1)}/m_{\rm E}$-term of Eq.~(\ref{pMGeff})
is now written as proportional to $(g^2_{\rm E}/m_{\rm E})^3$
[Eq.~(\ref{l1b})] instead of $g_s^3$ [Eq.~(\ref{l1a})]. 
Changes under the variation of $\mu_{\rm M}=\mu_m$
are included as in Fig.~\ref{pEMGeffvsTco}, and
$\delta_{\rm E} = \delta_{\rm G} = 0$ is kept. The curves vary
by the identical amounts, at any given $T$, 
as in Fig.~\ref{pEMGeffvsTco} under the changes 
$\delta_{\rm E}=\pm k_2^{(0)}$ and $\delta_{\rm G}=\pm 5$.}
\label{pEMGeffmvsTco}
\end{figure}

Let us now summarize our findings. We first collect the main
results for the optimal approximants (which have been presented
in detail in Figs.~\ref{pEMGvsT1eff}-\ref{pEMGeffmvsTvsc})
and combine them in Figs.~\ref{pEMGeffvsTco}
and \ref{pEMGeffmvsTco}, thereby
focussing on the crucial region of low temperatures. For comparison
we also include the predictions of lattice calculations taken from
Fig.~4(b) of Ref.~\cite{Karsch:2001vs} 
(which includes lattice results of Refs.~\cite{Karsch:2000ps}).
These figures thus present our resummation
results when Borel-Pad\'e ${\rm BP[1/2]}(a(\mu_{\rm E}))$ is
applied to expansion (\ref{pE2}) for $R_{\rm E}^{\rm can}$, 
and Pad\'e $[0/3](g_{\rm E}^2/m_{\rm E})$ to expansion
(\ref{pMGeff}) for ${\widetilde p}_{\rm M+G}$.
The central values of the renormalization scales
were taken $\mu_{\rm E} = 2 \pi T$ and 
$\mu_{\rm M} = \mu_m = m_{\rm E}^{(0)}(T)$ [Eq.~(\ref{m0T})],
and the central factorization scale was 
$\Lambda_{\rm E} = ( \mu_{\rm E} \mu_{\rm M} )^{1/2}$.
The EQCD parameters $g_{\rm E}^2$ and $m_{\rm E}^2$ were
calculated (resummed) as Pad\'e ${\rm P[1/1]}(g_s(\mu_m))$,
at low-energy renormalization scale $\mu_{\rm M} = \mu_m$
($\sim m_{\rm E} \sim g_s T$).
In addition, the
effect of replacing the Pad\'e-resummed ${\rm P}[2/3](a)$
$\beta$ function by the simple TPS $\beta$ is displayed as well.
For the lattice results, we used the values of critical temperature
$T_c(n_f\!=\!3) \approx 154$ MeV \cite{Karsch:2001vs},
and $T_c(n_f\!=\!0) \approx 267$ MeV. The latter
value is obtained from the result $T_c/\sqrt{\sigma} = 0.629$
\cite{Boyd} and $\sqrt{\sigma} = 425$ MeV
\cite{Karsch:2001vs}. 

The two Figures (\ref{pEMGeffvsTco} and \ref{pEMGeffmvsTco})
differ by the different treatment of the $\lambda_{\rm E}^{(1)}$-term: 
In Fig.~\ref{pEMGeffvsTco},
the $\lambda_{\rm E}^{(1)}/m_{\rm E}$-term of expansion
(\ref{pMGeff}) is added separately as the leading
order term of expansion (\ref{l1a}), i.e., as a term
proportional to $g_s^3(\mu_m)$.
In Fig.~\ref{pEMGeffmvsTco}, 
on the other hand, the $\lambda_{\rm E}^{(1)}/m_{\rm E}$-term 
is added separately as the leading
order term of expansion (\ref{l1b}), i.e., as a term
proportional to $(g_{\rm E}^2/m_{\rm E})^3$ [with the EQCD parameters
$g_{\rm E}^2$ and $m_{\rm E}^2$ resummed as ${\rm P[1/1]}(a(\mu_m))$].
In the latter case, the dependence on the low-energy renormalization
scale $\mu_{\rm M} = \mu_m$ becomes appreciably weaker,
as was seen in Fig.~\ref{pEMGeffmvsTvsc} [cf.~also the discussion
following Eqs.~(\ref{l1a})-(\ref{l1b})].

In both cases, the lattice results differ from these
curves by about $20\%$ and $10\%$ when $n_f=3,0$, respectively.
Taking into account that all lattice results should be taken 
with an error of $10$-$15\%$, we see 
that our resummed results come at low temperatures
reasonably close to the lattice results,
although our resummations are based only on perturbation
expansions. Further, in contrast to the various TPS evaluations,
the dependence of our results on the unknown parameters
$\delta_{\rm G}$ and $\delta_{\rm E}$ and on the
high-energy renormalization scale $\mu_{\rm E}$ ($\sim 2 \pi T$) 
is quite weak as shown in Fig.~\ref{pEMGeffvsTco}. 
The dependence on the low-energy renormalization
scale $\mu_{\rm M}=\mu_m$ ($\sim m_{\rm E}(T)$) is 
in Fig.~\ref{pEMGeffvsTco} by about a
factor of 2-3 weaker than in the analogous TPS case, and is thus
at low temperatures still quite significant. It has its origin
primarily in the  $\mu_m$-dependence of
the QCD term $g_s^3(\mu_m)$ of Eq.~(\ref{l1a}).
In Fig.~\ref{pEMGeffmvsTco}, the variation of the curves
when $\delta_{\rm G}$, $\delta_{\rm E}$ and $\mu_{\rm E}$
are changed, at any given $T$, are identical to the variations
of the corresponding curves of Fig.~\ref{pEMGeffvsTco}, i.e.,
quite weak. The dependence on the low-energy
renormalization scale $\mu_m=\mu_{\rm M}$ is in Fig.~\ref{pEMGeffmvsTco}
appreciably weaker than in Fig.~\ref{pEMGeffvsTco}.

Let us recall also, that the specific choices for the approximants
which entered in Figs.~\ref{pEMGeffvsTco} and \ref{pEMGeffmvsTco}
were based on the following reasoning:
a TPS at a given order, in principle, allows for various
Pad\'e and Borel-Pad\'e approximants. We chose those approximants
which give results reasonably stable under the variation of the corresponding
renormalization scales $\mu_{\rm E}$ and $\mu_{\rm M} = \mu_m$,
and of the factorization scale $\Lambda_{\rm E}$. This physical
criterion led us to two possible approximants: BP[0/3] and
P[0/3] for ${\widetilde p}_{\rm M+G}$; in both cases
BP[1/2] for $R_{\rm E}^{\rm can}$. They differed from each
other significantly in the low temperature regime,
cf.~Fig.~\ref{pEMGvsTB03}(b). 
In order to further eliminate one of them (BP[0/3]),
we had to apply an additional physically motivated criterion,
namely that the predicted pressure should be smaller than the
ideal gas value if the temperature $T$ is sufficiently near
the critical value $T_c$. In addition, two numerical arguments
were provided which also support the elimination of the
approxmant BP[0/3] for ${\widetilde p}_{\rm M+G}$. Thereby we ended
up with the approximants BP[1/2] for $R_{\rm E}^{\rm can}$
($\Rightarrow p_{\rm E}$) and P[0/3] for ${\widetilde p}_{\rm M+G}$
($\Rightarrow p_{\rm M+G}$) which are 
considered as our ``best approximants''.

\begin{figure}[htb]
 \centering\epsfig{file=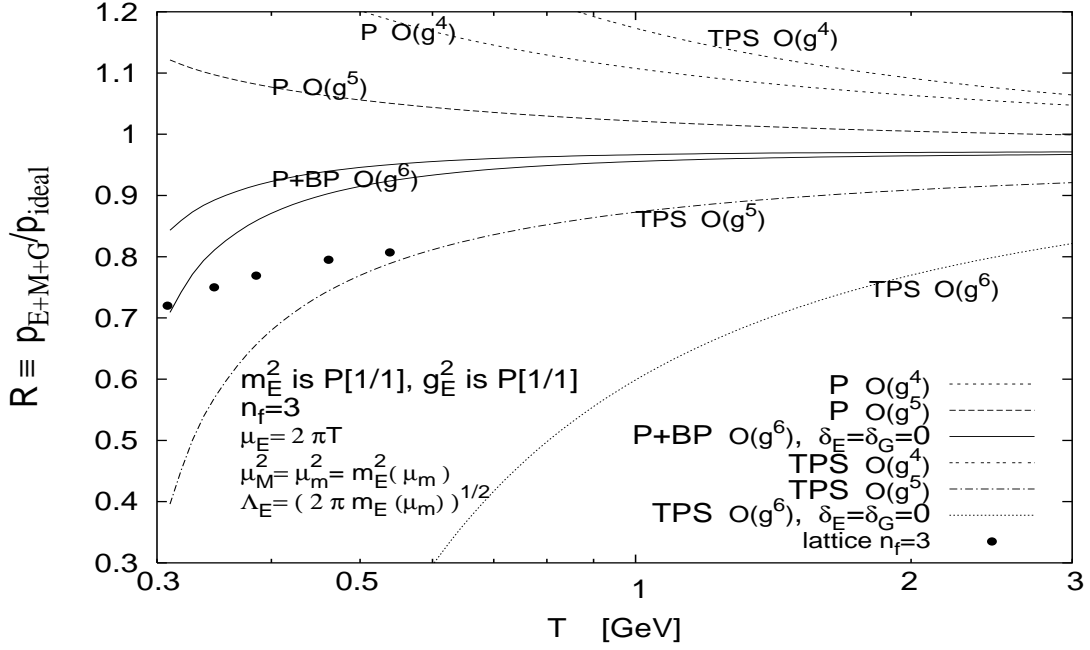,width=15.cm,height=8.8cm}
\caption{\footnotesize The convergence behavior of our
(Borel-)Pad\'e curves for $p/p_{\rm ideal}$ as a function
of temperature, at $n_f=3$, for orders ${\cal O}(g_s^4)$,
${\cal O}(g_s^5)$ and ${\cal O}(g_s^6)$, and the behavior
of the coresponding TPS's. The upper P+BP ${\cal O}(g_s^6)$
curve is for the case when the 
$\lambda_{\rm E}^{(1)}/m_{\rm E}$-term of Eq.~(\ref{pMGeff})
is now written as proportional to $(g^2_{\rm E}/m_{\rm E})^3$
instead of $g_s^3$.
Details are given in the text.}
\label{pEMGconv}
\end{figure}

Since both are of Pad\'e type, we expect that they also have a better 
convergence behavior than the ordinary truncated perturbation
theories. This, in fact, can be 
tested explicitly and the results are presented
in Fig.~\ref{pEMGconv}, where we demonstrate the 
behavior of the Pad\'e and Borel-Pad\'e resummed curves, for $n_f=3$, 
when the order of expansions (\ref{pE2}) and (\ref{pMGeff})
for $R_{\rm E}^{\rm can}$ and ${\widetilde p}_{\rm M+G}$
increases, and compare them with the corresponding TPS evaluations.
For all the curves here, including the TPS curves, 
the EQCD parameters $m_{\rm E}^2$
and $g_{\rm E}^2$ were evaluated as ${\rm P[1/1]}(a(\mu_m))$,
as in Figs.~\ref{pEMGvsTTPS1b} and \ref{pEMGeffvsTco}-\ref{pEMGeffmvsTco}. 
The ${\cal O}(g^4)$ P-curve
uses ${\rm P[1/1]}(a(\mu_{\rm E}))$ for $R_{\rm E}^{\rm can}$
and ${\rm P[0/1]}(g_{\rm E}^2/m_{\rm E})$ 
for ${\widetilde p}_{\rm M+G}$ of expansion
(\ref{pMGeff}); the ${\cal O}(g^5)$ P-curve
uses P[1/1] for $R_{\rm E}^{\rm can}$
and $P[0/2]$ for ${\widetilde p}_{\rm M+G}$.
The ${\cal O}(g^6)$ (P+BP)-curves use
BP[1/2] for $R_{\rm E}^{\rm can}$ and
P[0/3] for ${\widetilde p}_{\rm M+G}$ -- the lower curve is the
central curve of Fig.~\ref{pEMGeffvsTco}, the 
upper is the central curve of Fig.~\ref{pEMGeffmvsTco}.
In Fig.~\ref{pEMGconv} we see that the TPS results change strongly
when the order of the TPS is increased; on the contrary the resummed
results suffer weaker changes, although a clear convergence
at $T < 1$ GeV still cannot be seen at these orders.

The crucial point of our approach was to treat separately the
short-distance ($p_{\rm E}$)
and long-distance ($m_{\rm E}^2$, $g_{\rm E}^2$, $p_{\rm M+G}$)
quantities -- using in them the renormalization 
(and factorization) scales which correspond roughly to the physical scales
of the considered quantities, 
and then performing either Pad\'e or Borel-Pad\'e 
resummation of each quantity.
In this way we arrive at predictions for pressure which have
weak dependence on the unknown parameters $\delta_{\rm E}$ and
$\delta_{\rm G}$, and on the high-energy
renormalization scale $\mu_{\rm E}$ ($\sim 2 \pi T$)
of $p_{\rm E}$. On the other hand, the dependence of the predicted pressure
on the renormalization scale $\mu_m = \mu_{\rm M}$ ($\sim m_{\rm E}$)
of $p_{\rm M+G}$ is still large at $T \alt 1$ GeV  when the EQCD
$\lambda_{\rm E}^{(1)}/m_{\rm E}$-term in the EQCD
expansion (\ref{pMGeff}) is added separately as a power of
the QCD coupling parameter $g_s(\mu_m)$ [Eq.~(\ref{l1a})].
This $\mu_m$-dependence becomes significantly weaker
when the EQCD $\lambda_{\rm E}^{(1)}/m_{\rm E}$-term is instead
added separately as a power of the first EQCD coupling
parameter $g_{\rm E}^2/m_{\rm E}$ [Eq.~(\ref{l1b})].
Further, the resummed results change reasonably slowly when
the order of the TPS's, on which they are based, is increased.
  
On the other hand, simple TPS evaluations do not yield
reasonable results for the long-distance
quantities at low temperatures, because the TPS's at
such low energy scales have strong dependence
on the unknown parameters and on the renormalization scales,
and change strongly when the order of
the TPS's is increased.

\begin{acknowledgments}
We thank E.~Laermann, M.~Laine, and H.~Satz for helpful communication.
This work was supported in part by FONDECYT (Chile) 
grant No.~1010094 (G.C.)
\end{acknowledgments}

\appendix

\section{Pad\'e and Borel-Pad\'e approximants}
\label{app:PBP}

In this Appendix, we describe the procedure
we used to obtain the Pad\'e and Borel-Pad\'e
approximants on the bases of perturbation
expansions (\ref{mE}), (\ref{gE}), (\ref{pMG}),
(\ref{pMGeff}) and (\ref{pE2}).
In expansions (\ref{mE}), (\ref{gE}) and (\ref{pE2}),
the expansion parameter was $a({\overline \mu})
= \left[ g_s({\overline \mu})/ 2 \pi \right]^2$.
The canonical expansions of this type have the form:
\begin{equation}
S(a) = a \left( 1 + \sum_{n=1}^{\infty} r_n a^n \right) \ .
\label{Sa}
\end{equation}
Pad\'e approximant of order $(N,M)$ is
${\rm P}[N/M]_S(a) = P_N(a)/P_M(a)$, i.e., the
ratio of polynomials $P_N(a)$ and $P_M(a)$ of order $N$ and $M$, respectively.
The coefficients of these polynomials are determined
by the requirement that the re-expansion of the
approximant in powers of $a$ reproduce the
terms up to (and including) $\sim a^{N+M}$
in the expansion (\ref{Sa}). For example,
\begin{equation}
{\rm P}[1/1](a) = \frac{a}{(1 - r_1 a)} \ .
\label{P11}
\end{equation}
The Borel-Pad\'e approximation ${\rm BP}[N/M]_S(a)$
is constructed from expansion (\ref{Sa}) by 
applying Pad\'e ${\rm P}[N/M]_{B_S}(b)$ to the
expansion of the Borel-transform
\begin{equation}
B_S(b) = 1 + \sum_{n=1}^{\infty} \frac{r_n}{n! \beta_0^n} b^n \ ,
\label{BSb}
\end{equation}
and then performing the Borel integration
\begin{equation}
{\rm BP}[N/M]_S(a) = \frac{1}{\beta_0} \int_0^{\infty}
db \; \exp \left( - \frac{b}{\beta_0 a} \right)
{\rm P}[N/M]_{B_S}(b) \ .
\label{BPS}
\end{equation}
In principle, the integration here is along the
positive $b$-axis.
In order to avoid any possible poles on the positive
$b$-axis, the integration is in general performed
along a ray in the $b$ plane, say: $b = r \exp(- i \phi)$,
with $\phi \not= 0$ arbitrary and fixed; then
$r = |b|$ is integrated from zero to infinity, and the real part
of the result is taken -- cf.~Ref.~\cite{Cvetic:2001sn}. This leads,
by Cauchy theorem, to the Principal Value (PV)
prescription for integral (\ref{BPS}).

On the other hand, expansions (\ref{pMG}) and (\ref{pMGeff})
for ${\widetilde p}_{\rm M+G}$ have the following form:
\begin{equation}
T(g) = 1 + \sum_{n=1}^{\infty} t_n g^n \ ,
\label{Tg}
\end{equation}
where $g \equiv g_s(\mu_{\rm M})$ or $g \equiv g_{\rm E}^2/m_{\rm E}$.
Pad\'e approximant
${\rm P}[N/M]_T(g) = P_N(g)/P_M(g)$
was in this case simply the Pad\'e applied to 
expansion (\ref{Tg}).  
The Borel-Pad\'e approximation ${\rm BP}[N/M]_T(g)$
was constructed from expansion (\ref{Tg}) by 
applying Pad\'e ${\rm P}[N/M]_{{\widetilde B}_T}(z)$ to the
(expansion) of the following Borel transform of $T$:
\begin{equation}
{\widetilde B}_T(z) = 1 + 
\sum_{n=1}^{\infty} \frac{t_n}{n!} z^n  \ ,
\label{tBTz}
\end{equation}
and then performing the corresponding Borel integration
\begin{equation}
{\rm BP}[N/M]_T(g) = \frac{1}{g}\int_0^{\infty}
dz \; \exp \left( - \frac{z}{g} \right)
{\rm P}[N/M]_{{\widetilde B}_T}(z) \ .
\label{BPT}
\end{equation}
Again, in order to avoid the possible poles at $z > 0$,
the integration is performed along a ray in the $z$-plane
and the real part is taken.

\end{document}